\documentclass{emulateapj}
\usepackage{psfig}

\newcommand{\aas}{A\&ASS}
\newcommand{\alli}{\ensuremath{20.7\pm0.5}}

\newcommand{\tab}[1]{Table~\ref{#1}}
\newcommand{\fig}[1]{Figure~\ref{#1}}
\newcommand{\eqn}[1]{equation~(\ref{#1})}

\begin{document}

\title{Optical Morphologies of  Millijansky Radio Galaxies Observed by
{\it HST}\footnote{Based on observations made with the NASA/ESA Hubble
Space Telescope, obtained from the Data Archive at the Space Telescope
Science   Institute,  which   is  operated   by  the   Association  of
Universities for Research in  Astronomy, Inc., under NASA contract NAS
5-26555.} and in the {\it VLA} FIRST Survey }

\author{J. Russell\altaffilmark{2}, R. E. Ryan Jr.\altaffilmark{2}, S. H. Cohen\altaffilmark{3}, R. A. Windhorst\altaffilmark{3,2}, and I. Waddington\altaffilmark{4}}
\altaffiltext{2}{Dept. of Physics, Arizona State University, Tempe, AZ 85287--1504, USA}
\altaffiltext{3}{School of Earth and Space Exploration, Arizona State University, Tempe, AZ 85287--1404, USA}
\altaffiltext{4}{Astronomy Centre, University of Sussex, Brighton, BN1 9QH, UK}

\shorttitle{Radio Galaxy Morphologies}
\shortauthors{Russell et al.}
\email{jq@asu.edu}

\begin{abstract}

We  report on  a statistical  study of  the 51  radio galaxies  at the
millijansky  flux level  from the  Faint Images  of the  Radio  Sky at
Twenty centimeters, including their optical morphologies and structure
obtained  with the  Hubble Space  Telescope.  Our  optical  imaging is
significantly  deeper ($\sim\!2$~mag) than  previous studies  with the
superior angular resolution of space-based imaging.  We that find 8/51
(16\%) of the radio sources have no optically identifiable counterpart
to ${\rm  AB}\!\sim\!24$~mag.  For the  remaining 43 sources,  only 25
are  sufficiently resolved  in the  HST  images to  reliably assign  a
visual classification: 15 (60\%)  are elliptical galaxies, 2 (8\%) are
late-type spiral galaxies,  1 (4\%) is an S0,  3 (12\%) are point-like
objects (quasars), and 4 (16\%) are merger systems.  We find a similar
distribution of optical types with measurements of the S\'ersic index.
The  optical  magnitude  distribution   of  these  galaxies  peaks  at
$I\!\simeq\!\alli$~AB  mag, which is  $\sim\!3$~mag brighter  than the
depth  of our  typical HST  field and  is thus  not due  to  the WFPC2
detection  limit.   This  supports  the  luminosity-dependent  density
evolutionary  model,  where  the  majority  of  faint  radio  galaxies
typically  have $L^*$-optical  luminosities and  a median  redshift of
$z\!\simeq\!0.8$  with   a  relatively  abrupt   redshift  cut-off  at
$z\!\gtrsim\!2$.   We  discuss  our  results  in the  context  of  the
evolution of elliptical galaxies and active galactic nuclei.

\end{abstract}


\section{Introduction} \label{intro}

Faint  radio galaxies  are good  probes of  typical galaxies  at large
distances since they lack  a significant contribution from non-thermal
light   or   strong   emission   lines  in   their   optical   spectra
\citep[eg.][]{kro85,ben93,ham95a}.   Furthermore,  their selection  is
relatively  unaffected  by dust  absorption  when  these galaxies  are
identified  by their  synchrotron  emission.  In  the local  Universe,
powerful  radio  sources  are  primarily  found  in  giant  elliptical
galaxies,  which suggests  that radio  selection may  be  an efficient
method  of identifying high-redshift  ellipticals.  However,  at lower
radio  powers,  typically below  the  break  in  the radio  luminosity
function, there  is an increasing fraction of  star-forming and active
galaxies      \citep[such       as      Seyferts      and      LINERS;
eg.][]{con89,ben93,low97,hop03,bes05,afo06,mai08}.    Often   of   the
galaxy identification  is not  based on deep,  high-resolution optical
imaging, but rather on  broadband optical colors.  Such classification
schemes are only reliable to the extent to which these colors directly
map onto the Hubble sequence  at a given redshift.  Therefore, in this
paper  we focus  on the  optical morphologies  and structure  of radio
galaxies      with      fluxes     of      1~mJy$\lesssim\!S_{1.4~{\rm
GHz}}\!\lesssim\!100$~mJy  (we  will  denote $S_{1.4}$  as  integrated
radio  flux at  1.4~GHz  throughout) obtained  from  the Hubble  Space
Telescope (HST).

Early studies indicate that the majority of millijansky radio galaxies
typically  have  extended  radio  emission  with  optical  colors  and
luminosities     comparable    to     local     elliptical    galaxies
\citep[][]{kro85,win85,ben93,ham95a}.   The remainder  of  these radio
sources are generally  quasars or very blue galaxies,  which are often
interpreted as  late-type, starburst systems.  Like  the most powerful
radio sources, millijansky galaxies have a broad redshift distribution
of  $0\!\lesssim\!z\!\lesssim\!1.5$ \citep{ove03},  which  reflect the
increase of  active galaxies and  radio emission from  $z\!\sim\!0$ to
$z\!\sim\!2$ \citep[eg.][]{hop06}.   The space density  of millijansky
radio galaxies evolves strongly with redshift, which can be reproduced
by a simple evolutionary model.  \citet{con89} argue that the majority
of these radio galaxies reside in a ``shell'' of radius $z\!\sim\!0.8$
with thickness  of $\delta  z\!\sim\!0.8$.  In addition  to supporting
this   model,  \citet{wad01}   find   a  deficit   of  high   redshift
($z\!\gtrsim\!2$) millijansky radio galaxies  in the Hercules field of
the Leiden-Berkley Deep Survey  (LBDS).  They interpret this result as
an  effective redshift ``cut-off,''  which varies  from $z\!\simeq\!2$
for the most  luminous radio sources, to $z\!\simeq\!1$  for the lower
luminosity galaxies.

Millijansky radio sources are intrinsically rather luminous at optical
wavelengths,  with absolute  magnitudes in  a relatively  narrow range
($\pm\!0.8$~mag)    centered    at   $M_{r^*}\!\simeq\!-22$~mag    and
$M_{r^*}\!\simeq\!-23$~mag  for blue  and  red galaxies,  respectively
\citep{mac00,ive02}.   From photographic plate  and deep  CCD imaging,
\citet{win84b} and \citet{wad00} found the that the apparent magnitude
distribution from the LBDS  peaks around $I\!\simeq\!22$~mag, which at
$z\!=\!0.8$ corresponds to $M_B\!\simeq\!-21.5$~mag.  While there is 
a good deal of deep ground-based imaging on millijansky radio sources, 
there is little high-resolution HST imaging.  Early works with HST 
are generally based on at most a  few objects \citep{pas96,wad99,wad02}.

In the microjansky flux range, \citet{afo06} identified 64 faint radio
galaxies of which  57 had optical counterparts in  the Advanced Camera
for  Surveys (ACS)  imaging of  the Great  Observatories  Origins Deep
Survey,  South   \citep[GOODS-S;][]{gia04}.   The  median  photometric
redshift of the optically identified microjansky radio sources is also
$z\!\sim\!0.8$.   Since the  redshift distribution  of  radio galaxies
changes    only     very    slowly    for     radio    fluxes    below
$S_{1.4}\!\simeq\!1$~Jy   \citep{win90},   this   median   photometric
redshift is  consistent with the evolutionary  model of \citet{con89}.
Based on  their $X$-ray  properties and limited  optical spectroscopy,
the  majority of these  microjansky radio  galaxies are  either active
galactic   nuclei  (AGN)   or  star-forming   galaxies  \citep{afo06}.
Furthermore,  the   optical  counterparts  generally   have  disturbed
morphologies, which suggests many are in the later stages of merging.

In this  paper, we investigate  the optical properties  of millijansky
radio  galaxies selected from  the Faint  Images of  the Radio  Sky at
Twenty centimeters  \citep[FIRST;][]{bec95}.  The FIRST  survey covers
$\simeq\!10^4$  square degrees  of the  north Galactic  cap to  a 95\%
completeness  limit  of   $\simeq\!2.8$~mJy.   In  order  to  reliably
determine   the  morphology   of   these  small   and  faint   optical
counterparts,  we  require   the  angular  resolution  of  space-based
imaging.  Therefore, our  sample is derived from the  overlap of FIRST
sources  with  the  existing   HST-WFPC2  Archive  as  of  2002.   All
magnitudes  quoted herein  are in  the AB  system  \citep{oke83}. This
paper  is  organized  as  follows:  In \S~\ref{proc}  we  outline  the
critical   steps  to   define   our  final   millijansky  sample,   in
\S~\ref{results}   we   discuss   our   four   primary   results,   in
\S~\ref{comment}  we provide general  comments on  individual objects,
and in \S~\ref{sum} we conclude  with a general summary and additional
discussion of our results.

\section{Procedure} \label{proc}
\subsection{Data Selection} \label{data}
We begin by correlating the  positions from the FIRST catalog with the
HST-WFPC2  Archive.  Out  of $\gtrsim\!10^4$~WFPC2  images,  we obtain
$\sim\!850$~potential frames  with FIRST sources.   These WFPC2 images
were stacked and cosmic-ray  cleaned following the method discussed by
\citet{coh03}  to  yield  a  sample of  177~optical  candidates.   The
HST-WFPC2  dataset contains  all the  archival F606W  ($V$)  and F814W
($I$) images with exposure times longer than 160~seconds, however only
a  single band  is  generally available  for  a given  target, as  the
primary goals of  the HST images.  Roughly 50\% of  the WFPC2 data are
observed as HST parallel fields,  which are images taken by WFPC2 when
a different instrument was  observing the primary target.  In general,
these millijansky radio sources are extremely difficult to image since
they   are  typically   optically  faint   ($I\!\gtrsim\!20$~mag)  and
distributed   over   degree   scales.    Therefore,   these   parallel
observations  provide  an efficient  method  to  optically image  this
elusive population.

\subsection{Secondary Selection Criteria}\label{ssc}

Since the motivation of this work is to study the optical morphologies
and structure of a well  defined, {\it random}, and complete sample of
weak radio galaxies  with the superior resolution of  HST, many of the
candidates must be eliminated from the sample.  There are four reasons
what we  could exclude a given  field: (1) Major  WFPC2 image defects,
such the majority of the pixels are saturated from nearby stars; (2) a
significant  fraction of  the  field-of-view is  covered  by a  nearby
object,  such as  an  NGC-type galaxy,  which  would prevent  reliable
optical flux  measurements and morphological  classifications; (3) the
WFPC2 image  was centered on a  dense star-field or  galaxy cluster to
eliminate ambiguity  in the optical identifications; or  (4) the radio
source was  {\it targeted} by  HST.  In this  way, we can  subselect a
complete  sample  of  background  objects  from a  non-random  set  of
foreground observations.   Furthermore, owing to  the relatively small
WFPC2  field-of-view,  most  FIRST  radio sources  are  generally  the
primary target.  In \tab{redsum}, we summarize  the eliminated fields.
The  selection effects  of  the  various HST  programs  is not  always
possible to  quantify based on the  HST target lists  and PI proposals
available  in the  STScI/HST Archive.   In \tab{target},  we  list the
targeted radio  sources that were removed to  reduce the contamination
of  preselected  objects in  constructing  our  random sample.   These
criteria provide  a compromise between the  resulting completeness and
reliability of  the final sample.   The majority of the  radio sources
are brighter than  the 95\% completeness limit of  the FIRST survey of
$S_{1.4}\!\simeq\!2.8$~mJy.  From the  177 FIRST candidates, our final
sample contains 51 radio galaxies with HST-WFPC2 observations.

\begin{table}
\caption{Sample Reduction Summary}\label{redsum}
\begin{tabular*}{0.48\textwidth}{@{\extracolsep{\fill}}lr}
\hline
\hline
\multicolumn{1}{c}{Reason} & \multicolumn{1}{c}{Number}\\
\hline
Radio Sources In Final Sample                                        &      51 \\
Excluded because of Pointed HST Observations                         &      43 \\
Excluded because NGC or Nearby Contamination                         &      32 \\
Excluded because Incomplete in Radio Flux Cutoff                     &      47 \\
Misc\tablenotemark{a}                                                &       4 \\ 
\hline
Total                                                                &     177 \\ 
\hline
\tablenotetext{a}{This category includes fields where the target is uncertain due to a galaxy group or cluster, saturated WFPC2 fields, and/or local WFPC2 defects precluding inclusion in the optical sample.}
\end{tabular*}
\end{table}

\begin{table*}
\caption{Targeted Sources Removed from Sample}\label{target}
\begin{tabular*}{0.98\textwidth}{@{\extracolsep{\fill}}lrrcl}
\hline
\hline
\multicolumn{1}{c}{Radio Source\tablenotemark{a}} & \multicolumn{1}{c}{RA J2000} & \multicolumn{1}{c}{Dec J2000} & \multicolumn{1}{c}{$\log{S_{1.4}}$} & \multicolumn{1}{c}{WFPC2 Target}\\
\multicolumn{1}{c}{$ $} & \multicolumn{1}{c}{(deg)} & \multicolumn{1}{c}{(deg)} & \multicolumn{1}{c}{(mJy)} & \multicolumn{1}{c}{(targname)\tablenotemark{b}}\\
\hline
J001303.7--012305  &    3.265625 & --1.384777  &  0.61  &   CL-ULIR00105-0139            \\               
J022448.1--000711  &   36.200542 & --0.120111  &  0.04  &   GAL-CLUS-022448-000717            \\         
J074123.0+320810  &  115.345963 &  32.136360  &  0.60  &   QDOTULIR073810+321511             \\         
J082209.5+470552  &  125.539925 &  47.098114  &  1.72  &   A646            \\                           
J082313.7+275120  &  125.807213 &  27.856083  &  1.03  &   QDOTULIR082010+280119            \\          
J083803.6+505509  &  129.515091 &  50.919613  &  0.88  &   QDOTULIR0838+5055             \\             
J090850.9+374818  &  137.211304 &  37.805500  &  3.06  &   3C217              \\                        
J094308.4+465700  &  145.785828 &  46.950359  &  0.20  &   GAL-CLUS-093942+47            \\             
J095254.3+435348  &  148.226196 &  43.896782  &  0.85  &   GAL-CLUS-0952+44             \\              
J095644.7+493034  &  149.187088 &  49.513168  &  0.68  &   A895               \\                        
J102326.2+124855  &  155.859283 &  12.815556  &  0.22  &   ABELL999-BCG            \\                   
J103141.7+350231  &  157.924042 &  35.044613  &  1.64  &   A1033            \\                           
J105648.8--033727  &  164.203873 & --3.623805  &  0.82  &   GAL-CLUS-105427-032125-POS1          \\
J105659.5--033727  &  164.248123 & --3.624416  &  1.09  &   GAL-CLUS-105427-032125-POS5            \\      
J110050.3+104654  &  165.211929 &  10.783639  &  1.06  &   AO1058+11             \\                      
J115106.8+550443  &  177.778549 &  55.078945  &  0.95  &   UGC06823              \\                      
J122906.7+020308  &  187.277954 &   2.052416  &  4.57  &   PG1226+023             \\                     
J123357.1+074206  &  188.488297 &   7.701722  &  0.87  &   NGC4526             \\                        
J125929.2+361713  &  194.875290 &  36.284283  &  2.17  &   6C1257+36            \\                       
J131128.9--012116  &  197.870880 & --1.354444  &  0.36  &   ABELL1689               \\                    
J131130.0--012028  &  197.875122 & --1.341333  &  0.94  &   A1689-10             \\                       
J131131.5--011931  &  197.880920 & --1.325666  &  1.63  &   ABELL1689              \\                      
J131132.6--011959  &  197.886169 & --1.333000  &  0.39  &   A1689-10              \\                      
J133238.3+503335  &  203.159988 &  50.560001  &  0.78  &   A1758G7              \\                       
J133239.4+503432  &  203.164673 &  50.575615  &  0.77  &   A1758G7              \\                       
J133557.2+544338  &  203.987244 &  54.726387  &  1.09  &   MS1333.9+5500             \\                  
J134337.1--001525  &  205.904648 & --0.257166  &  0.74  &   GAL-CLUS-134339-001349              \\         
J134447.4+555410  &  206.197723 &  55.903053  &  0.10  &   IR13428+5608.0              \\                
J134733.3+121724  &  206.889297 &  12.290195  &  3.69  &   PKS1345+12-PSF              \\                 
J135609.9+290535  &  209.041580 &  29.093473  &  1.03  &   NGWULIR1353+2920              \\              
J140318.1+542157  &  210.824783 &  54.365974  &  1.14  &   NGC5457-FLD1             \\                    
J141721.3+132429  &  214.339127 &  13.408417  &  0.37  &   MS1414.9+1337             \\                  
J142357.6+383247  &  215.990982 &  38.546196  &  1.02  &   FSC142118+3845              \\                 
J142553.5+374805  &  216.473129 &  37.801472  &  0.27  &   A1914              \\                          
J143242.8+245614  &  218.172867 &  24.921638  &  2.09  &   B2+1430+25              \\                     
J144516.4+095836  &  221.318619 &   9.976666  &  3.38  &   1442+101              \\                      
J151002.9+570243  &  227.512466 &  57.045555  &  2.41  &   GB1508+5714              \\                    
J160218.2+155912  &  240.576035 &  15.986777  &  0.87  &   ABELL2147-BCG             \\                  
J162124.6+381008  &  245.353226 &  38.169422  &  0.75  &   R1621+38              \\                      
J162439.6+234524  &  246.162979 &  23.753334  &  3.15  &   GAL-CLUS-3C336-POS2             \\             
J162548.7+264658  &  246.453186 &  26.782833  &  0.99  &   Q1623+268              \\                      
J164658.9+454824  &  251.745468 &  45.806778  &  0.74  &   QDOTULIR1645+4553             \\               
J231713.1--110034  &  349.305084 &--11.009694  &  0.82  &   CL-ULIR23146-1117              \\ 
\hline
\tablenotetext{a}{The name of the source as per IAU recommendations, as used by FIRST.  FIRST Jhhmmss.s+ddmmss in which the coordinates are equinox J2000.0 and are truncated \citep{bec95}.}
\tablenotetext{b}{This is the header keyword for the field target from the WFPC2 images.}
\end{tabular*}
\end{table*}

\subsection{Astrometric Correction of HST Fields} \label{astro}

\begin{figure}
\epsscale{1.0}
\plotone{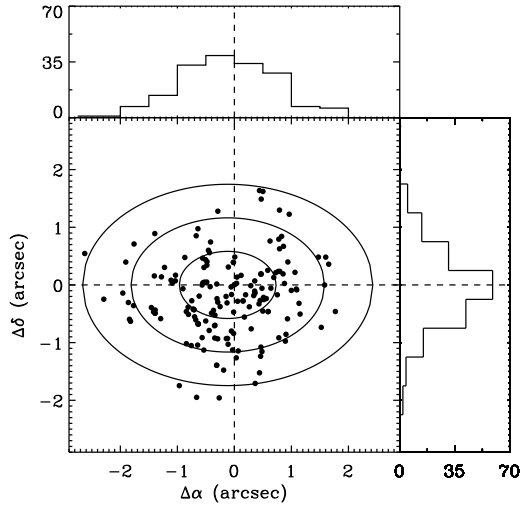}
\caption{Optical  astrometric  corrections   in  RA  and  Dec.   These
relative offsets are computed as the positional difference between the
HST-WFPC2 and  USNO coordinates for  multiple stars in a  given field.
Both distributions  are roughly Gaussian with mean  values of $(\Delta
\alpha,\Delta    \delta)\!=\!(-0\farcs1,0\farcs0)$,    with   standard
deviations       $(\sigma_{\alpha}^{\rm      opt},\sigma_{\delta}^{\rm
opt})\!=\!(0\farcs85,0\farcs55)$.    The   ellipses    represent   the
$1\sigma$,  $2\sigma$,  and   $3\sigma$  contours.   These  systematic
offsets  and  uncertainties are  used  to  determine  the most  likely
optical    identification   for    radio    galaxies   discussed    in
\S~\ref{imprep}.}\label{scatter}
\end{figure}

Due  to technical  issues in  the  the world  coordinate system  (WCS)
generated in  the HST-WFPC2 pipeline,  the astrometry keywords  in the
FITS  headers  of WFPC2  fields  prior to  1997  September  15 may  be
unreliable \citep{bir00}.  To better calibrate the astrometry of these
frames,  we match the  objects in  each field  with the  United States
Naval  Observatory (USNO)  A2.0  catalog \citep{mon96},  which is  the
standard  in the  FK5 system.   We  derive shifts  in Right  Ascension
($\Delta \alpha$) and Declination ($\Delta\delta$) from the positional
differences between  the HST-WFPC2 images  and USNO catalog,  which we
show  in \fig{scatter}.   These  offsets are  individually applied  to
their respective  WFPC2 frame.  However, several  fields lacked usable
USNO  sources, and  were  left uncorrected.   In  general, these  were
high-latitude  fields or  contained only  saturated stars,  which made
accurate  determination  of   their  centers  unreliable.   While  the
dispersion  in  $\Delta\delta$ is  not  significantly  higher for  the
pre-1997  data, it  does increase  by a  factor of  $\sim\!4$  for the
dispersion  in $\Delta\alpha$.   The applied  shifts roughly  follow a
normal   distribution   centered   on   $\left<(\Delta   \alpha,\Delta
\delta)\right>\!=\!(-0\farcs1,0\farcs0)$  with standard  deviations of
$(\sigma_{\alpha}^{\rm                        opt},\sigma_{\delta}^{\rm
opt})\!=\!(0\farcs85,0\farcs55)$.  This mean offset gives a systematic
uncertainty   of  $\simeq\!0\farcs1$,   while  the   $1\sigma$  random
uncertainty is $\simeq\!1\farcs0$.

\subsection{WFPC2 Image Preparation and Investigation} \label{imprep}

\begin{figure*}
\centerline{\hbox{
\psfig{file=  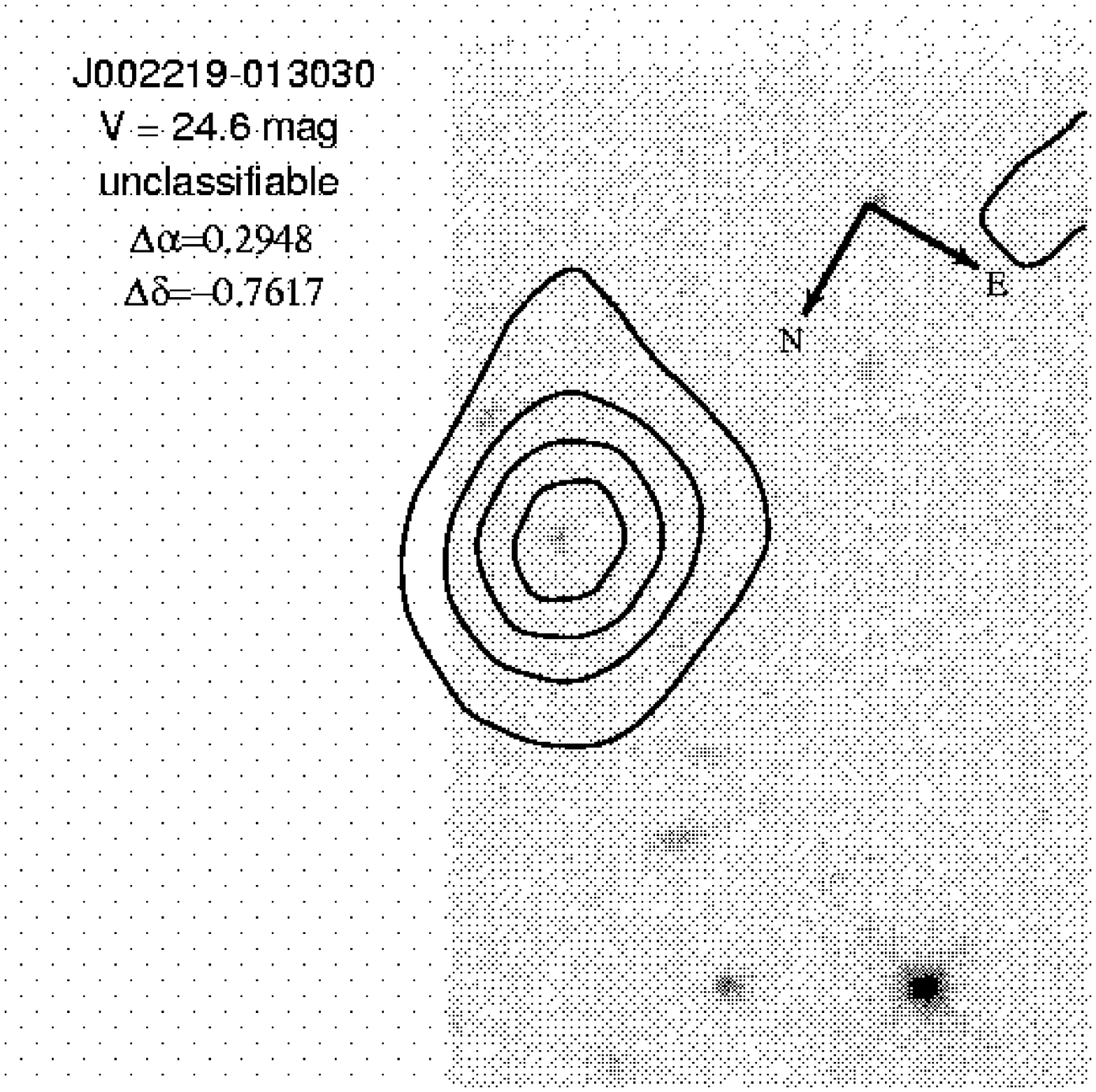 ,width=2.6truein}
\psfig{file=  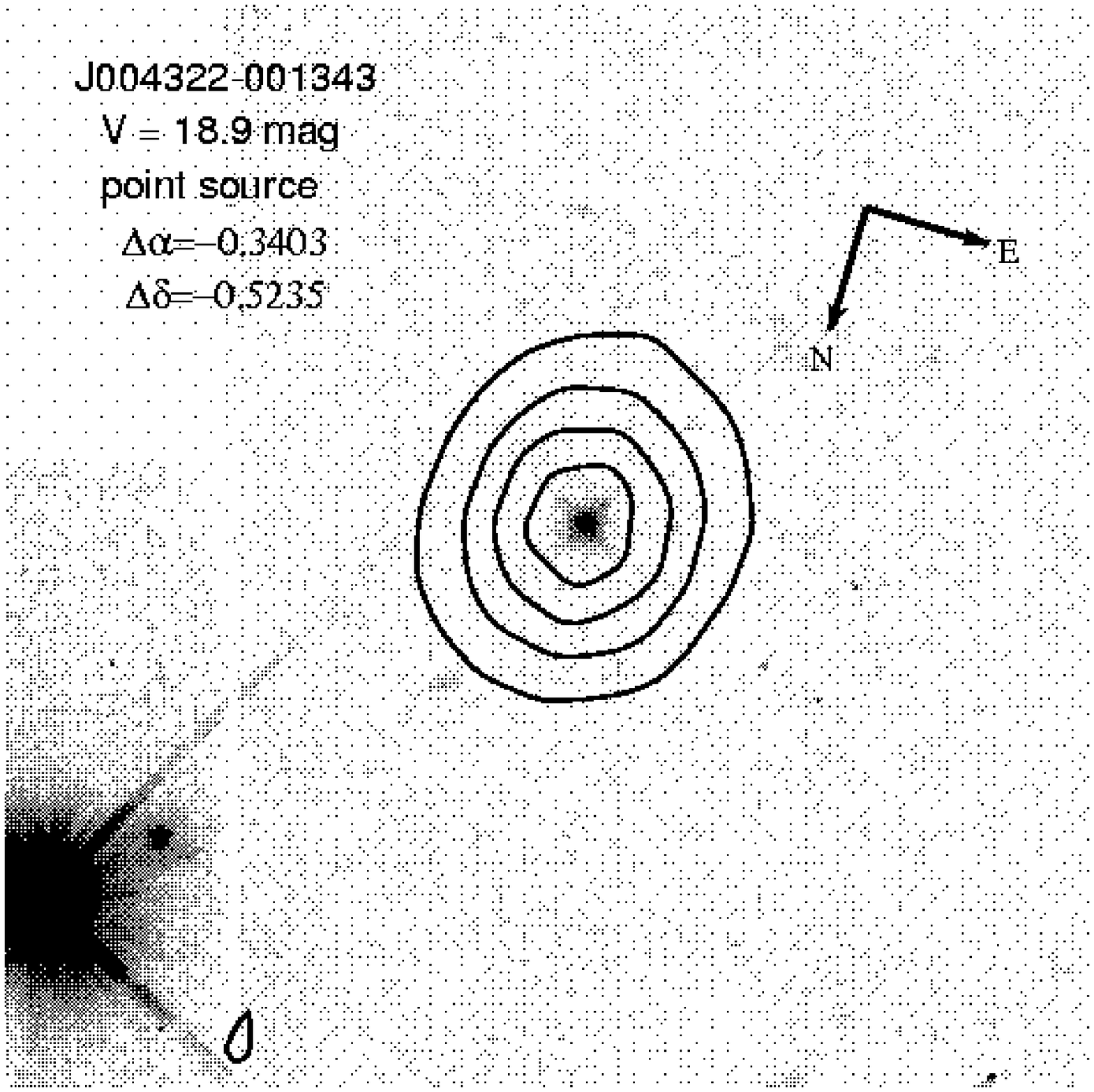 ,width=2.6truein}}}
\centerline{\hbox{
\psfig{file=  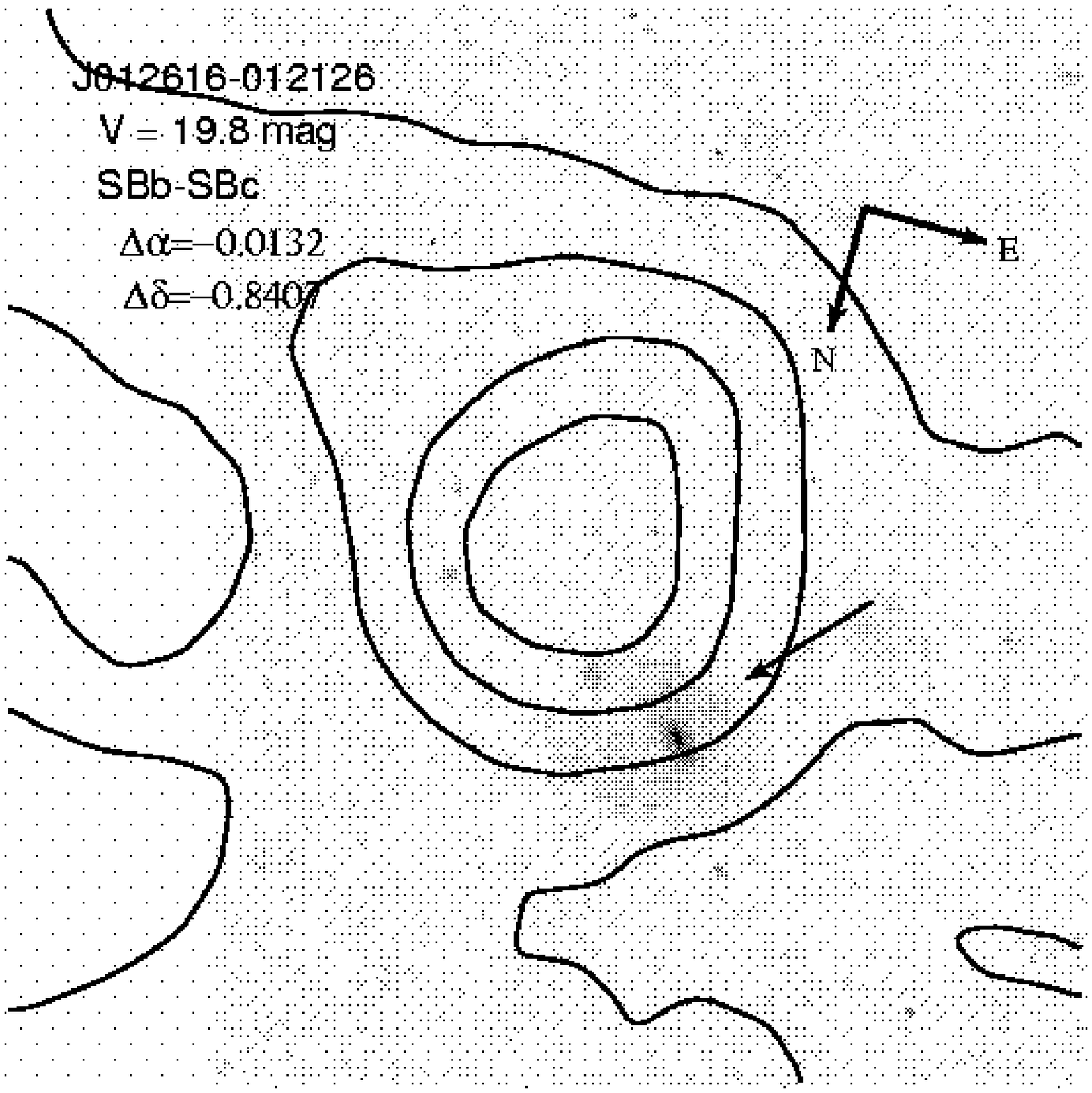 ,width=2.6truein}
\psfig{file=  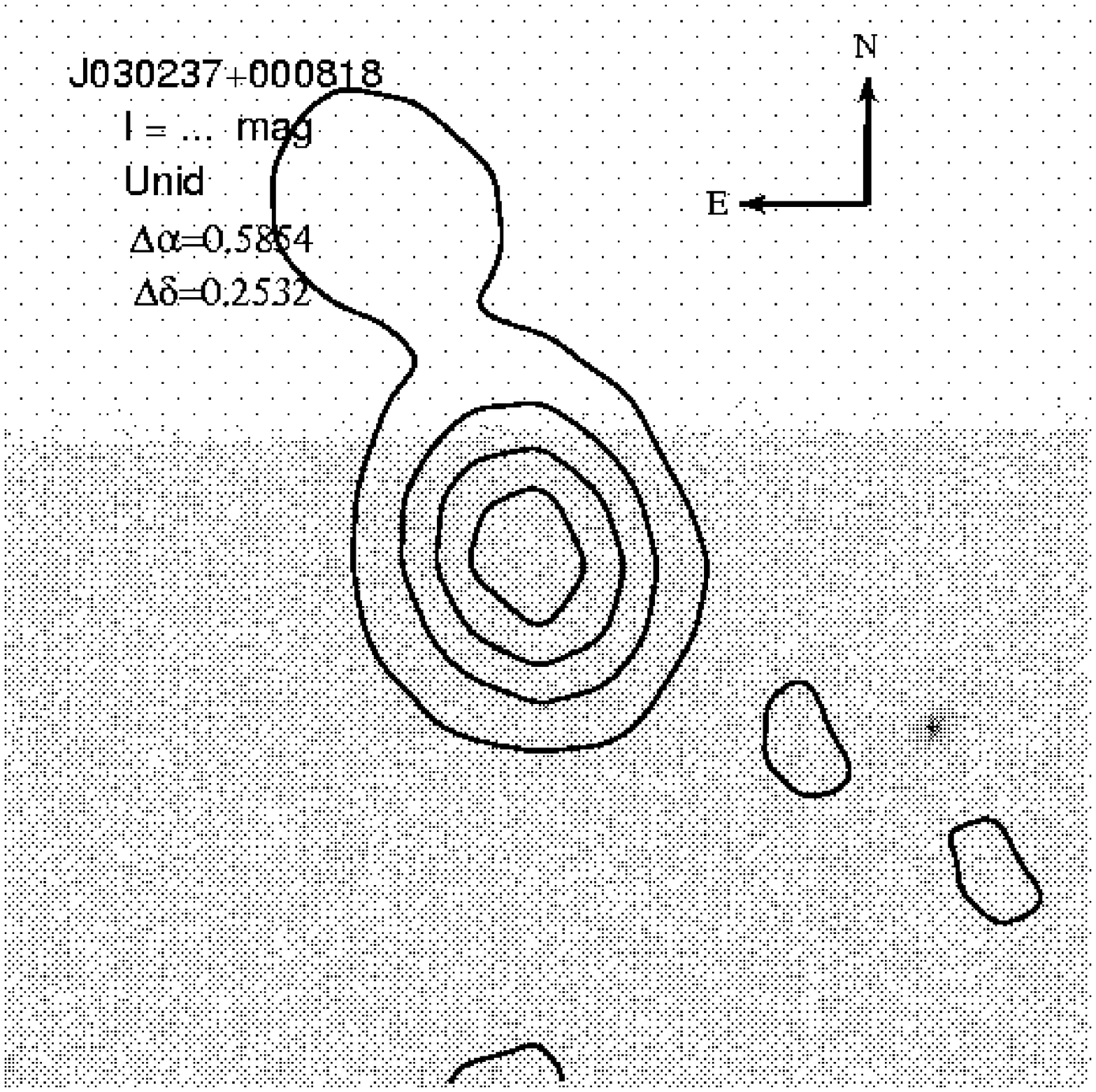 ,width=2.6truein}}}
\centerline{\hbox{
\psfig{file=  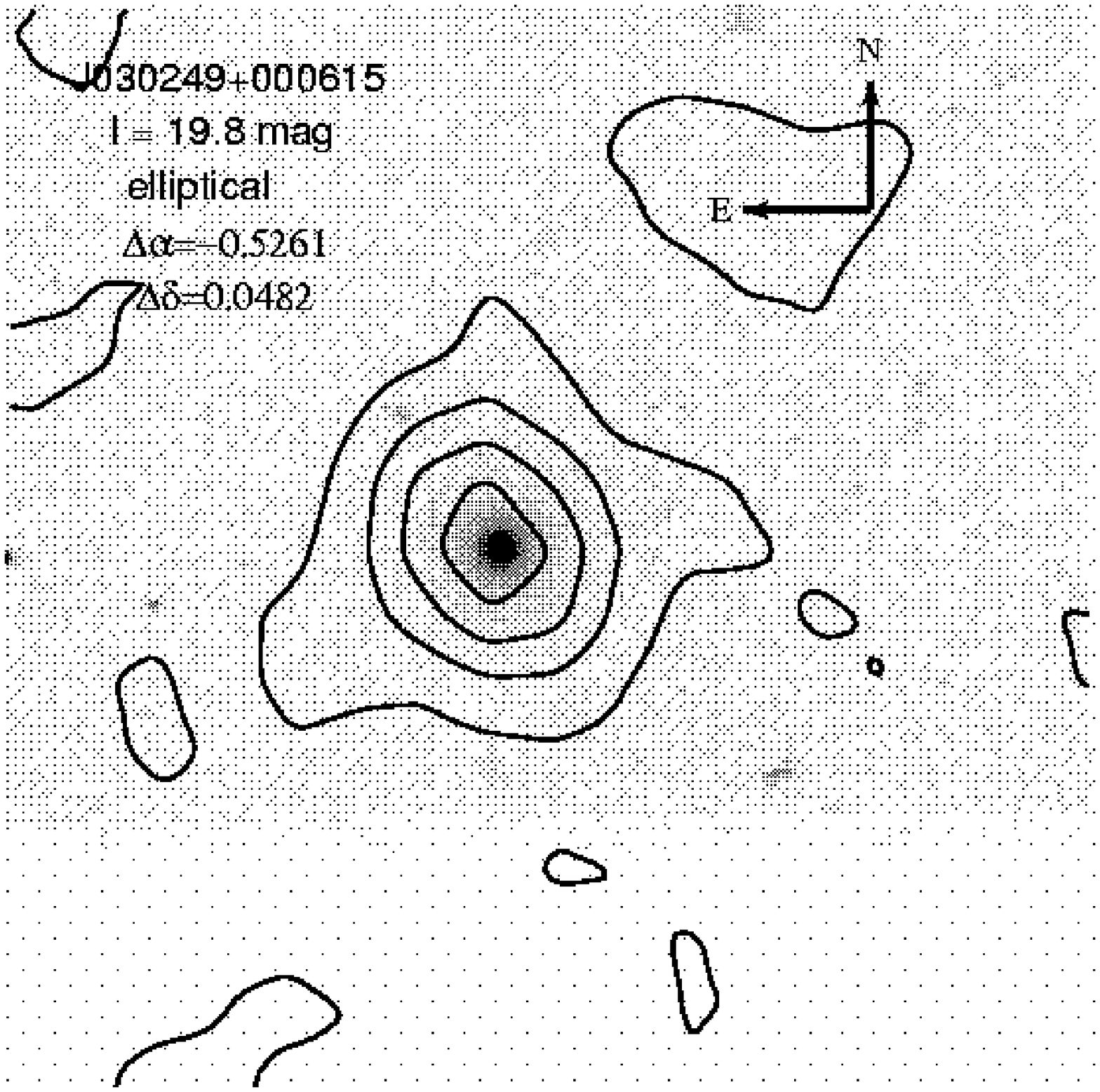 ,width=2.6truein}
\psfig{file=  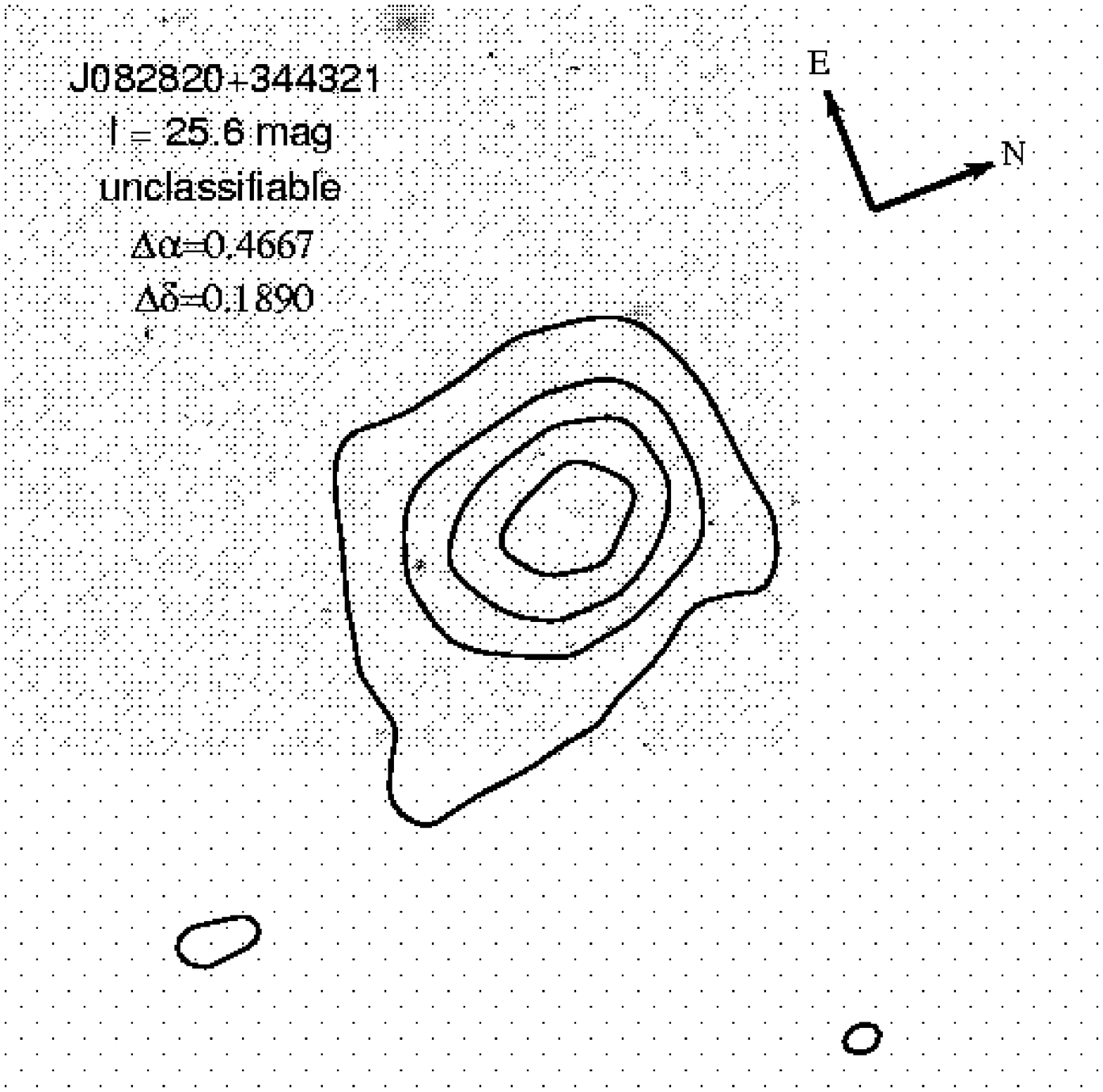 ,width=2.6truein}}}
\caption{WFPC2 images  of our  60 millijansky radio  sources.  Objects
not located  in the middle of  the radio contours are  indicated by an
arrow.  In  the upper left  of each stamp,  we list the name  from the
FIRST catalog, the HST-WFPC2  optical magnitude in the given bandpass,
the  visually-defined morphology, and  shifts in  RA and  Dec ($\Delta
\alpha$ and $\Delta\delta$) defined  in \S~\ref{astro}.  Each stamp is
$30''\times30''$.  {\bf  Higher  resolution  images are  available  by
request.} \label{stamps}}
\end{figure*}

We  excise  300$\times$300~pixel  ``postage  stamp'' images  from  the
astrometrically corrected  and stacked WFPC2 mosaics  centered on each
FIRST source position.   In \fig{stamps}, we show the  51 WFPC2 stamps
with FIRST radio  contours placed at $2^n$ multiples  of 0.31~mJy, and
list  the relevant  photometric and  morphological information  in the
upper-left corner.   Optical counterparts  to the given  radio sources
are identified by the spatial  alignment between the optical image and
the center of  the radio contours. Since several  objects have complex
radio  morphologies  (such  as  elongated or  double  isophotes),  the
identification of  each most likely  optical counterpart was  for such
objects done  on a case-by-case basis.   For the FIRST  sources with a
double-lobed morphology, we adopt  the optical object which is nearest
to the geometric center of the lobes as the most likely identification
\citep[eg.][]{mac71}.  If the  optical counterpart is considerably off
the line which connects the two lobes, then we use the the orientation
and outer isophotes to aide in the identification (such as in the case
of  J094930.7+295938).   Both  lobes   of  double  radio  sources  are
confirmed to not separately contain optical counterparts.  Stamps with
no optical object within the radio contours (complete to 2.8~mJy) were
deemed  {\it  unidentified} radio  sources  (hereafter,  Unid) to  the
limiting optical magnitude cataloged in \tab{unidtab}.  For identified
optical counterparts, we assign visual types and measure their surface
brightness  profiles (see  \S~\ref{morph}), however  the  faintest and
smallest    HST    objects     are    generally    not    classifiable
\citep{ode96,coh03}.

\begin{figure*}
\centerline{\hbox{
\psfig{file=  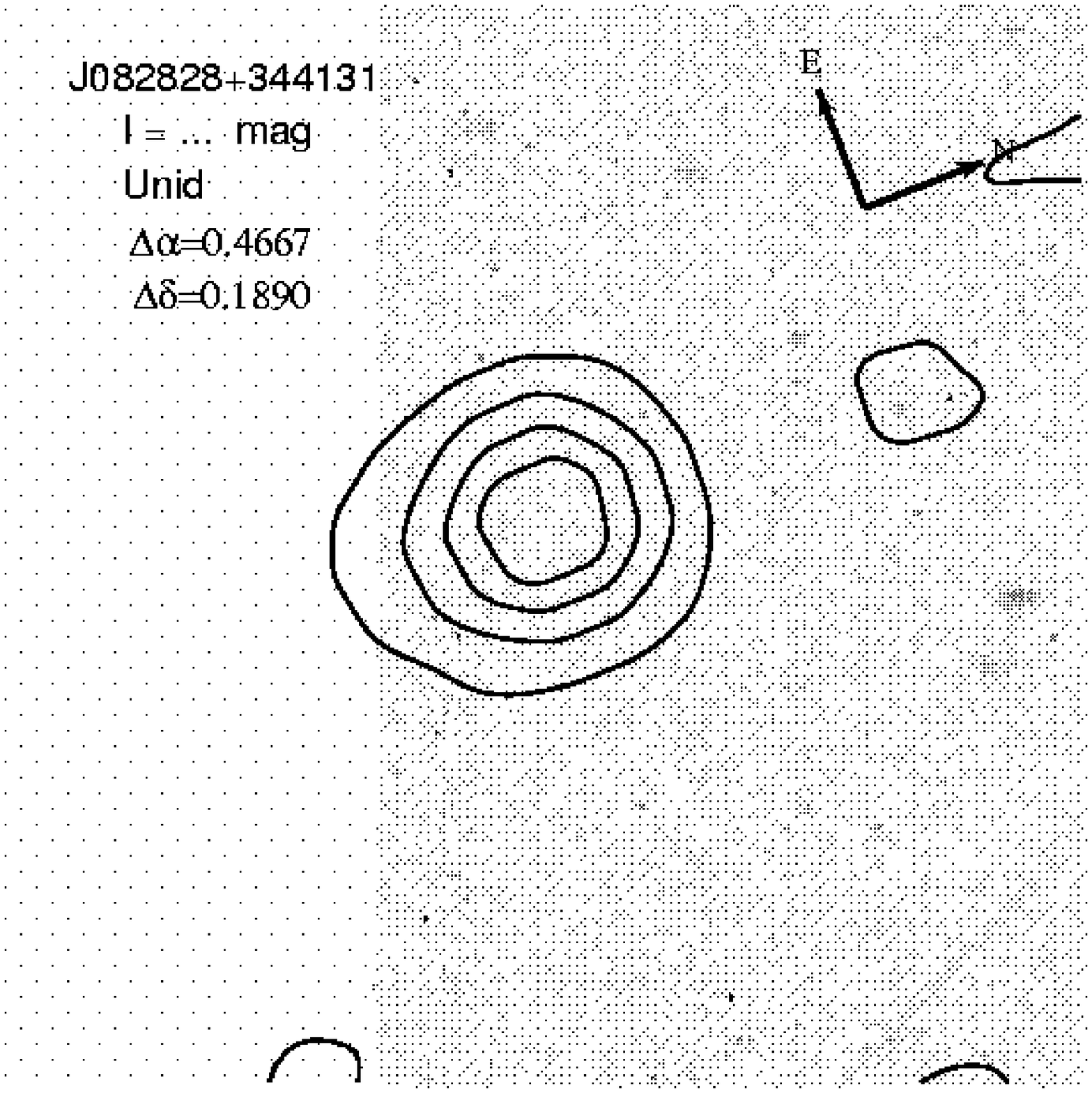 ,width=2.6truein}
\psfig{file=  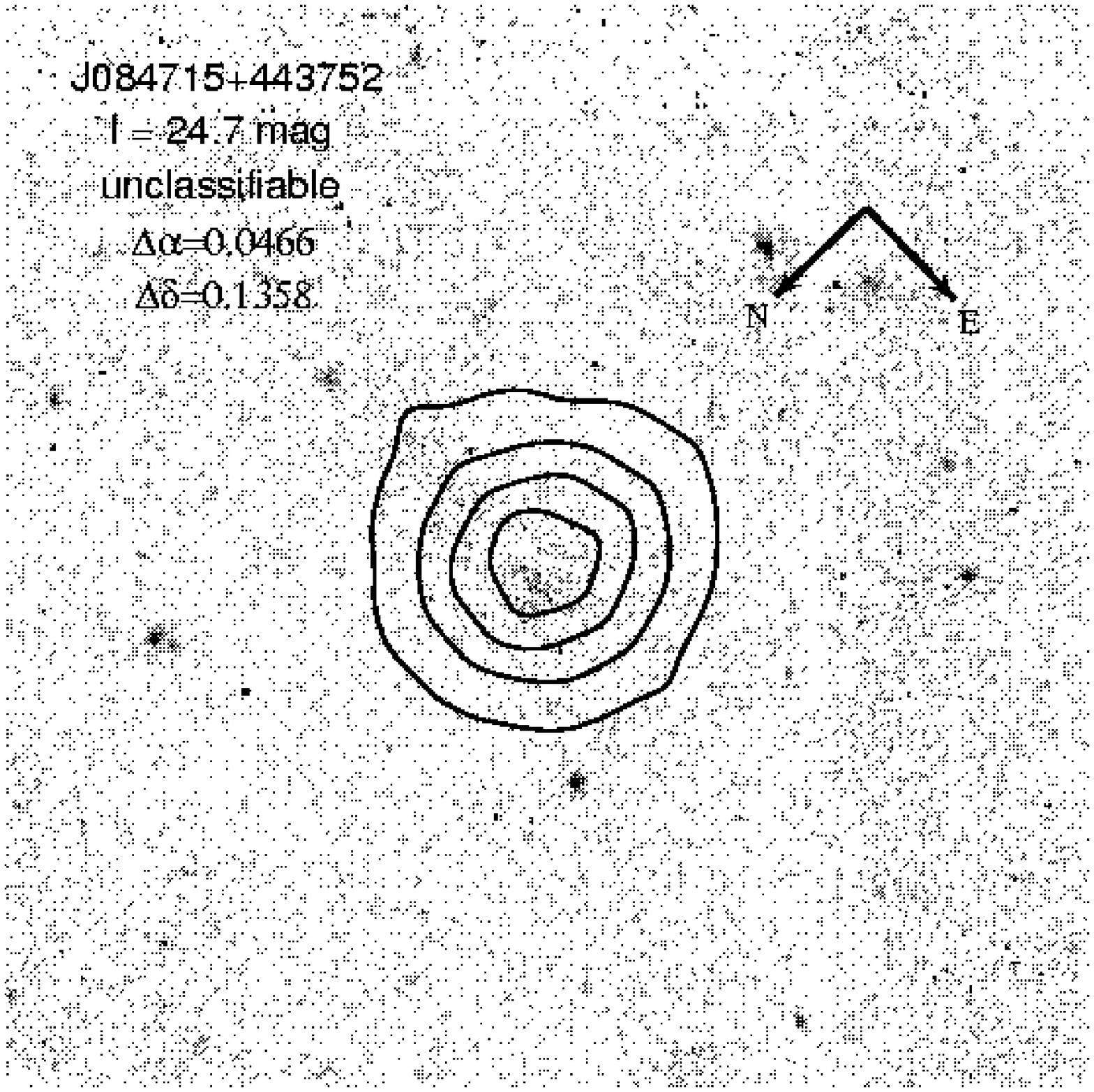 ,width=2.6truein}}}
\centerline{\hbox{
\psfig{file=  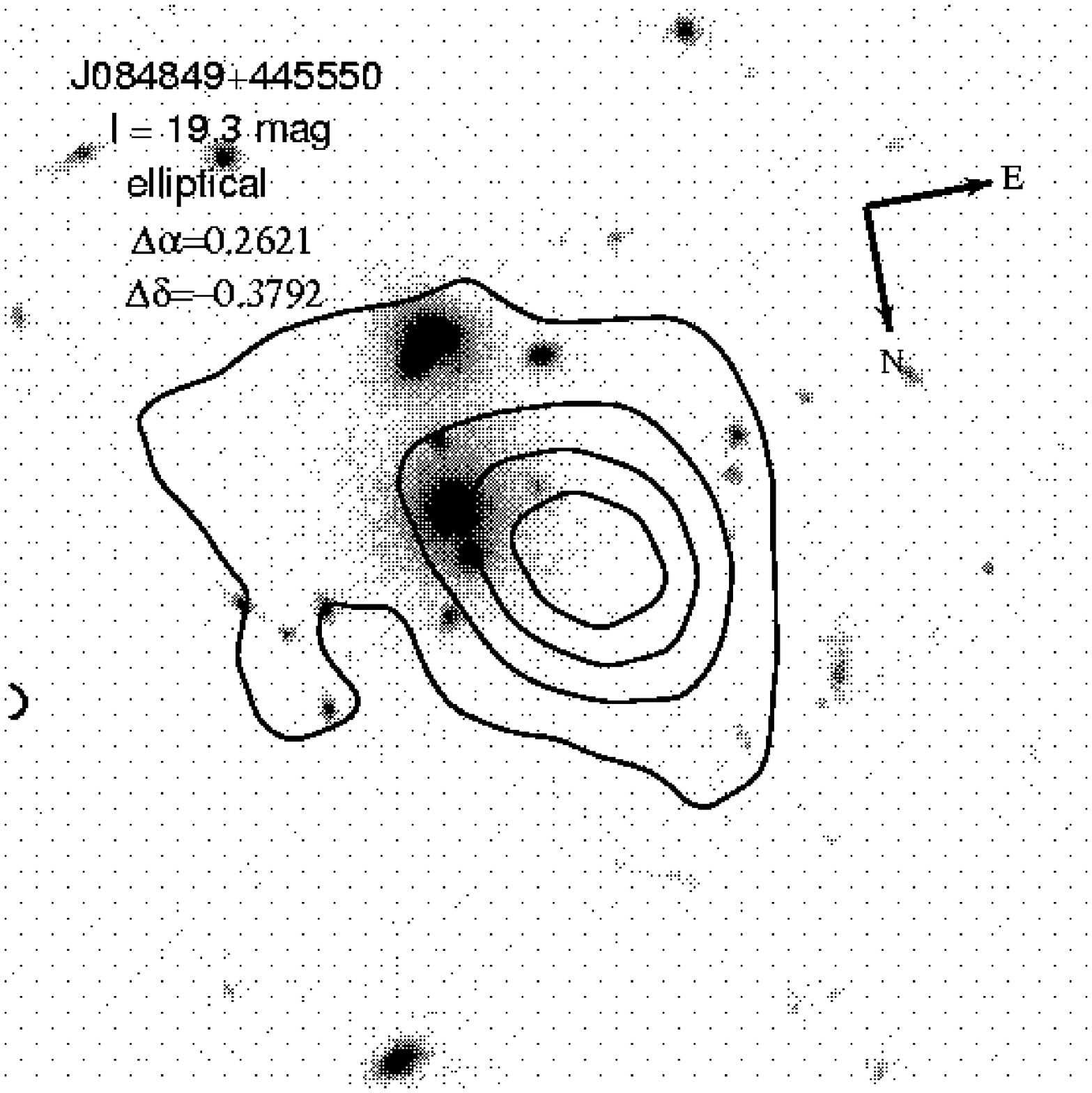 ,width=2.6truein}
\psfig{file=  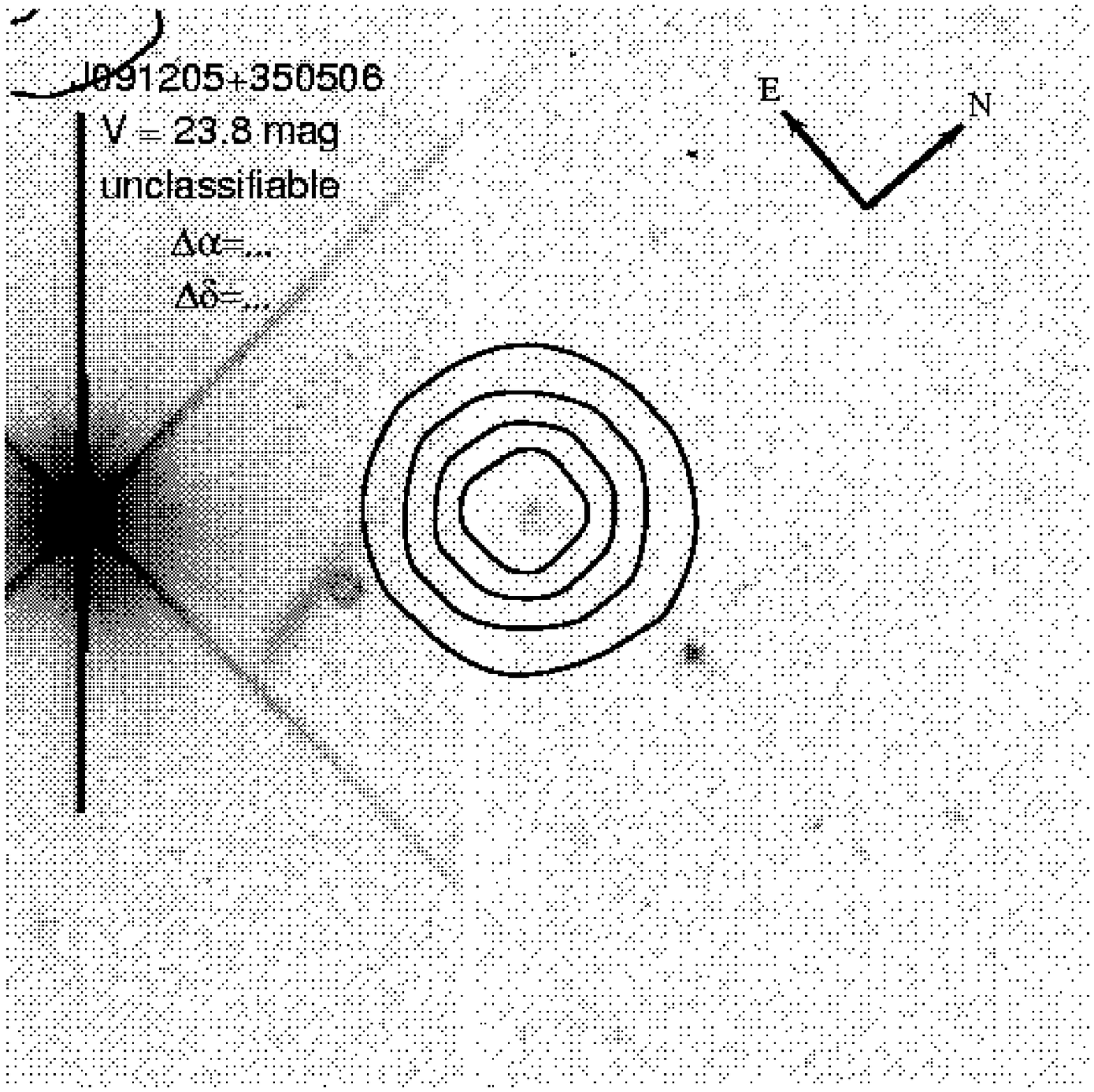 ,width=2.6truein}}}
\centerline{\hbox{
\psfig{file=  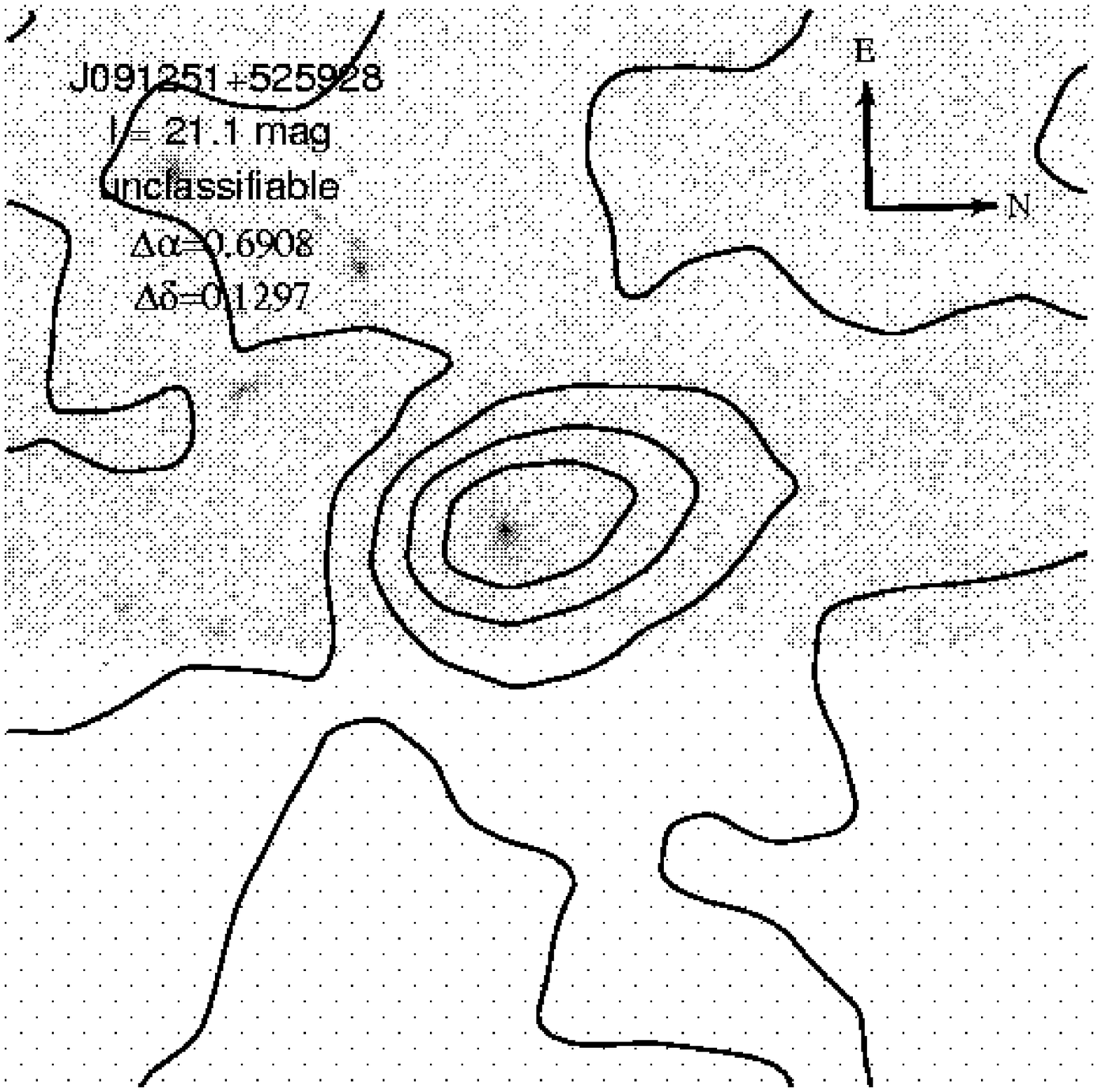 ,width=2.6truein}
\psfig{file=  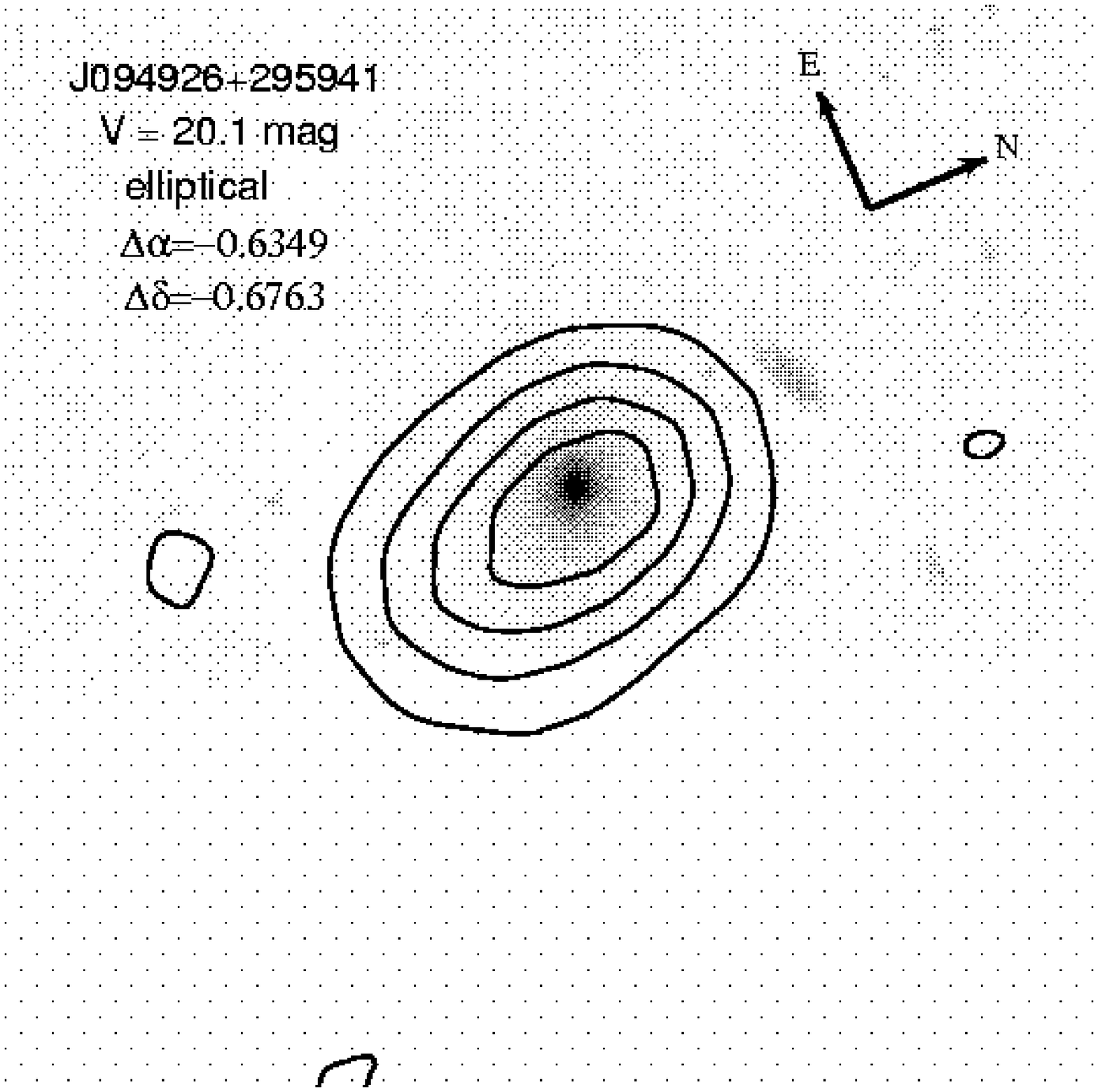 ,width=2.6truein}}}
\end{figure*}
\begin{figure*}
\centerline{\hbox{
\psfig{file=  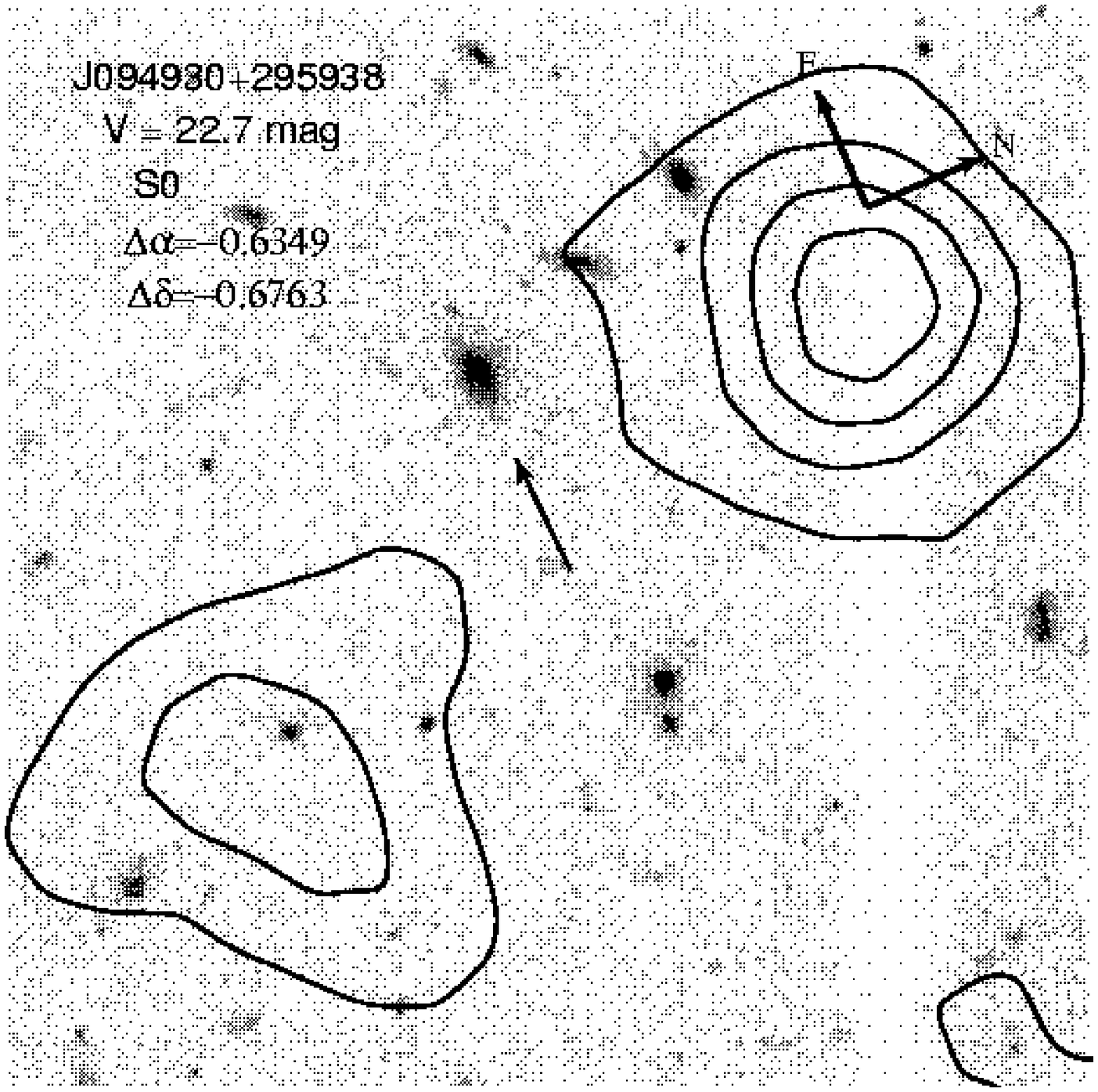 ,width=2.6truein}
\psfig{file=  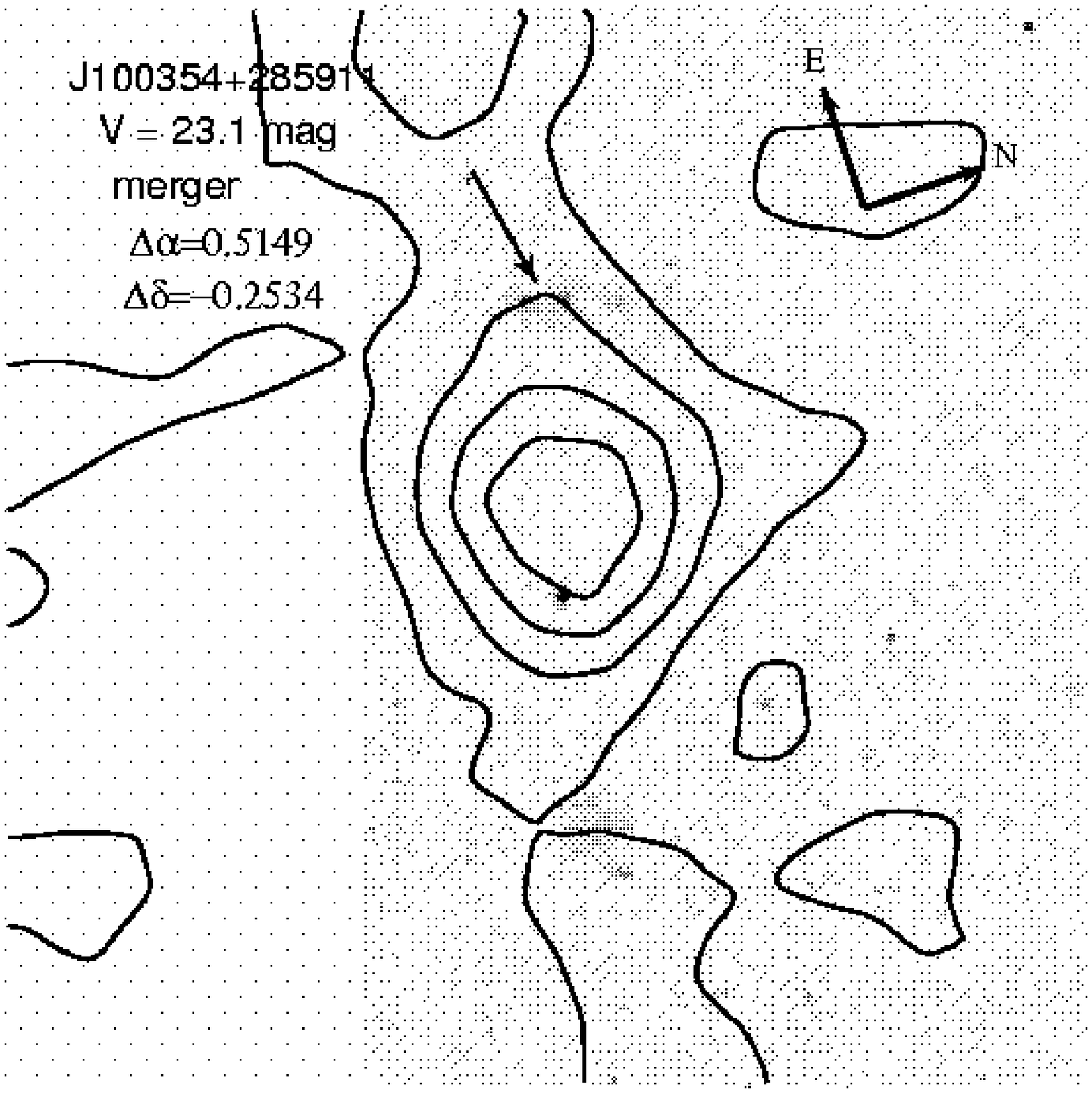 ,width=2.6truein}}}
\centerline{\hbox{
\psfig{file=  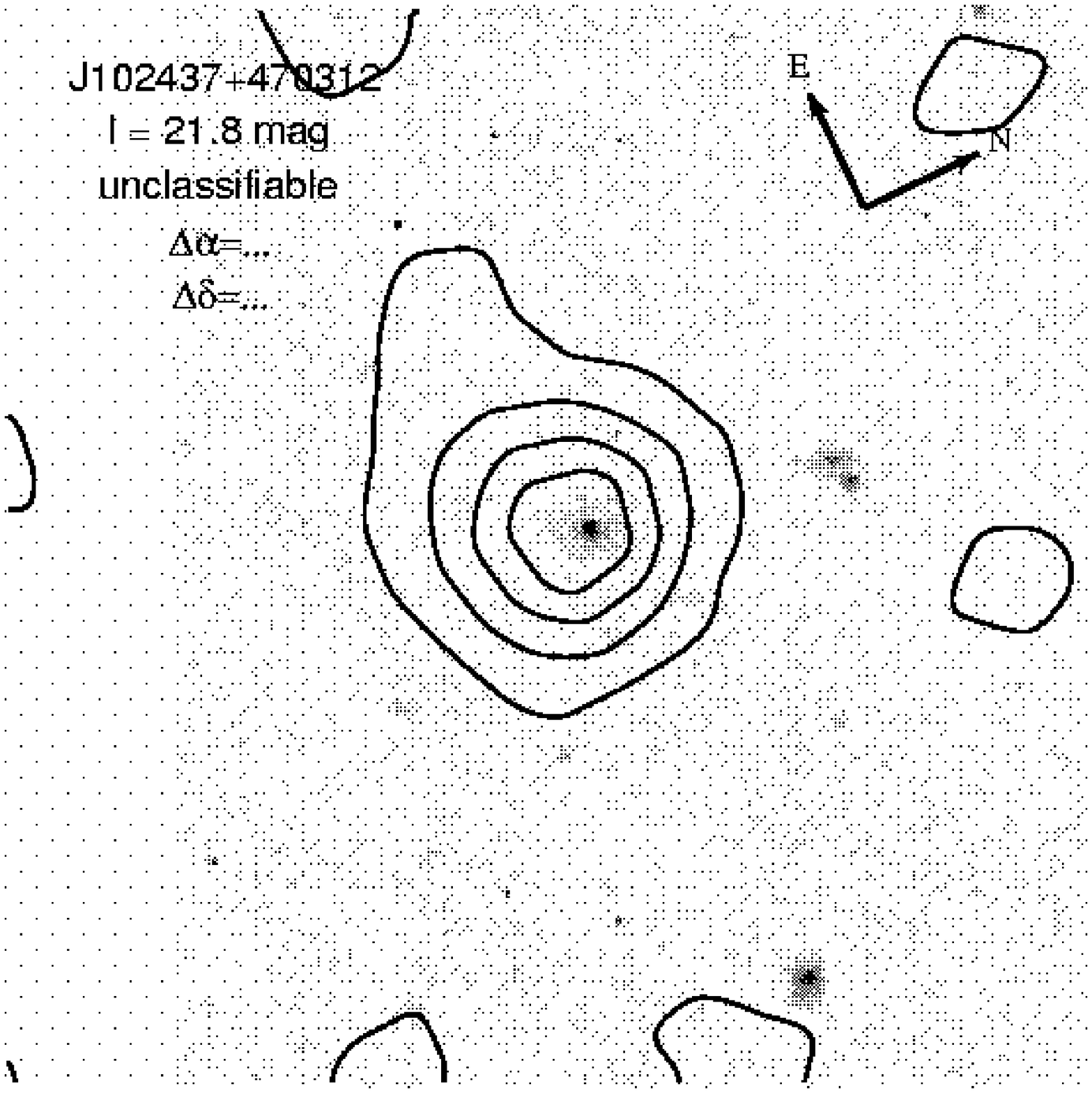 ,width=2.6truein}
\psfig{file=  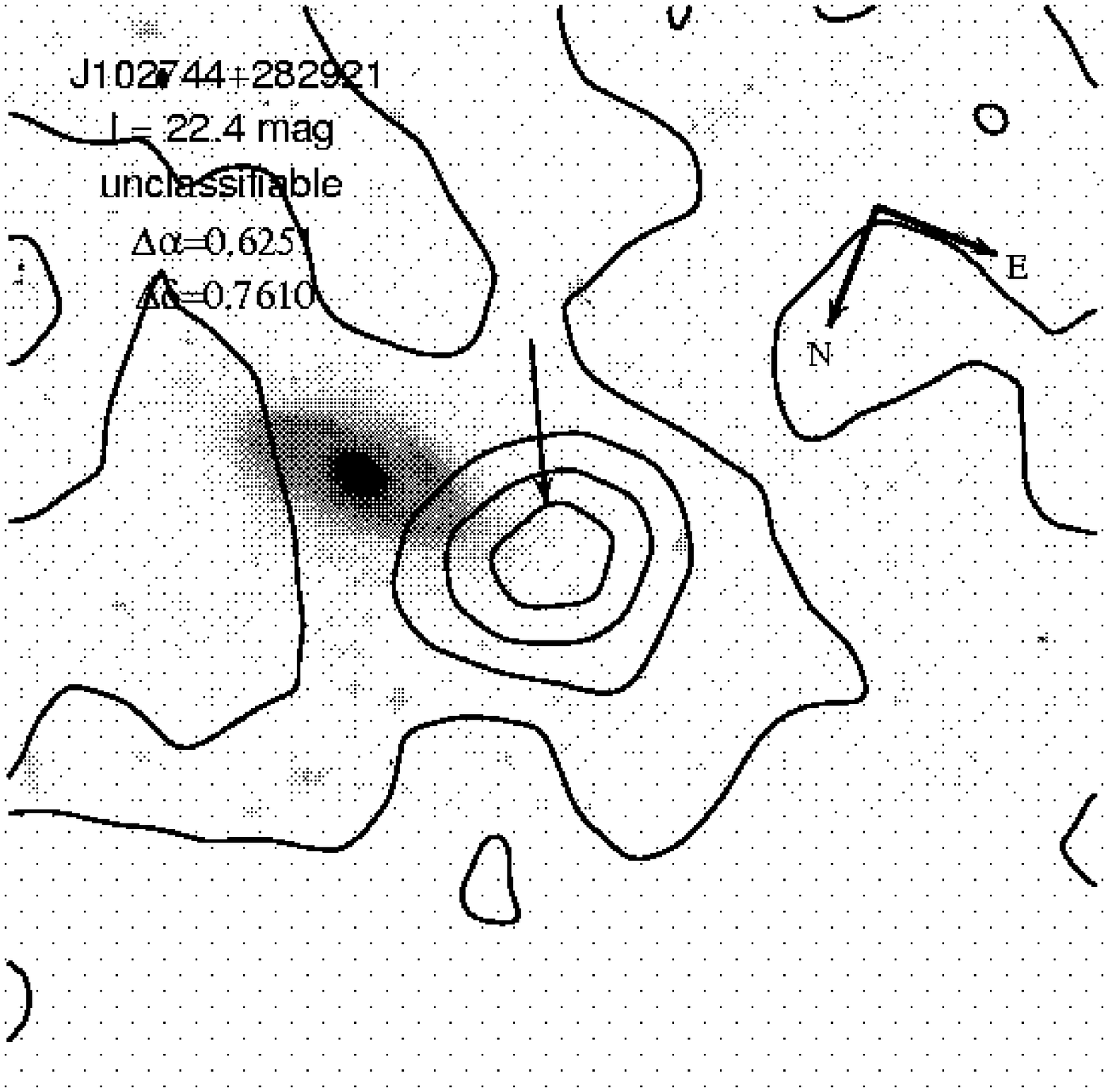 ,width=2.6truein}}}
\centerline{\hbox{
\psfig{file=  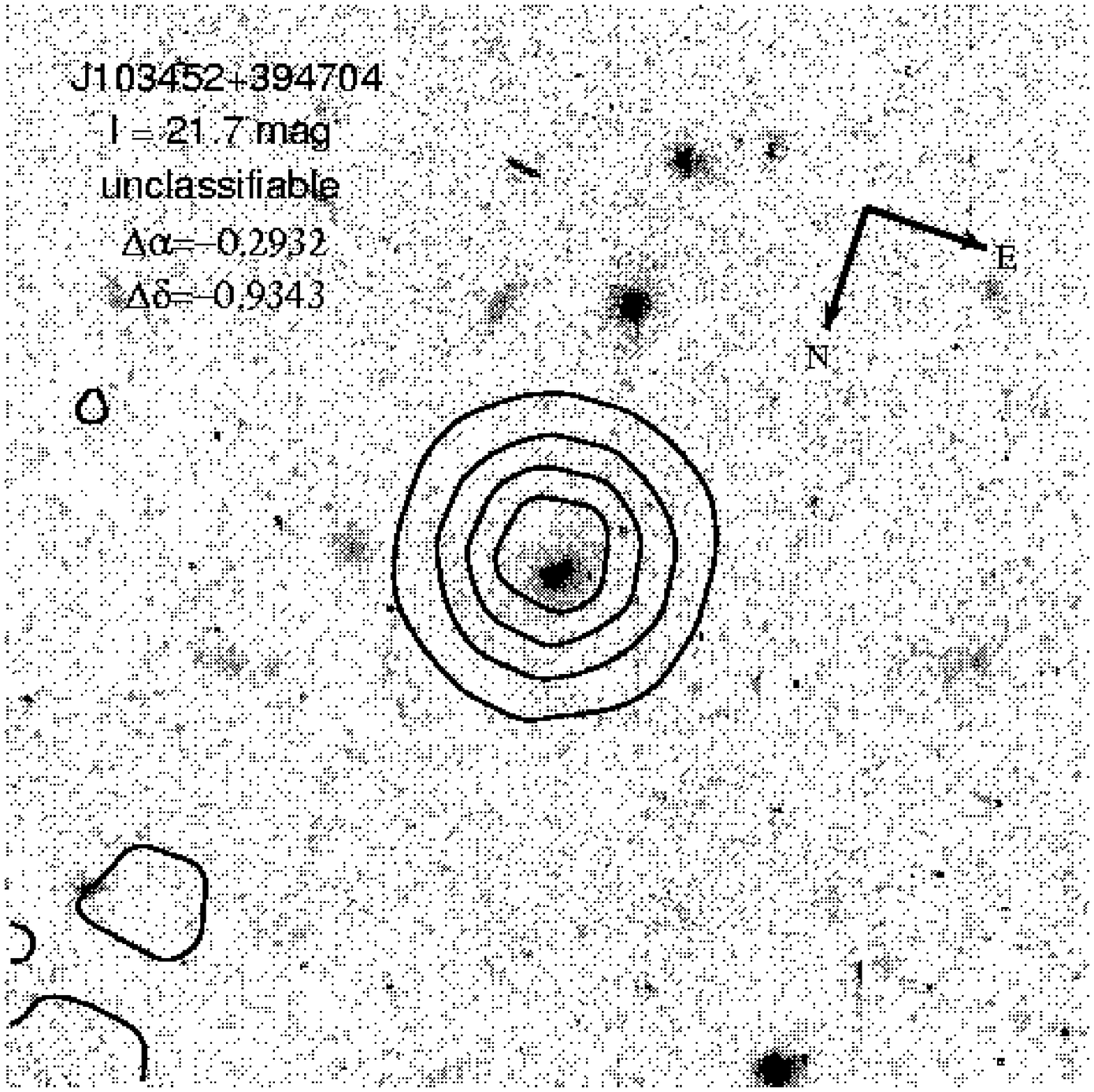 ,width=2.6truein}
\psfig{file=  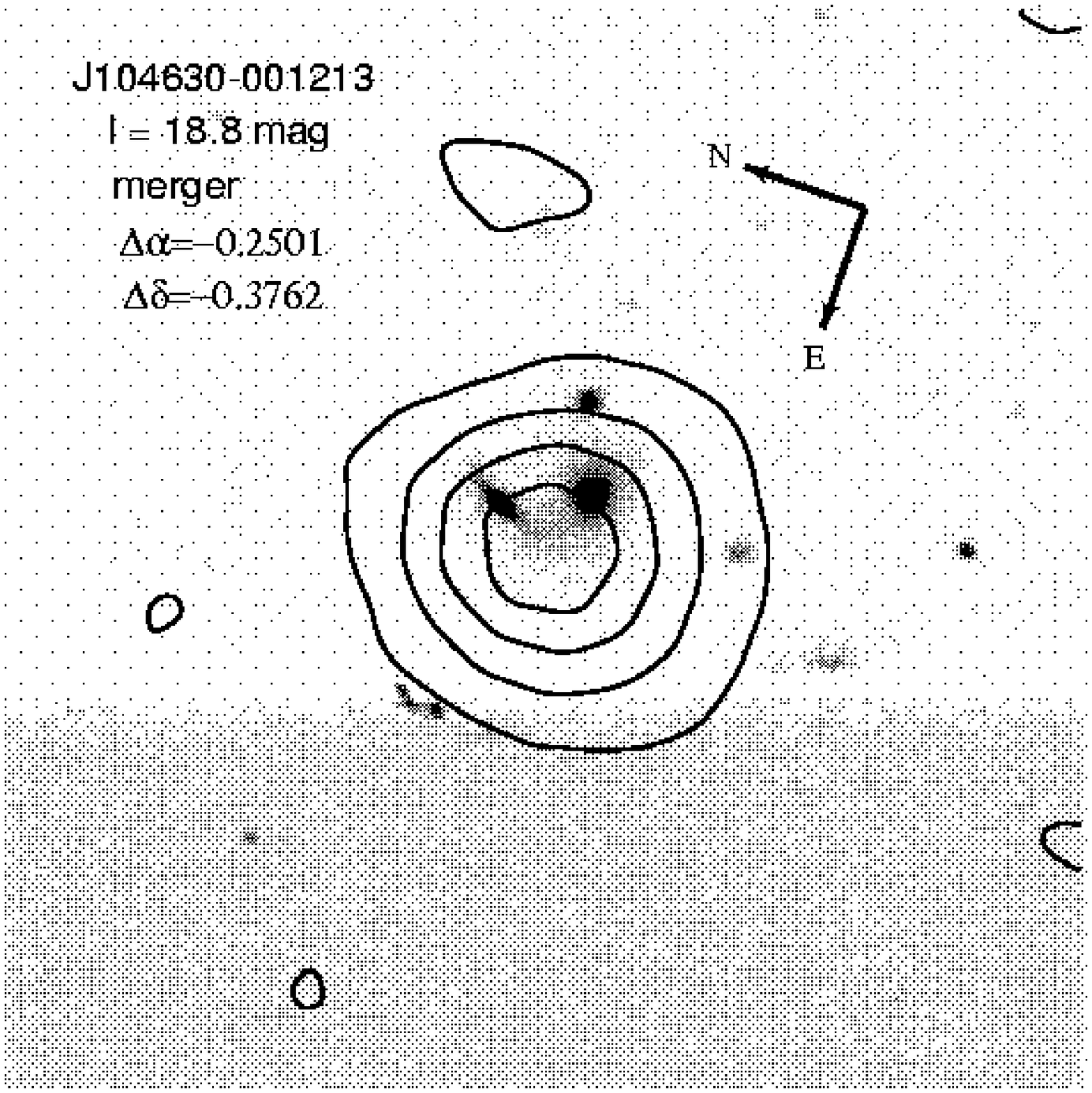 ,width=2.6truein}}}
\end{figure*}
\begin{figure*}
\centerline{\hbox{
\psfig{file=  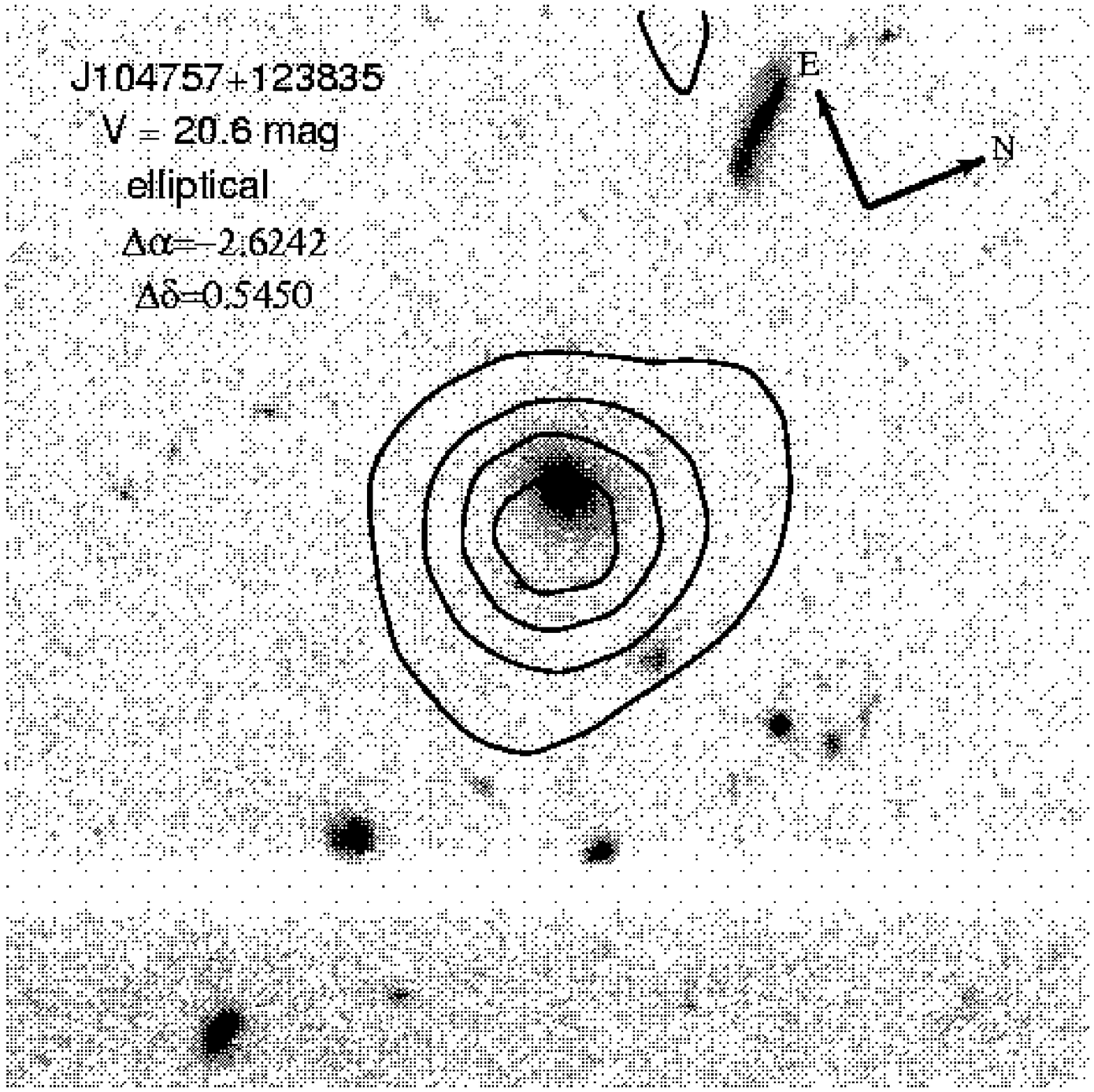 ,width=2.6truein}
\psfig{file=  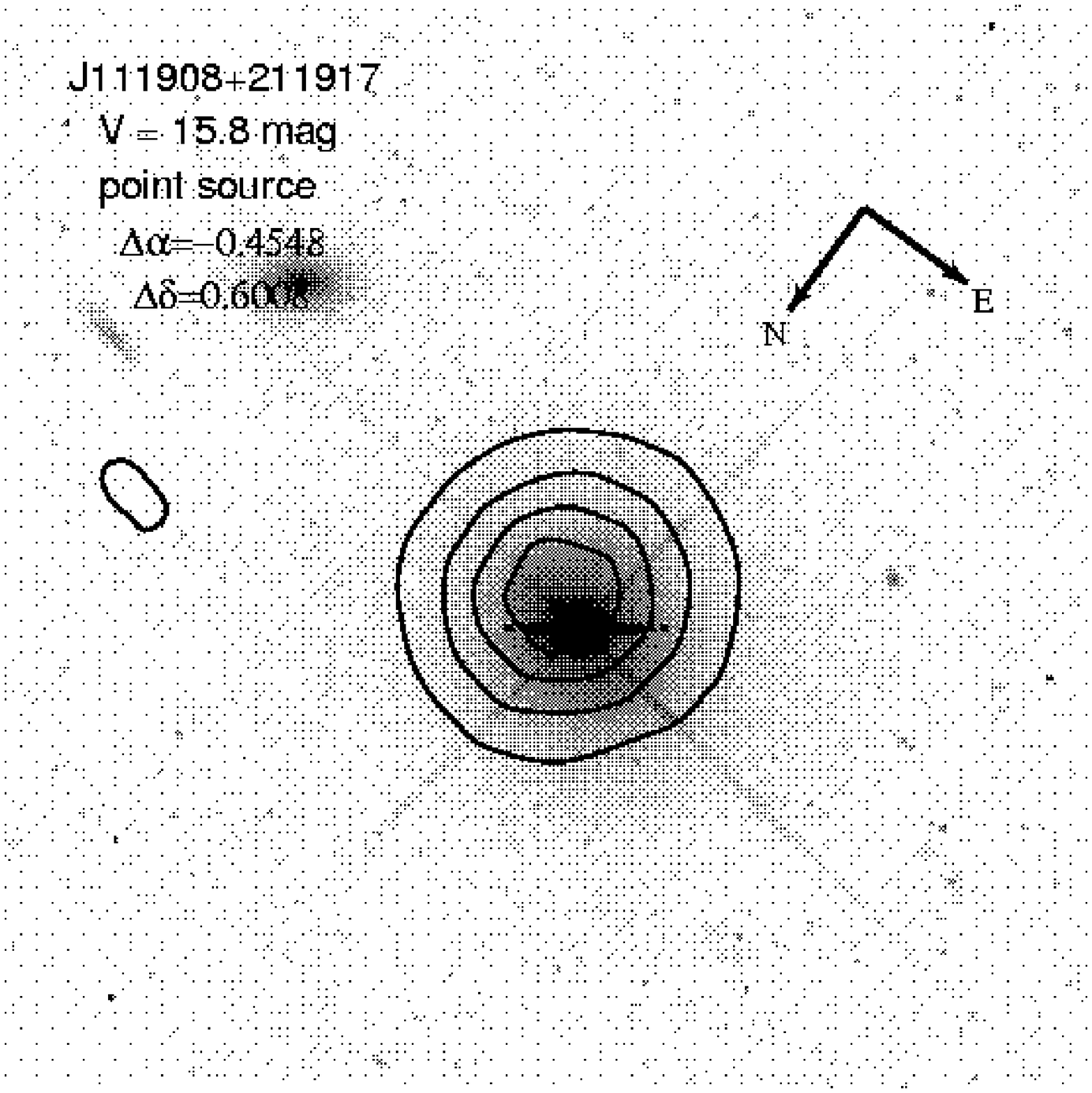 ,width=2.6truein}}}
\centerline{\hbox{
\psfig{file=  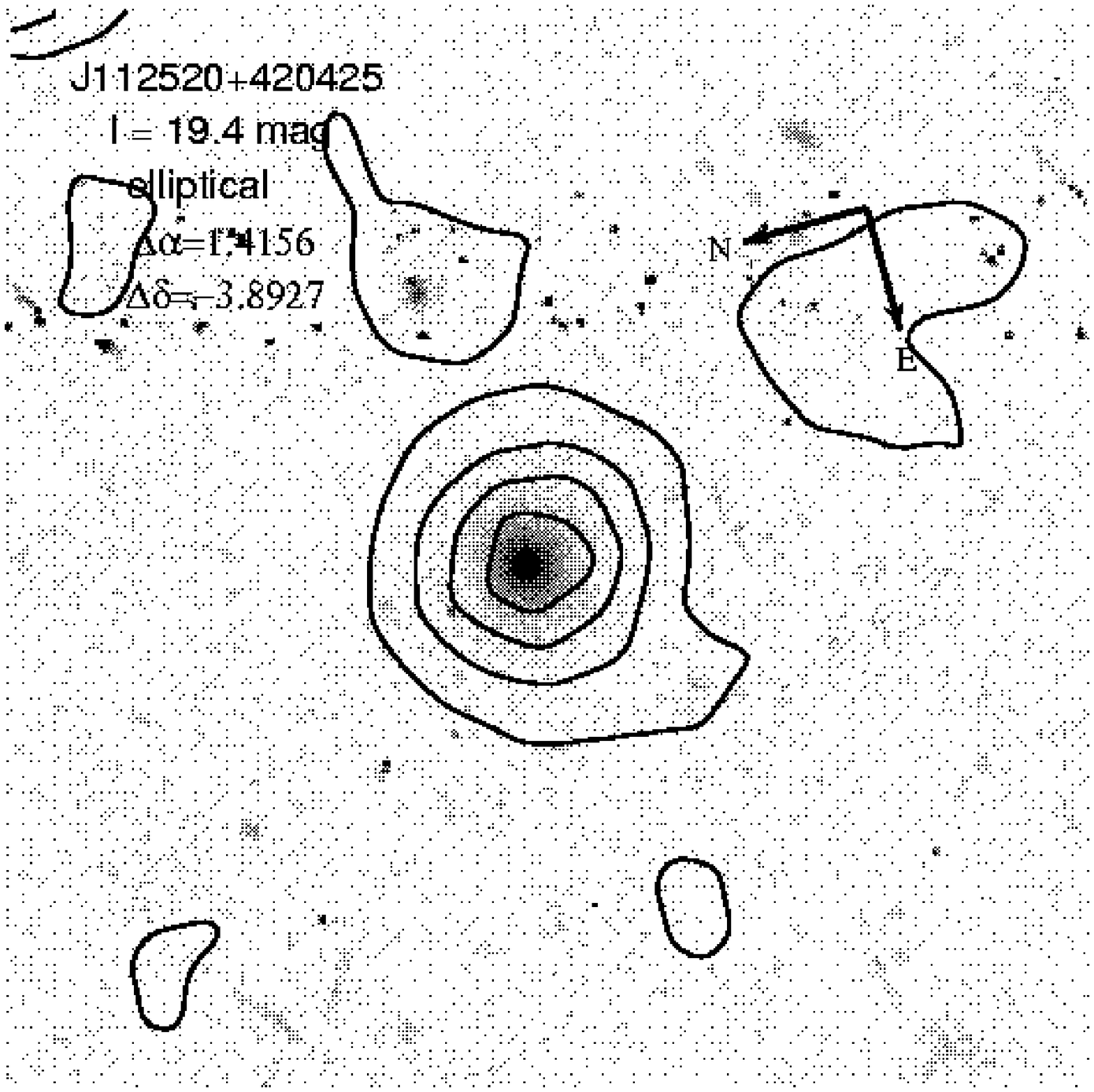 ,width=2.6truein}
\psfig{file=  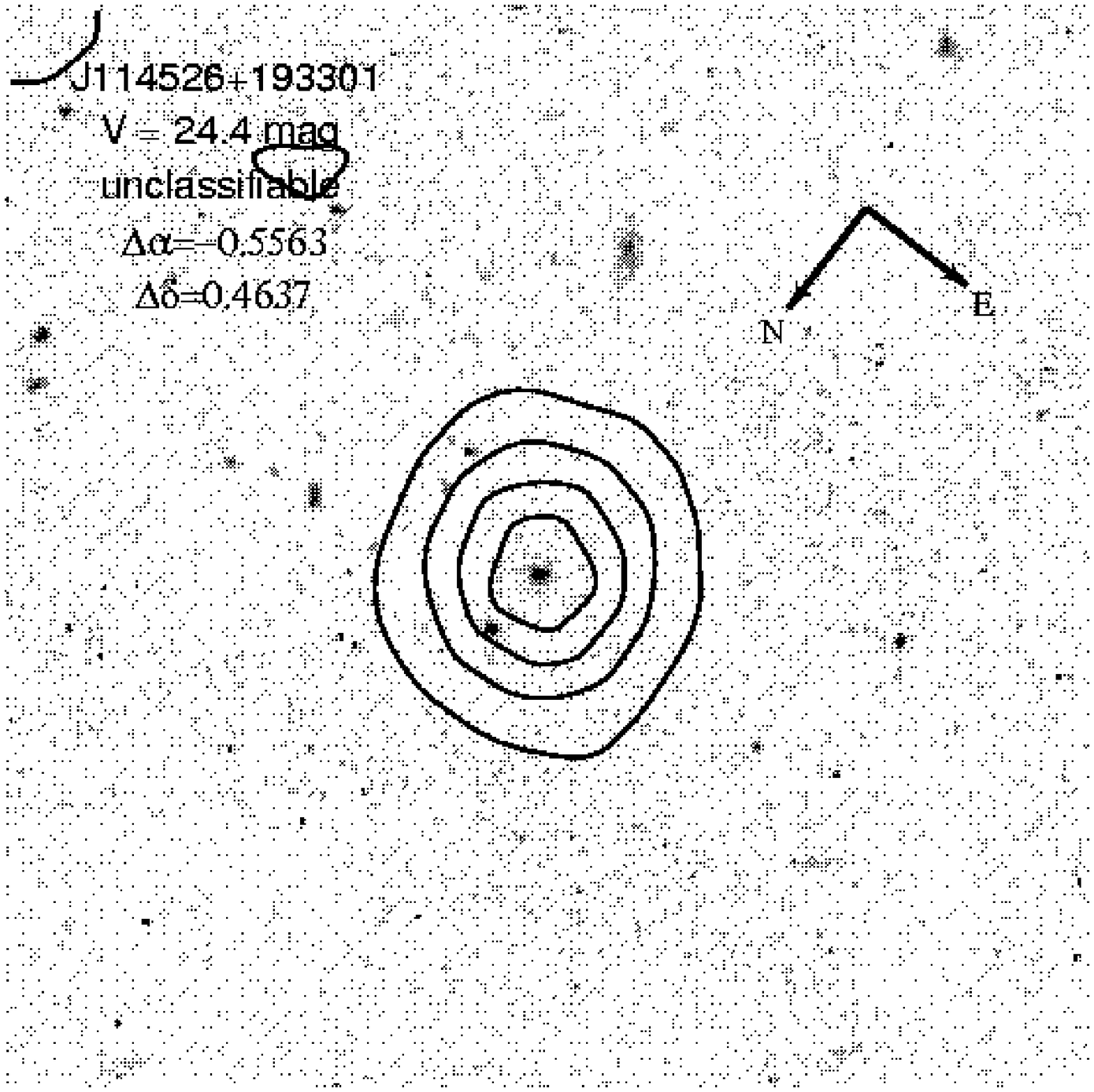 ,width=2.6truein}}}
\centerline{\hbox{
\psfig{file=  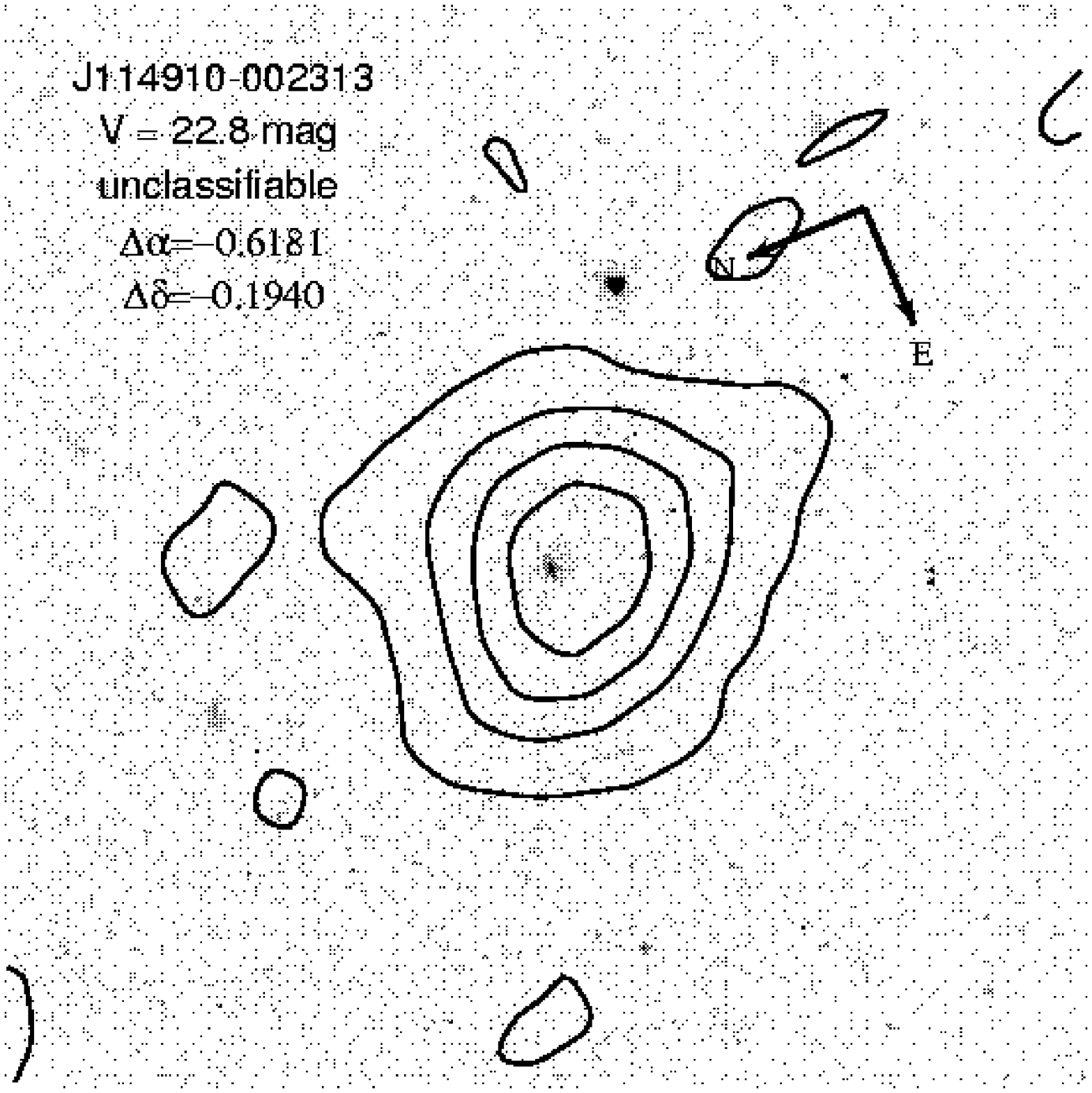 ,width=2.6truein}
\psfig{file=  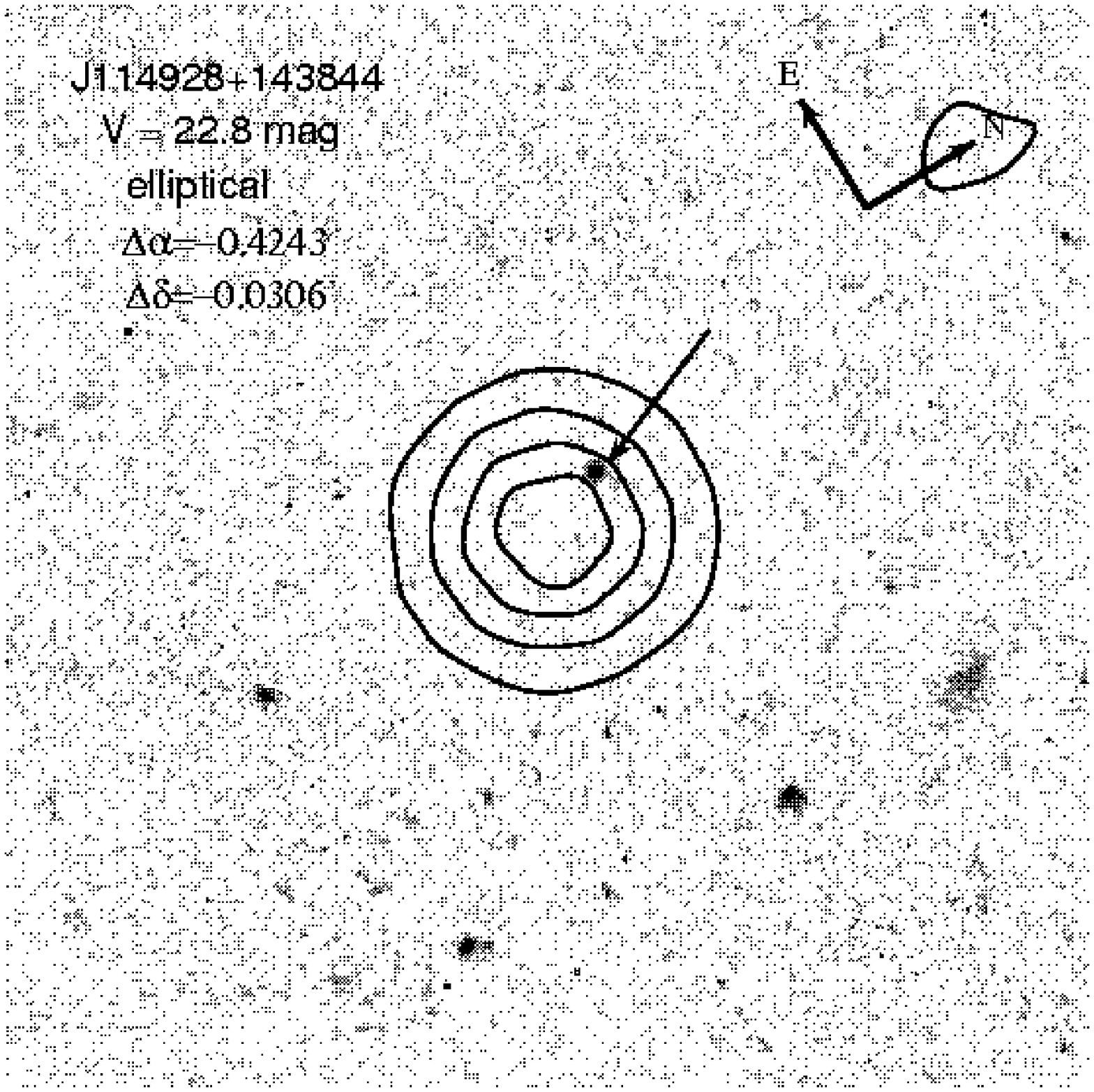 ,width=2.6truein}}}
\end{figure*}
\begin{figure*}
\centerline{\hbox{
\psfig{file=  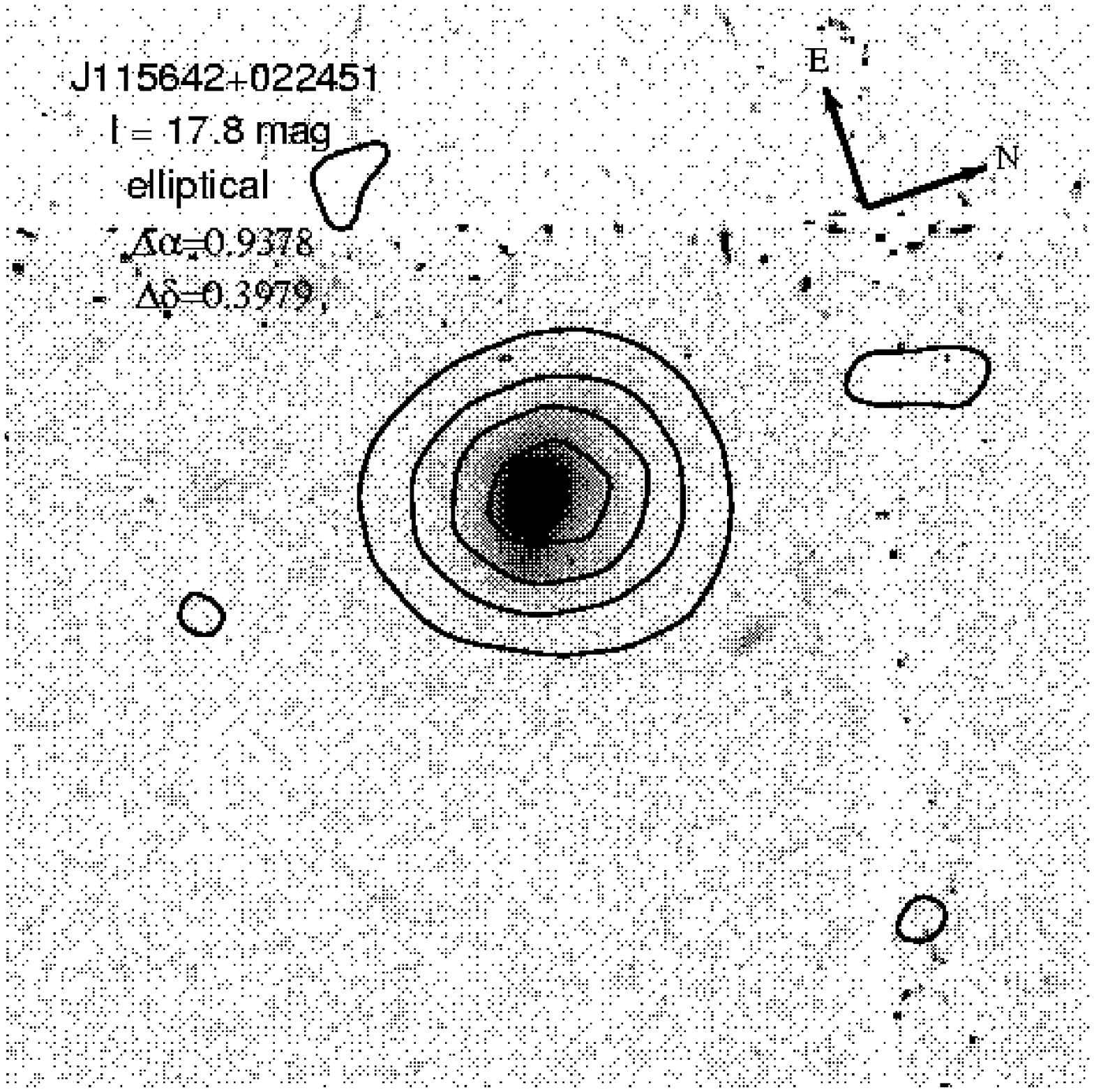 ,width=2.6truein}
\psfig{file=  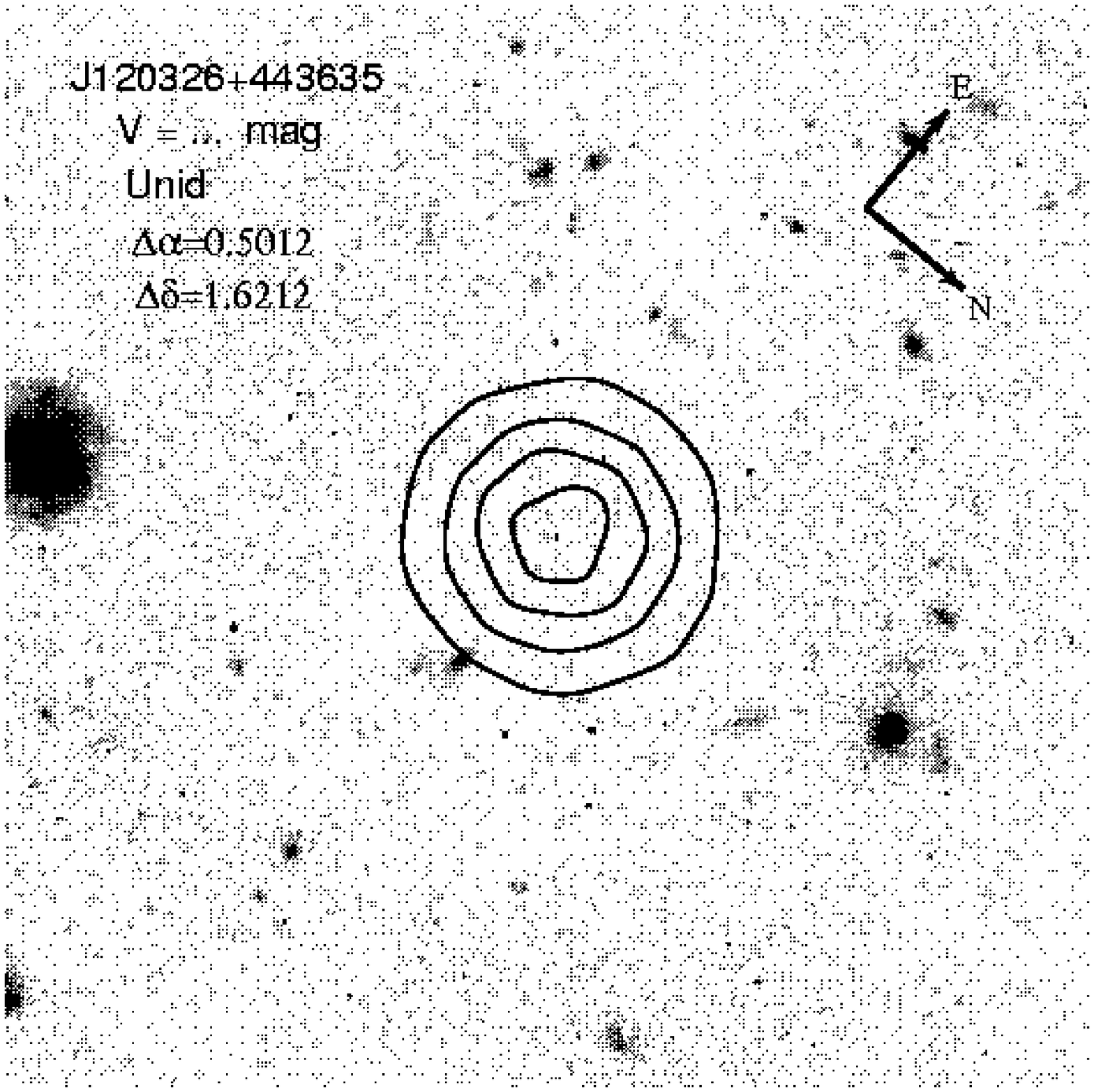 ,width=2.6truein}}}
\centerline{\hbox{
\psfig{file=  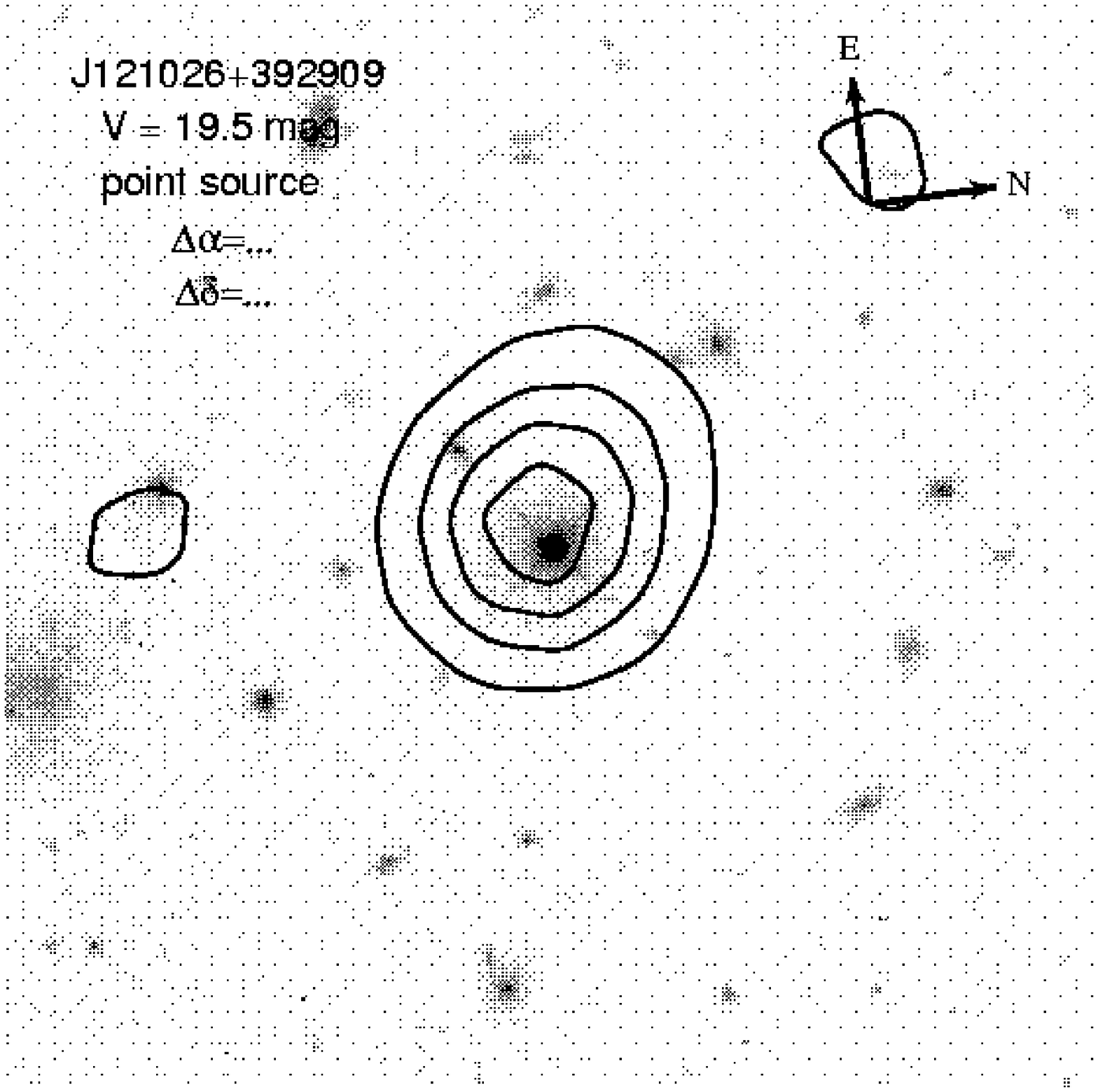 ,width=2.6truein}
\psfig{file=  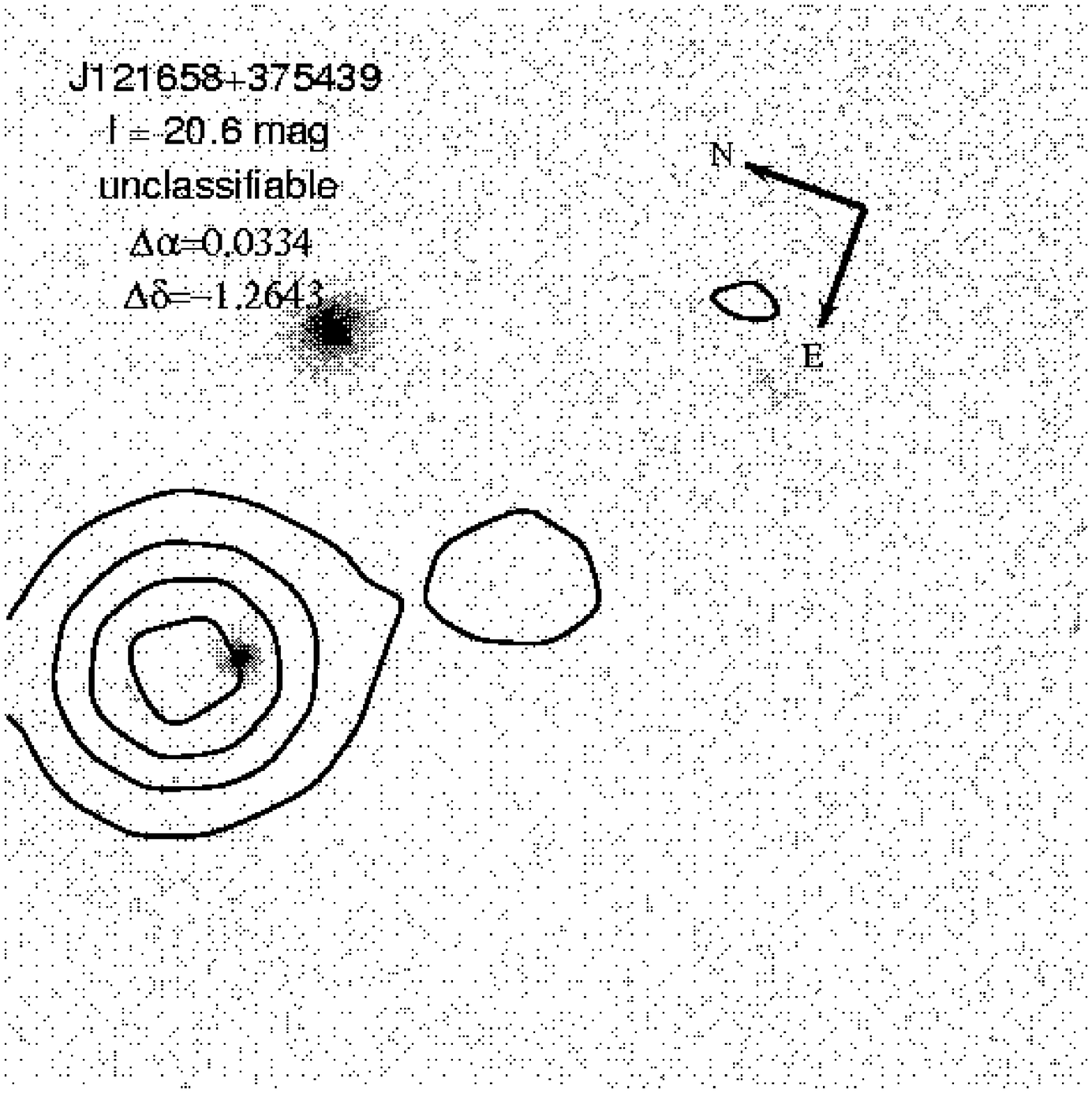 ,width=2.6truein}}}
\centerline{\hbox{
\psfig{file=  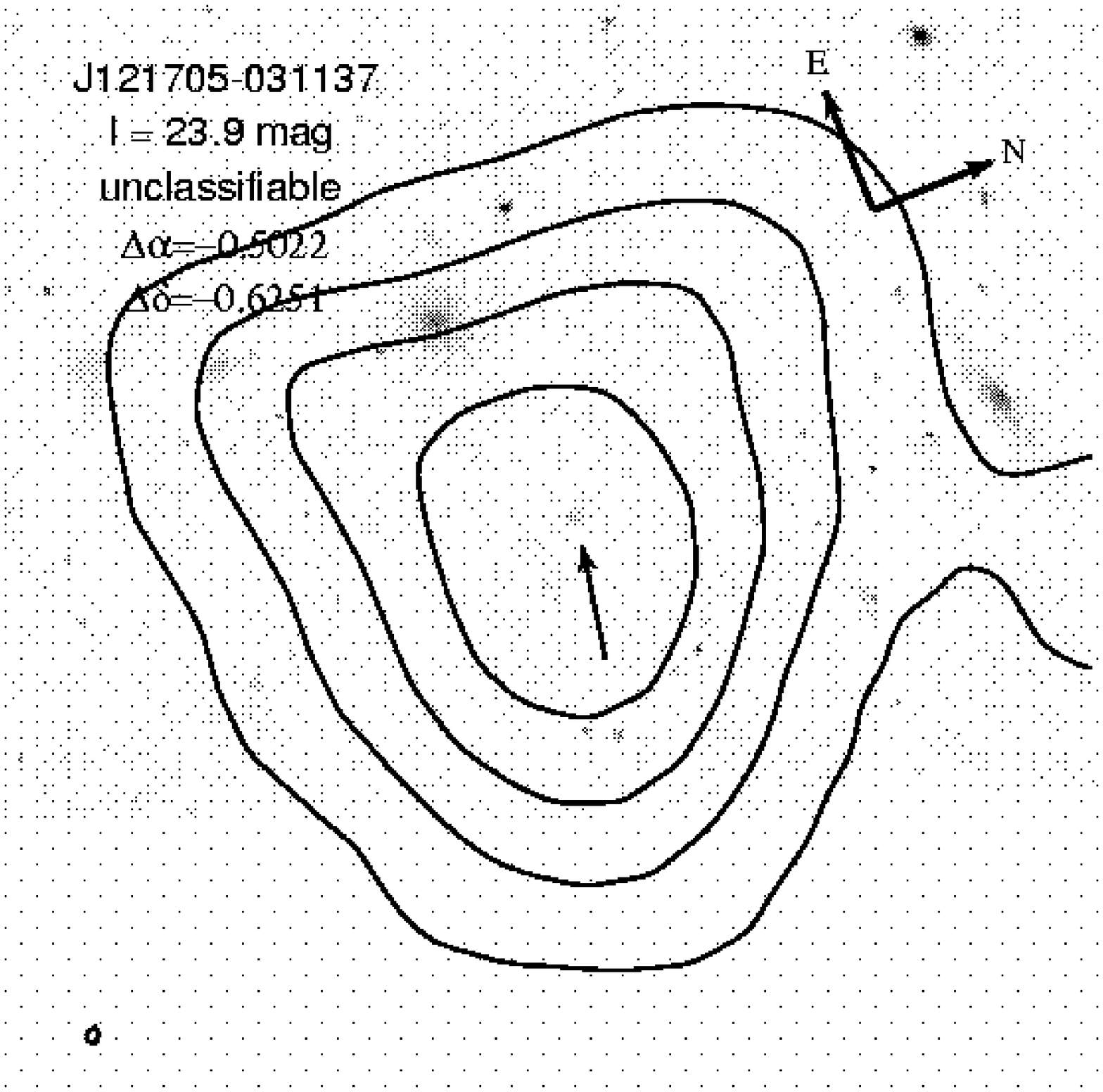 ,width=2.6truein}
\psfig{file=  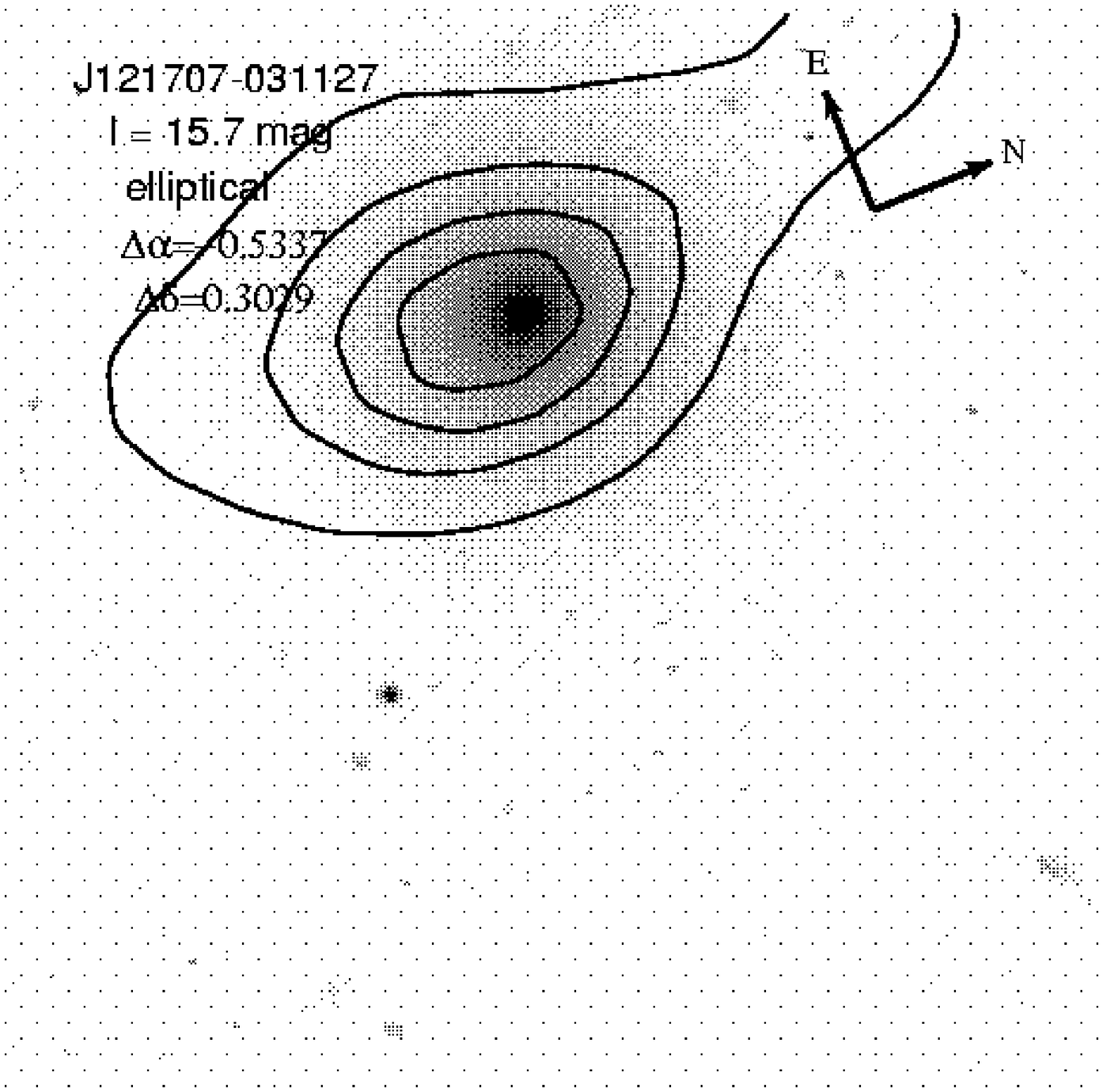 ,width=2.6truein}}}
\end{figure*}
\begin{figure*}
\centerline{\hbox{
\psfig{file=  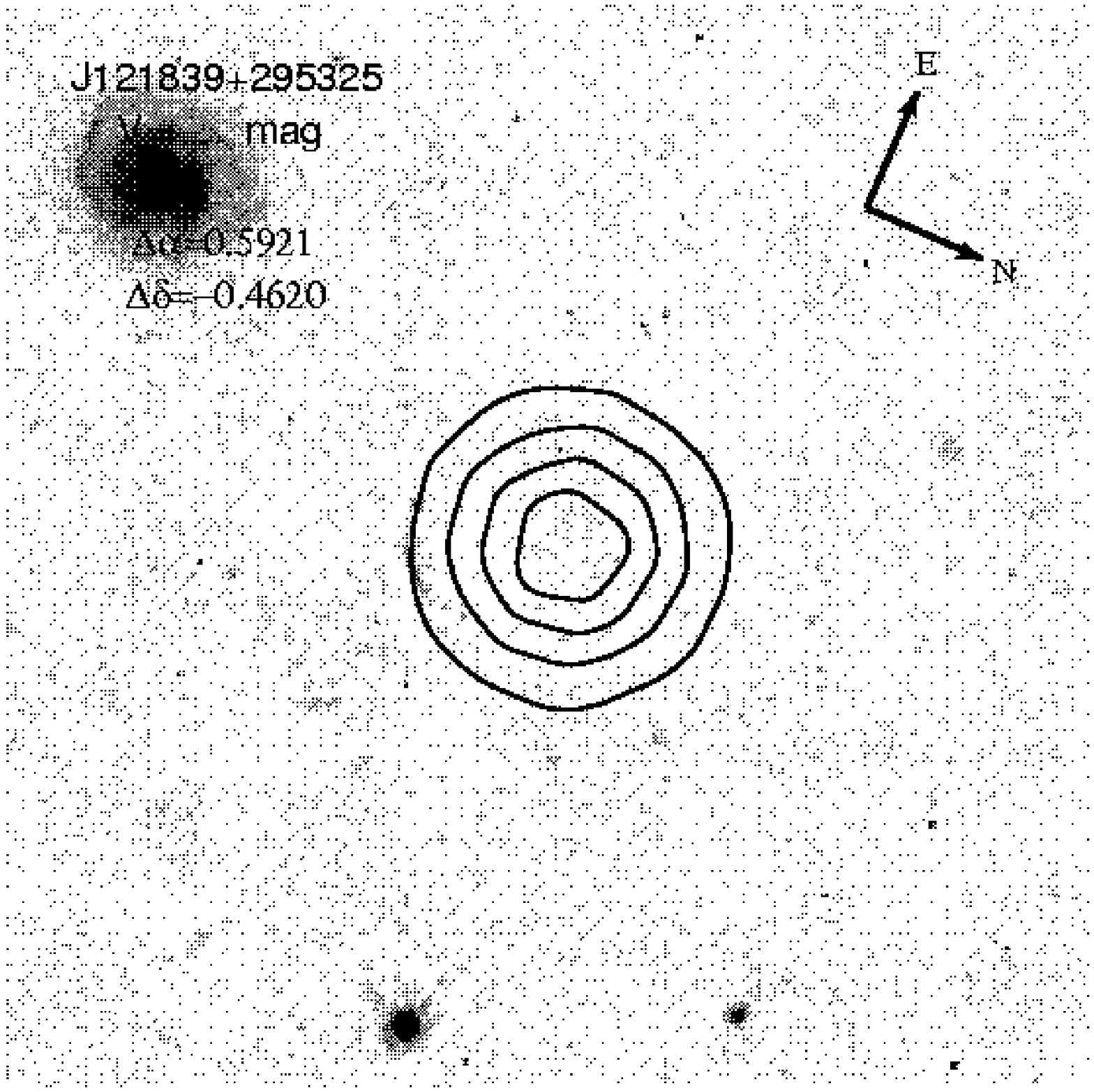 ,width=2.6truein}
\psfig{file=  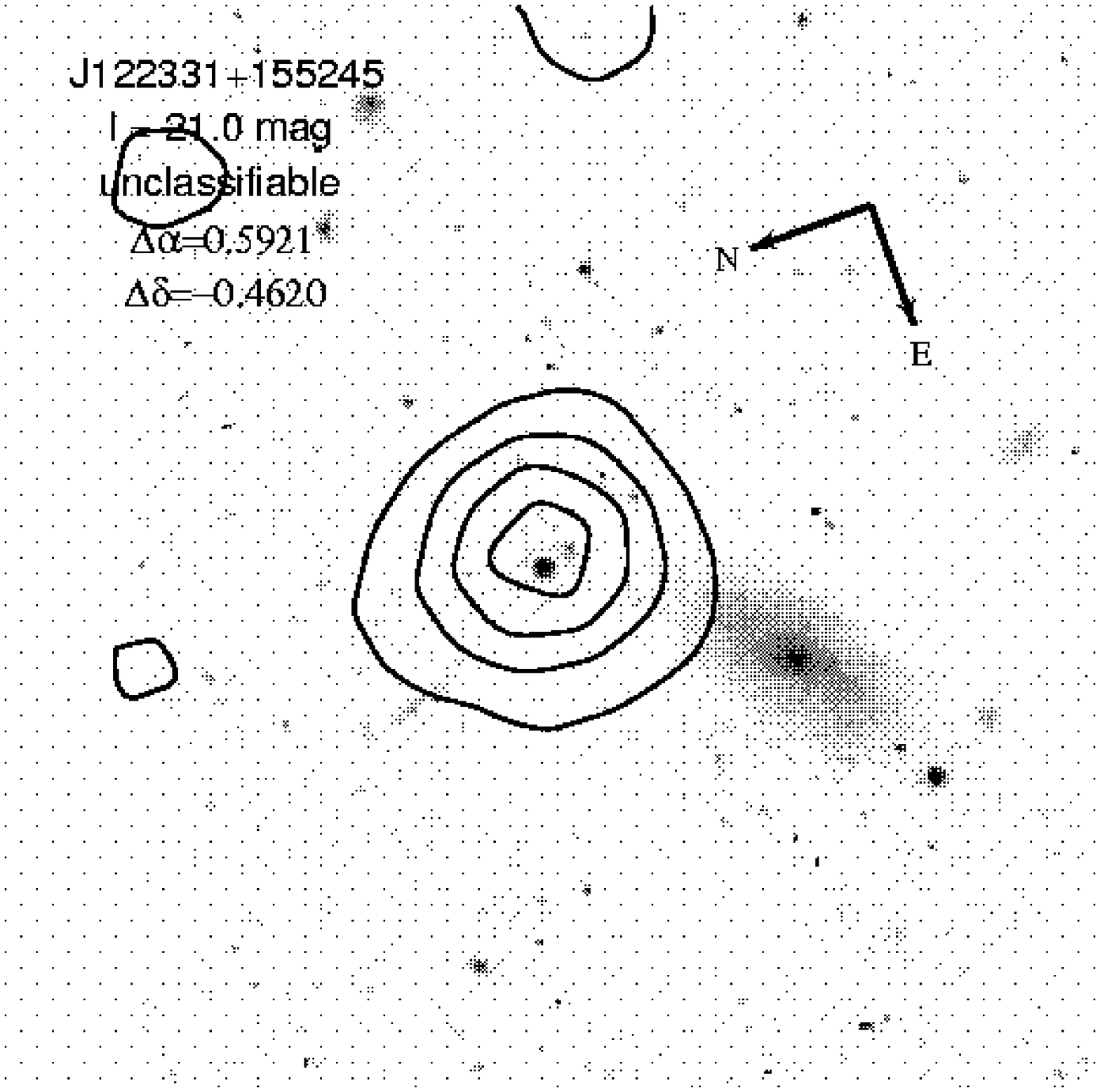 ,width=2.6truein}}}
\centerline{\hbox{
\psfig{file=  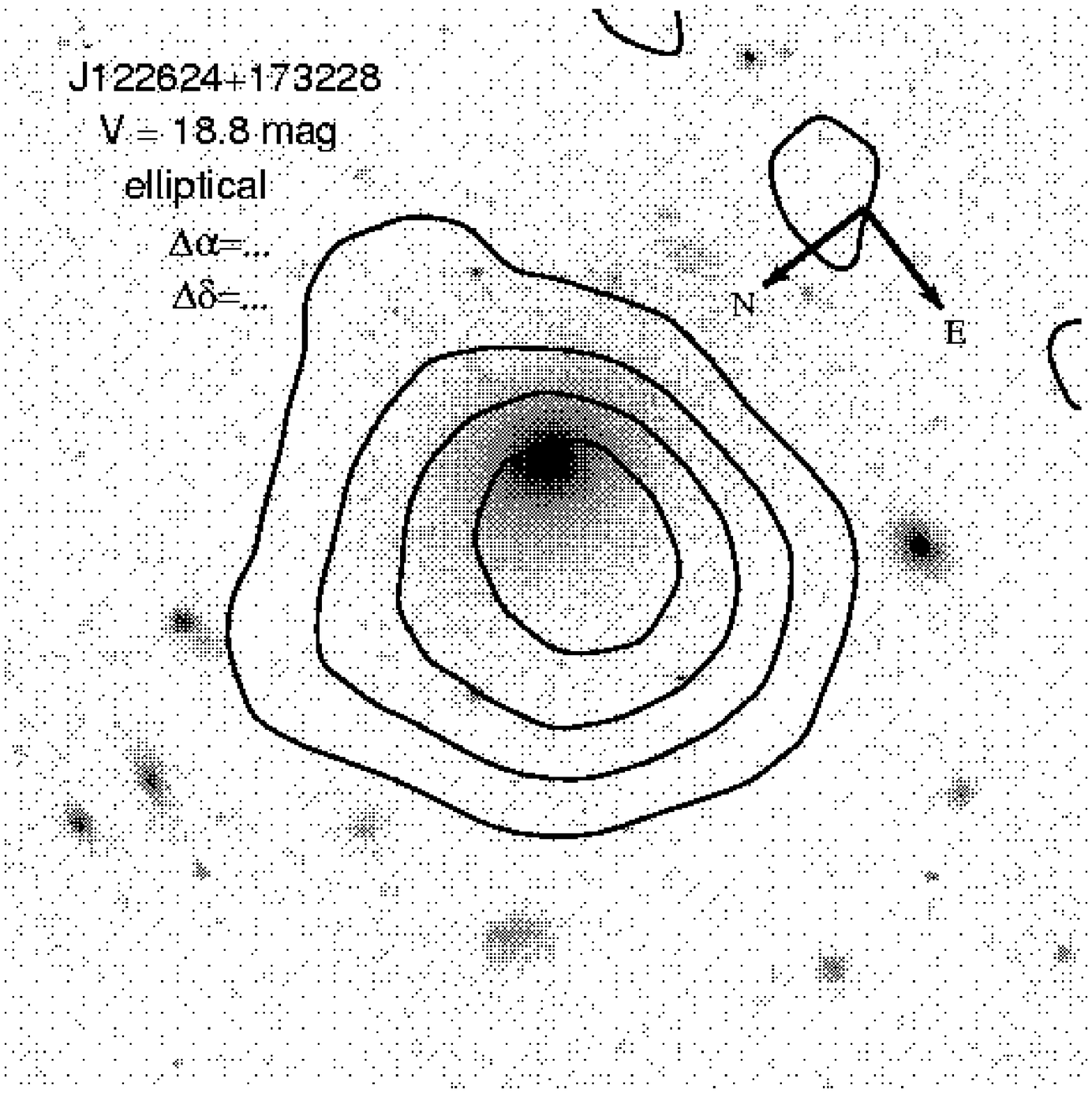 ,width=2.6truein}
\psfig{file=  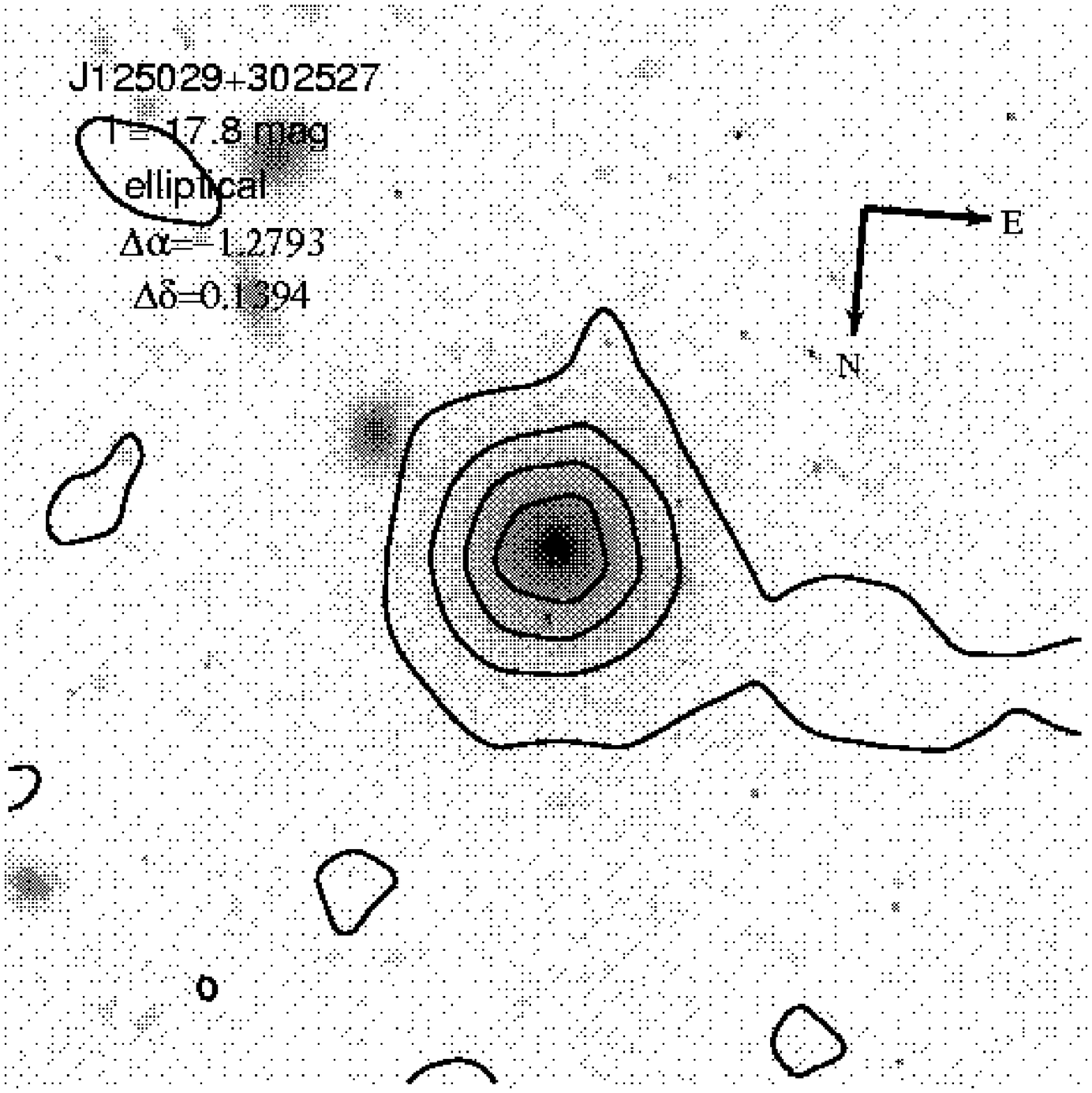 ,width=2.6truein}}}
\centerline{\hbox{
\psfig{file=  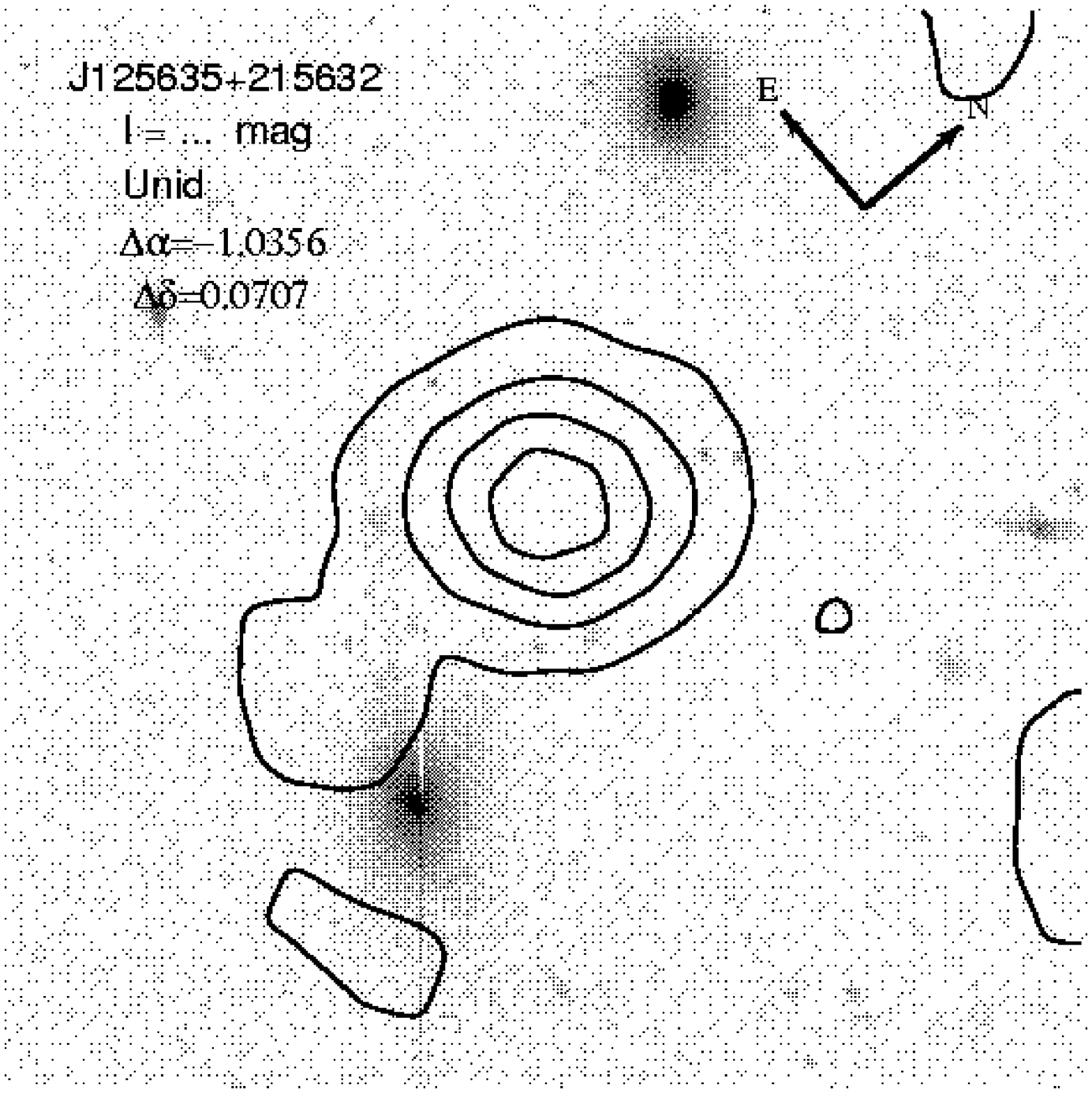 ,width=2.6truein}
\psfig{file=  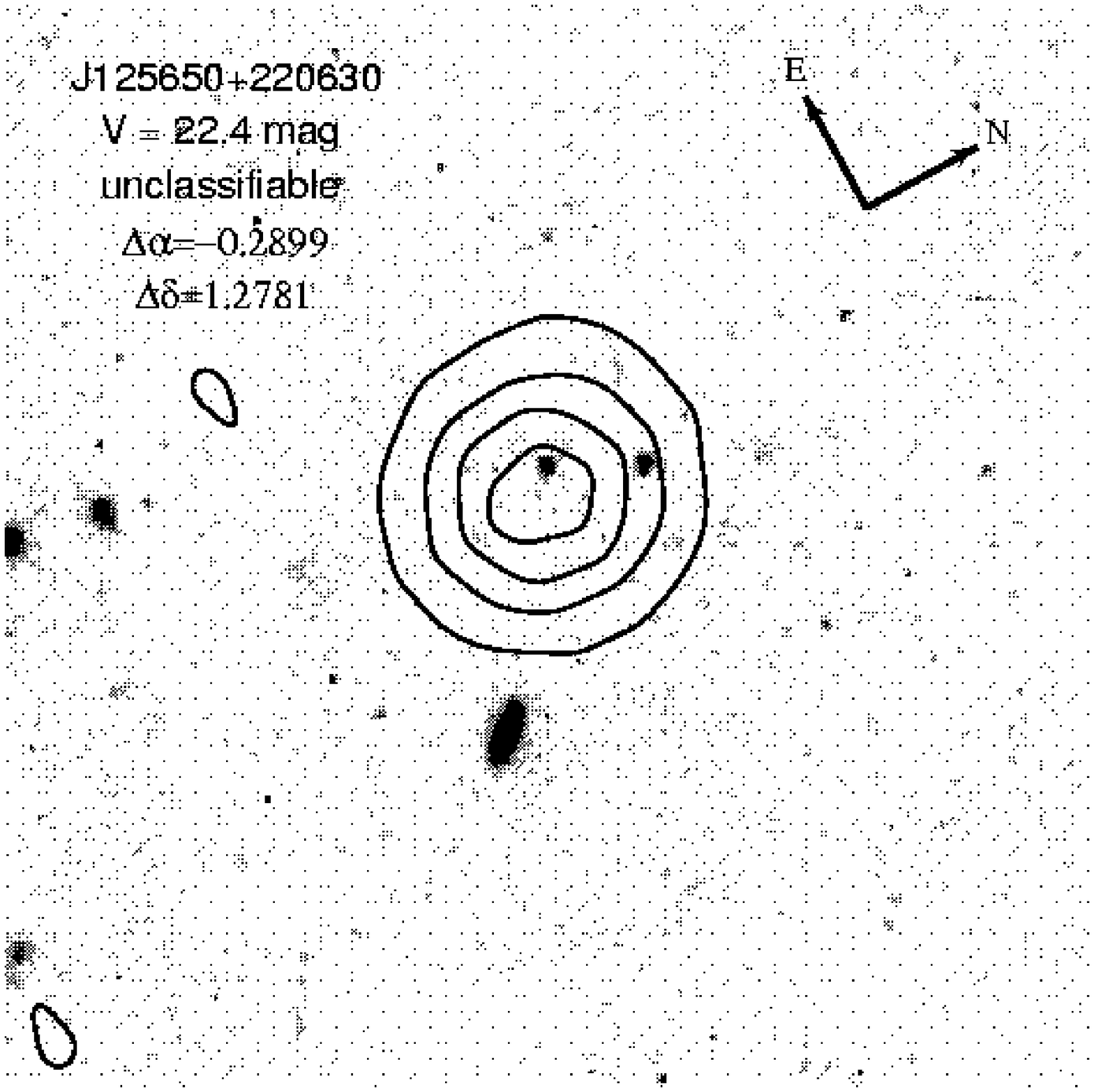 ,width=2.6truein}}}
\end{figure*}
\begin{figure*}
\centerline{\hbox{
\psfig{file=  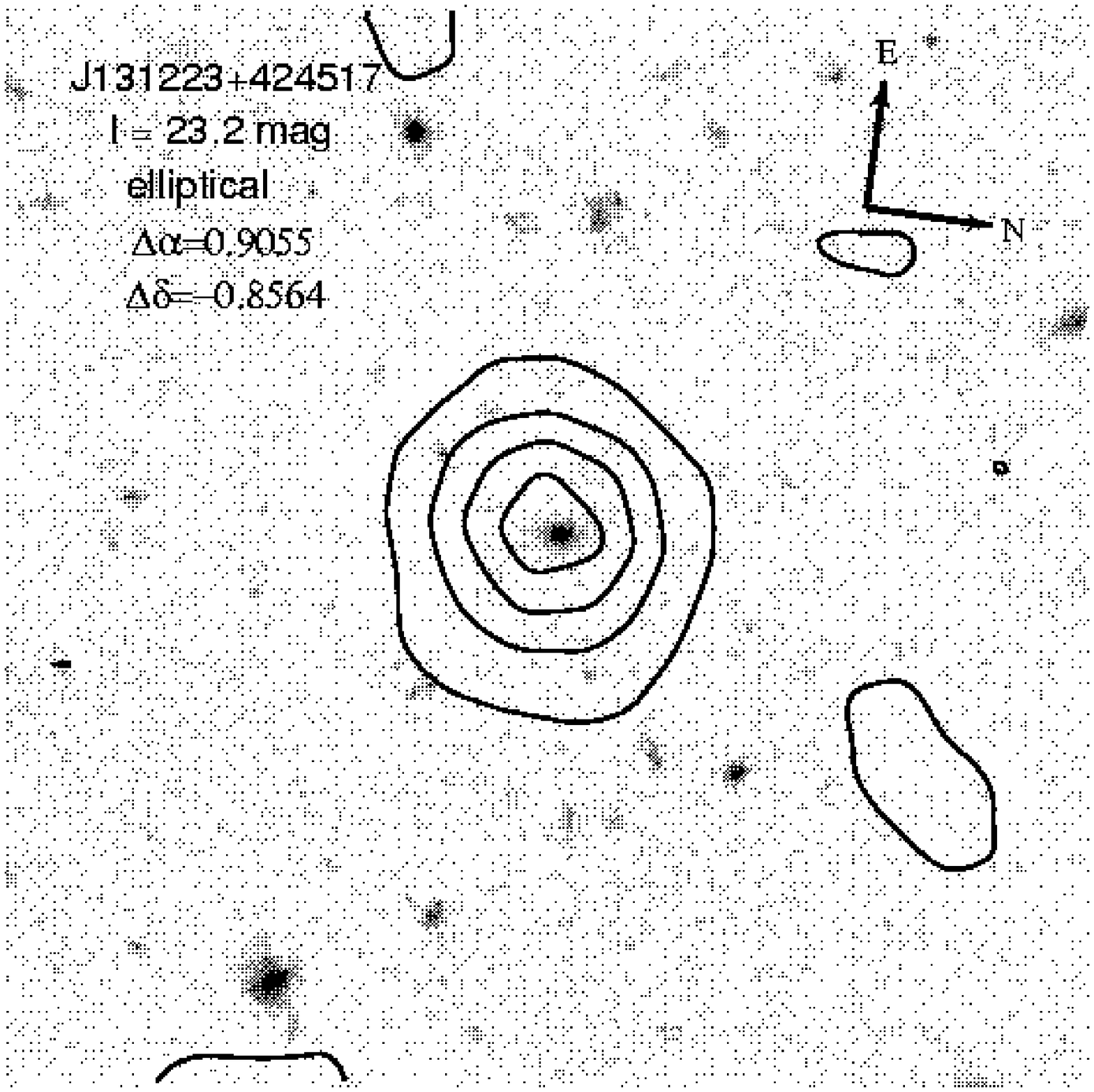 ,width=2.6truein}
\psfig{file=  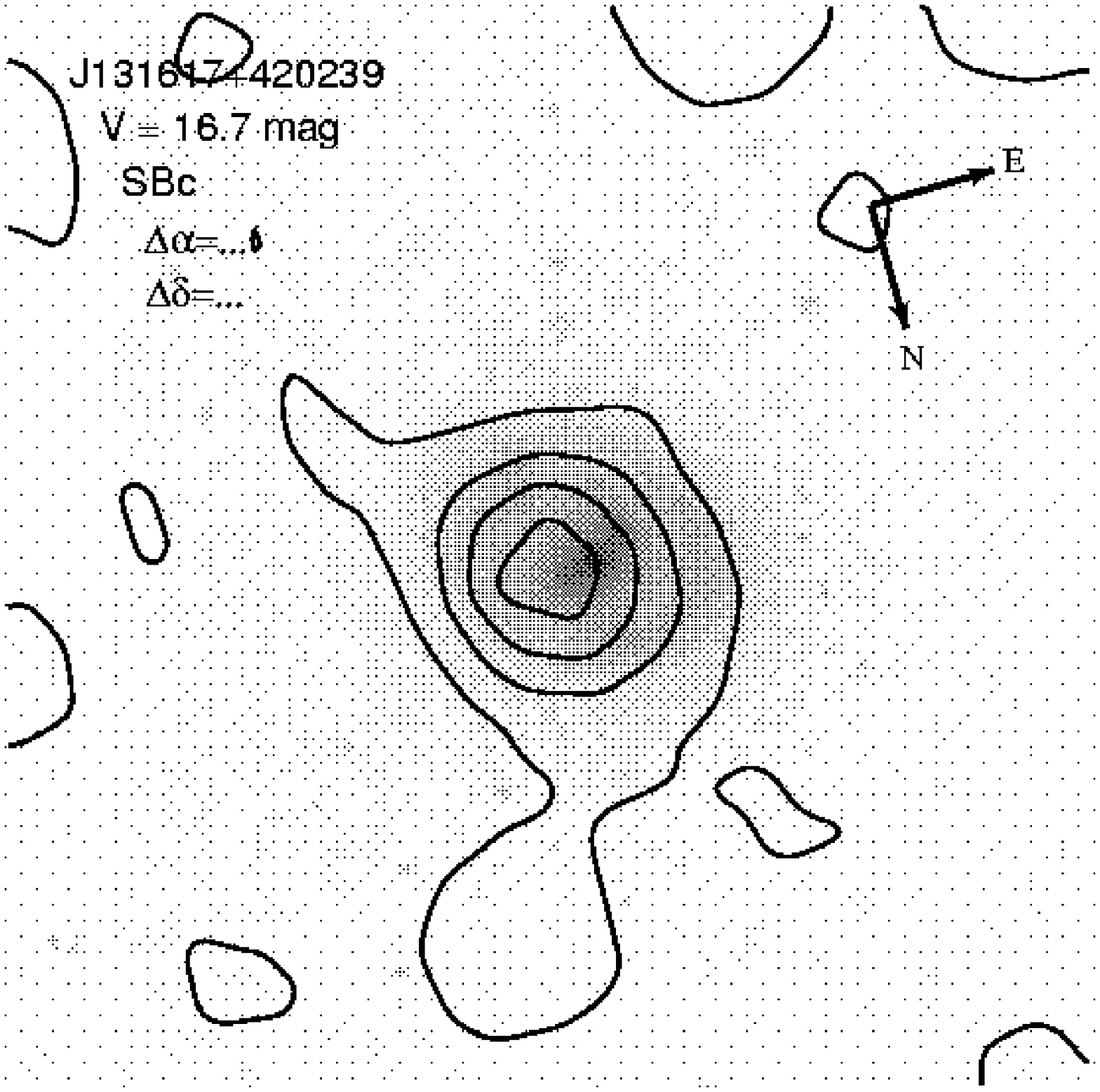 ,width=2.6truein}}}
\centerline{\hbox{
\psfig{file=  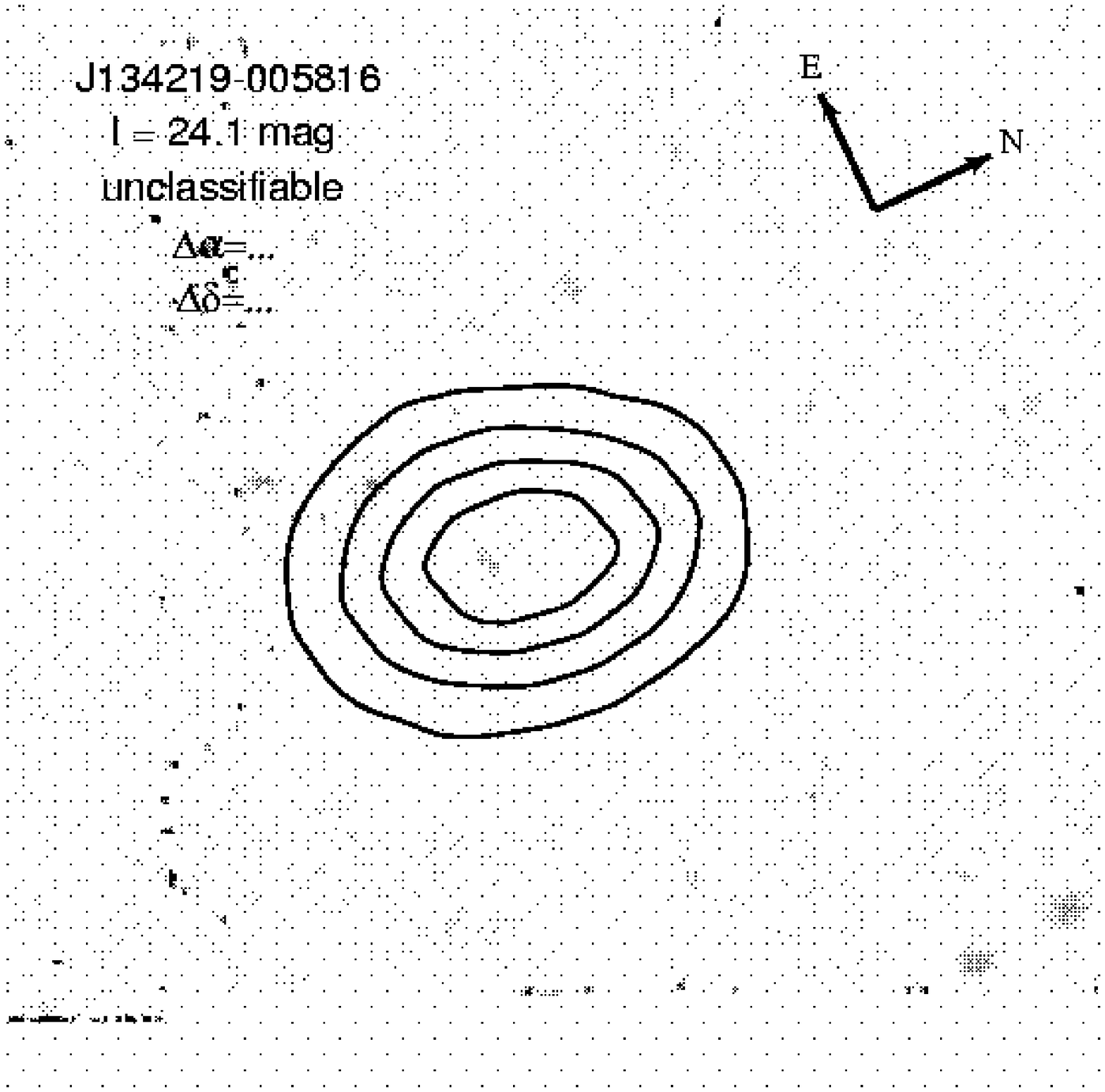 ,width=2.6truein}
\psfig{file=  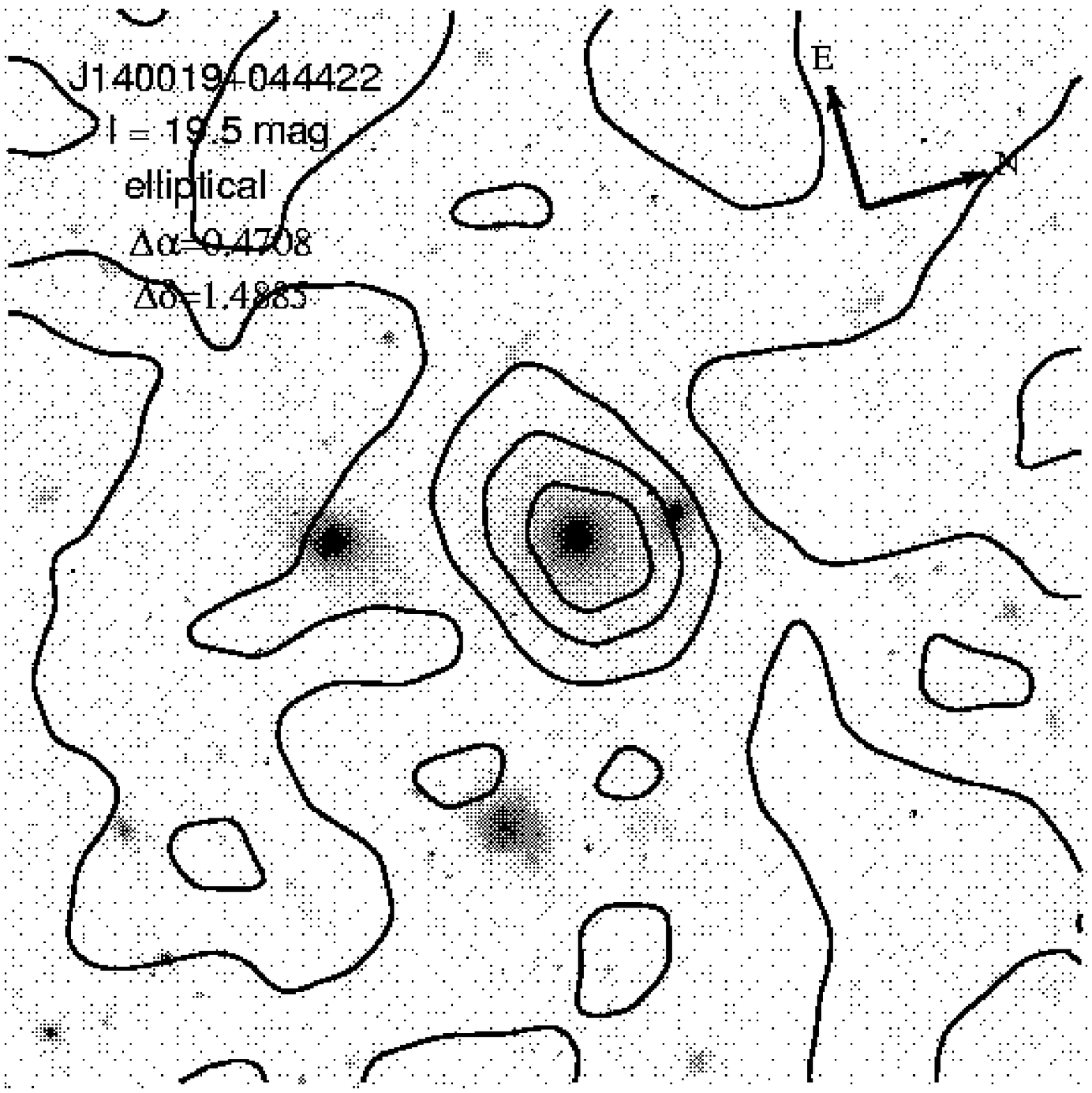 ,width=2.6truein}}}
\centerline{\hbox{
\psfig{file=  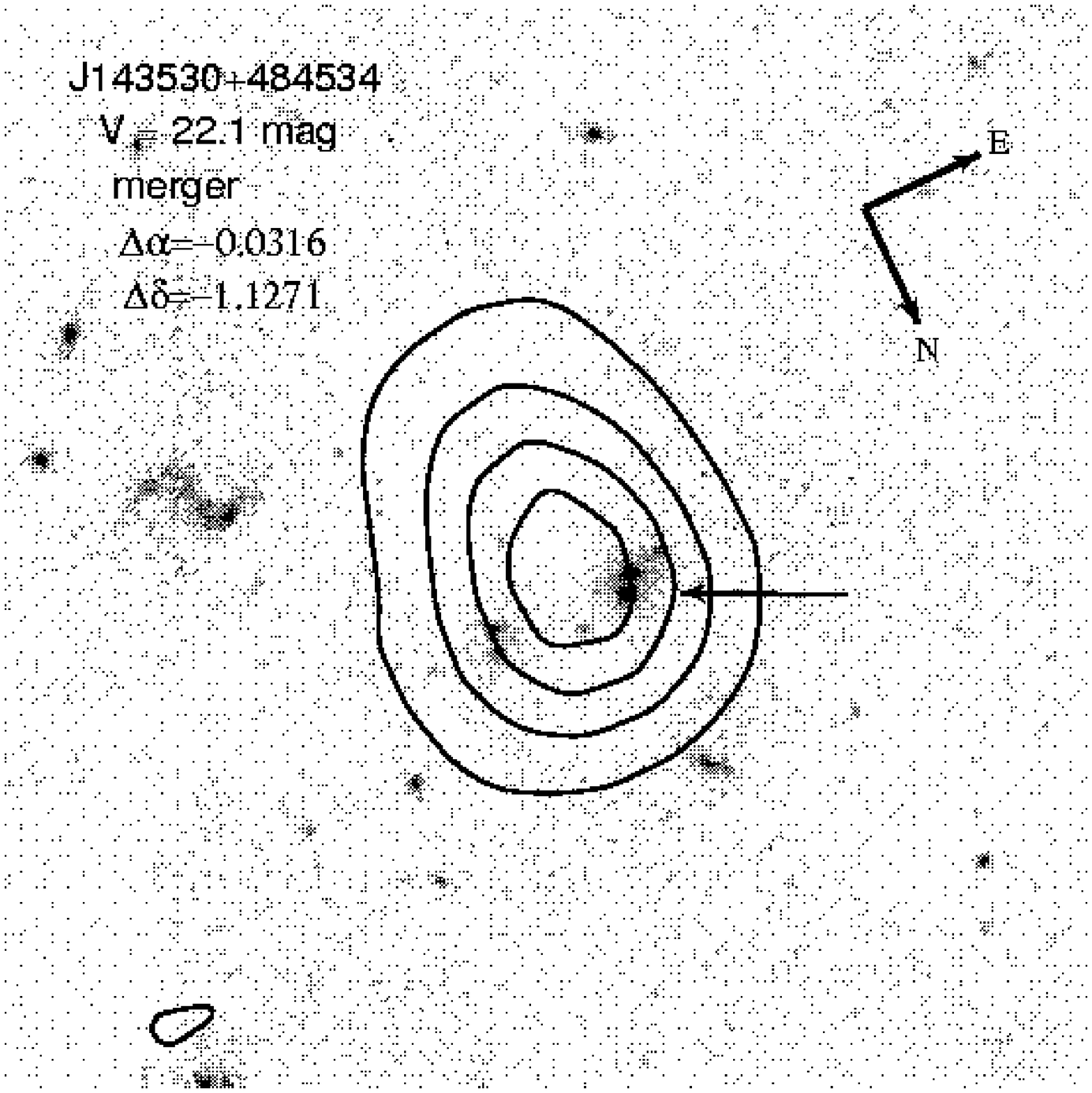 ,width=2.6truein}
\psfig{file=  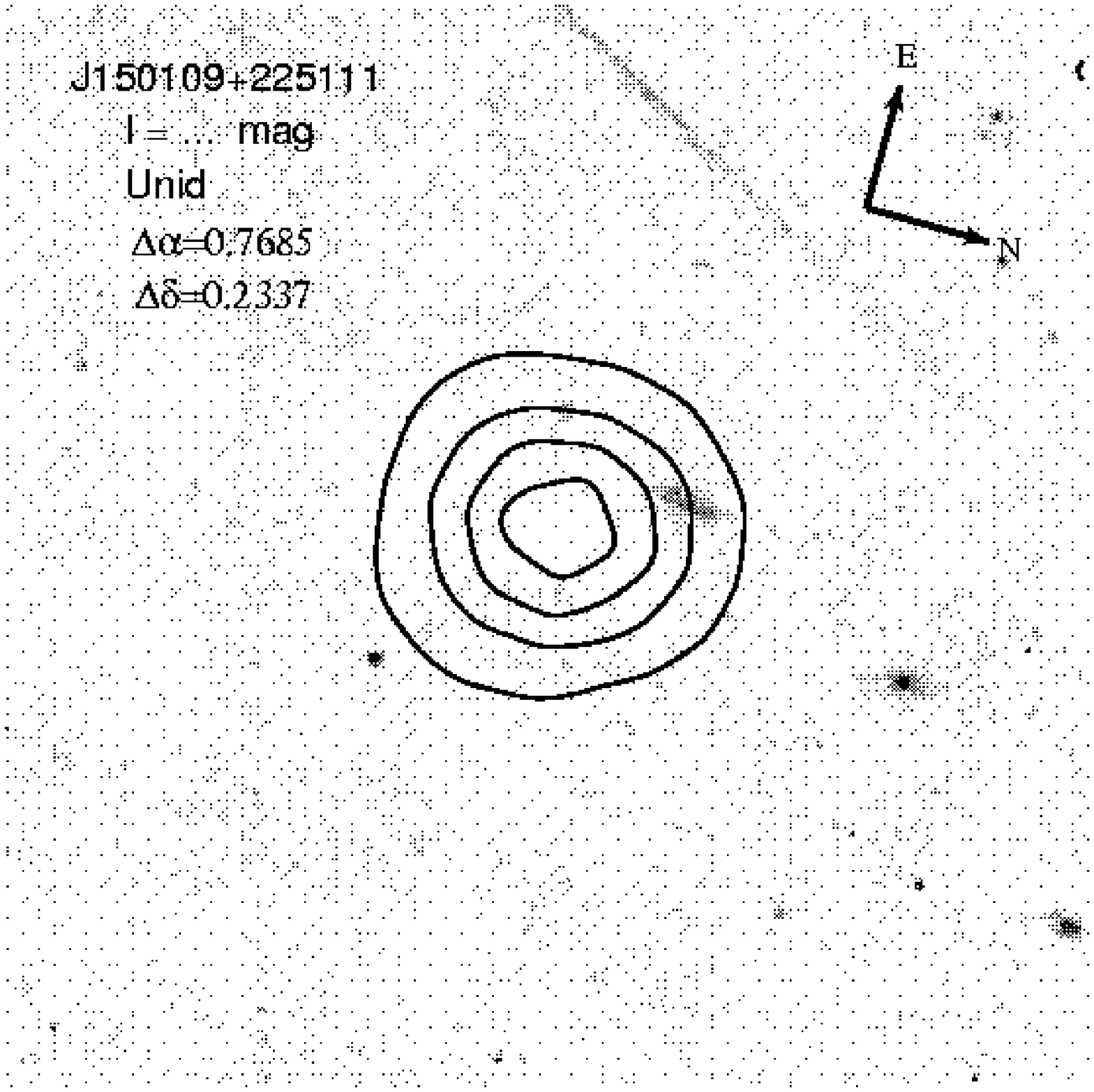 ,width=2.6truein}}}
\end{figure*}
\begin{figure*}
\centerline{\hbox{
\psfig{file=  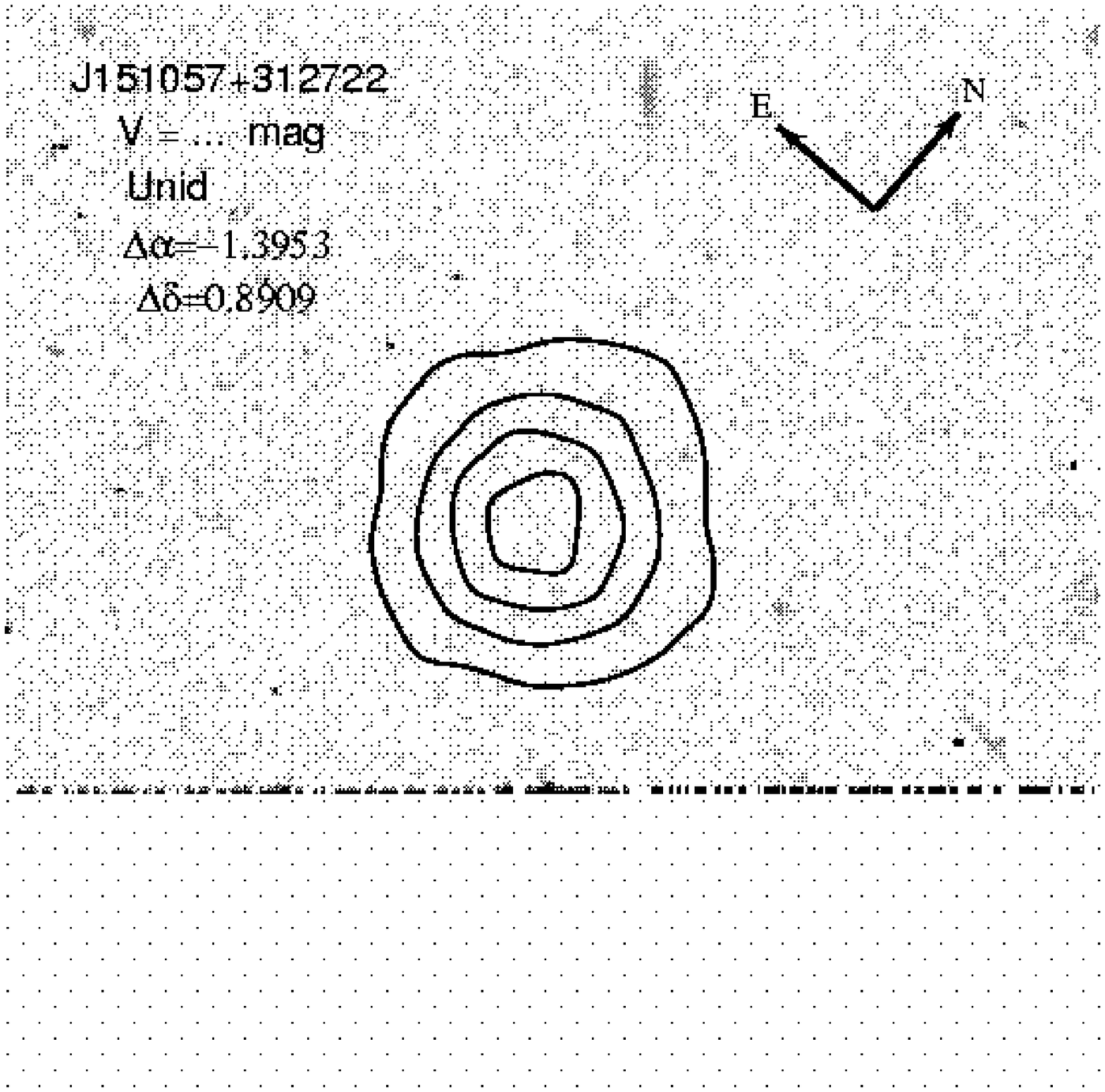 ,width=2.6truein}
\psfig{file=  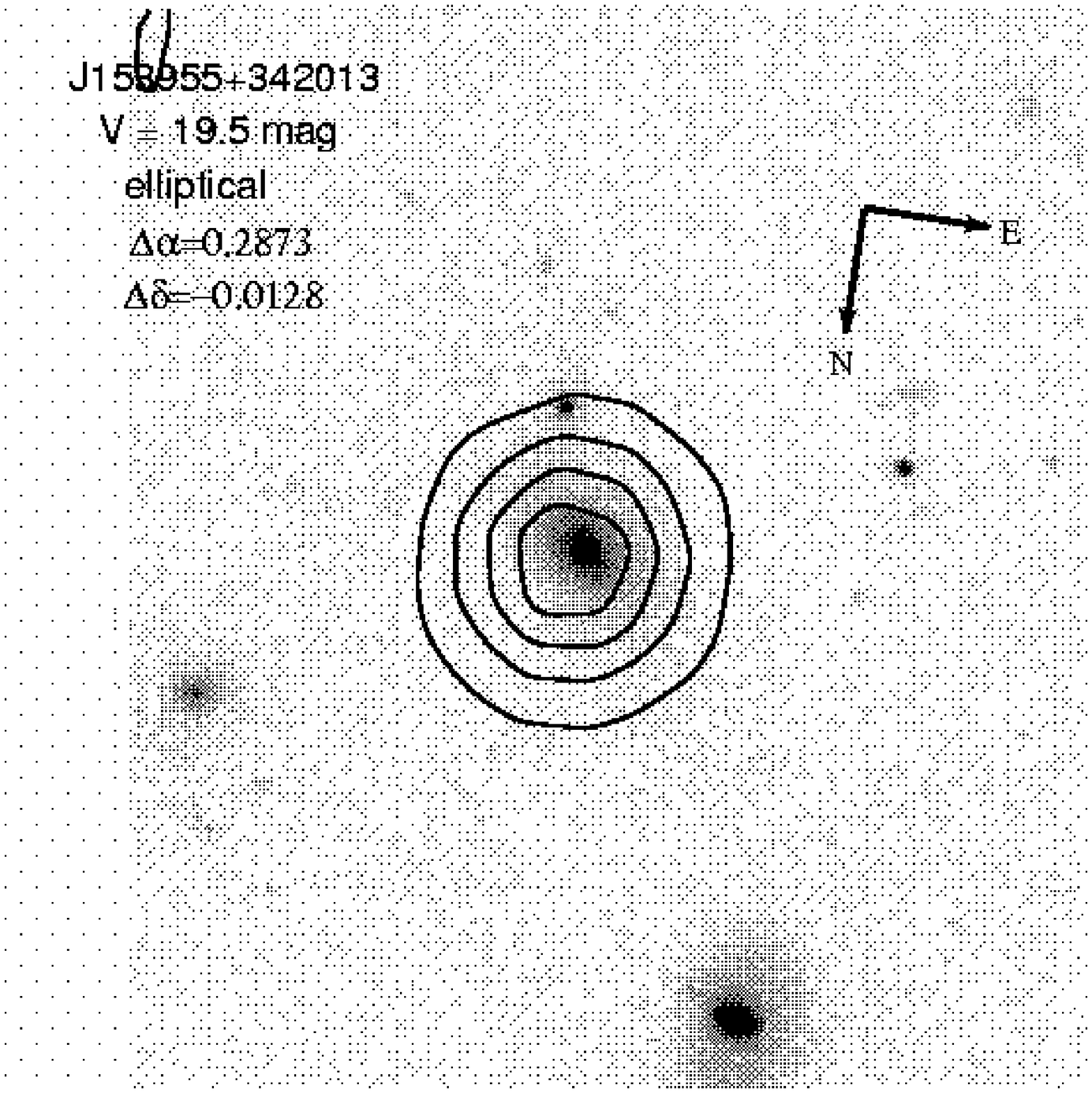 ,width=2.6truein}}}
\centerline{\hbox{
\psfig{file=  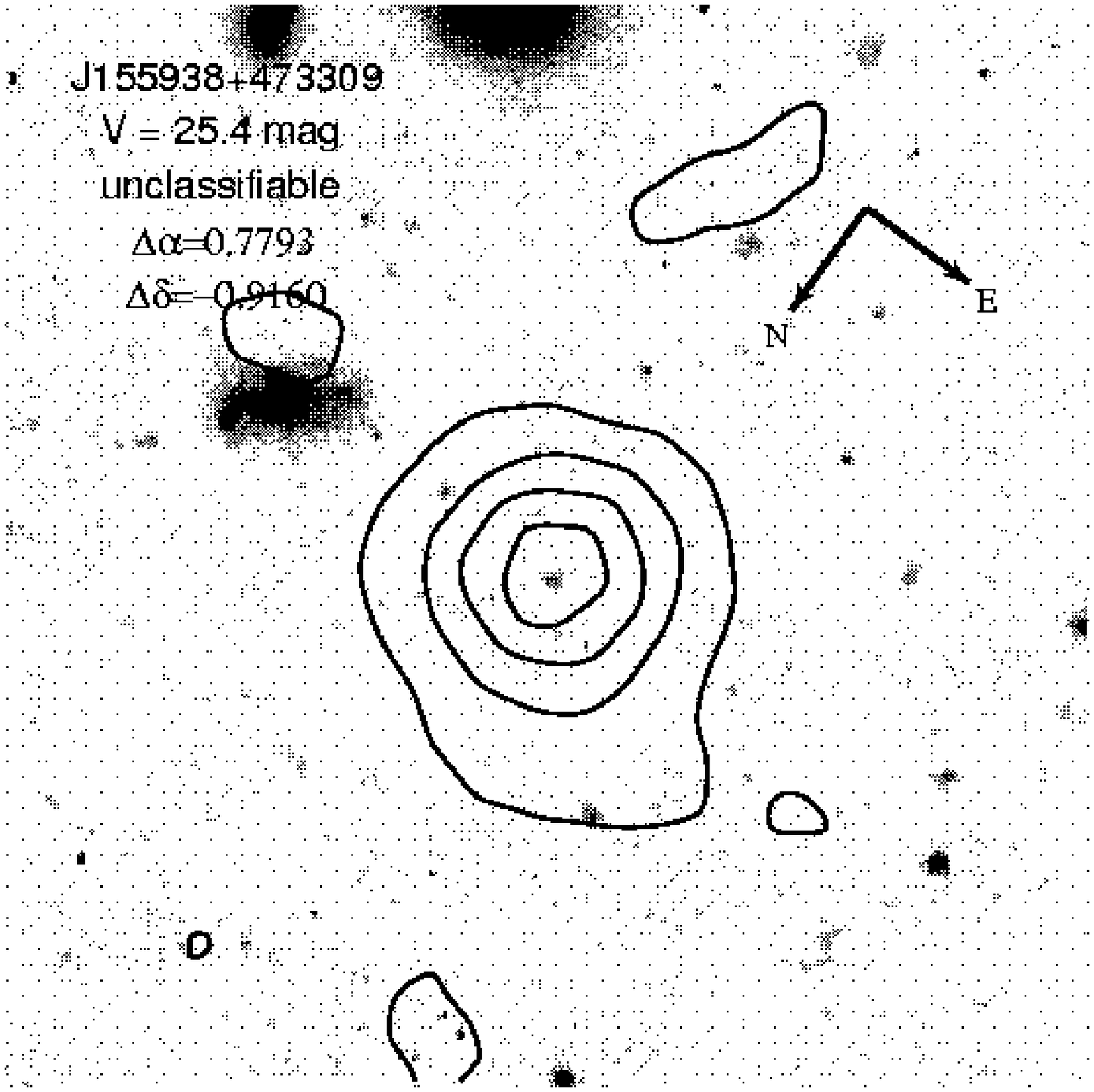 ,width=2.6truein}
\psfig{file=  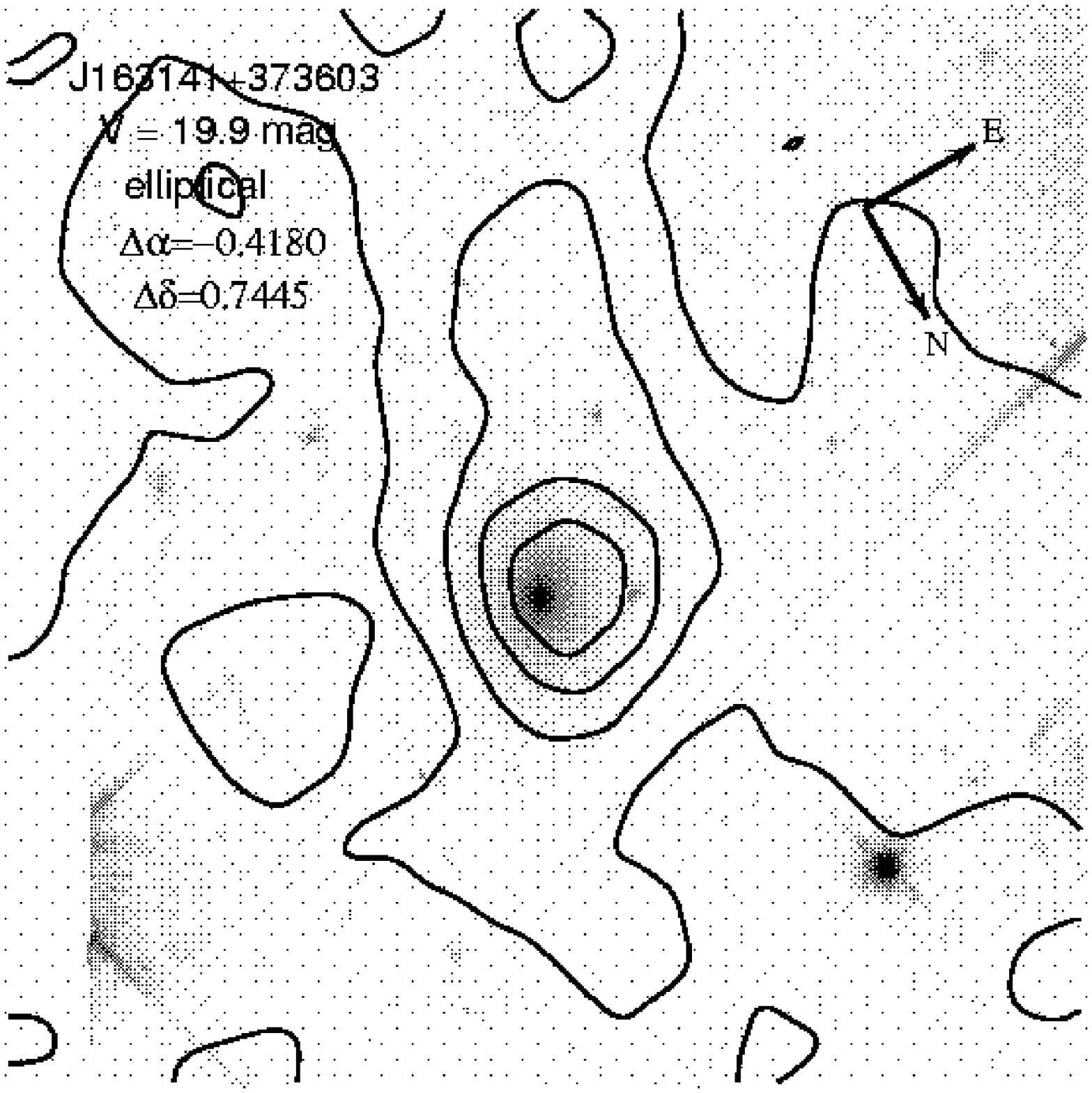 ,width=2.6truein}}}
\centerline{\hbox{
\psfig{file=  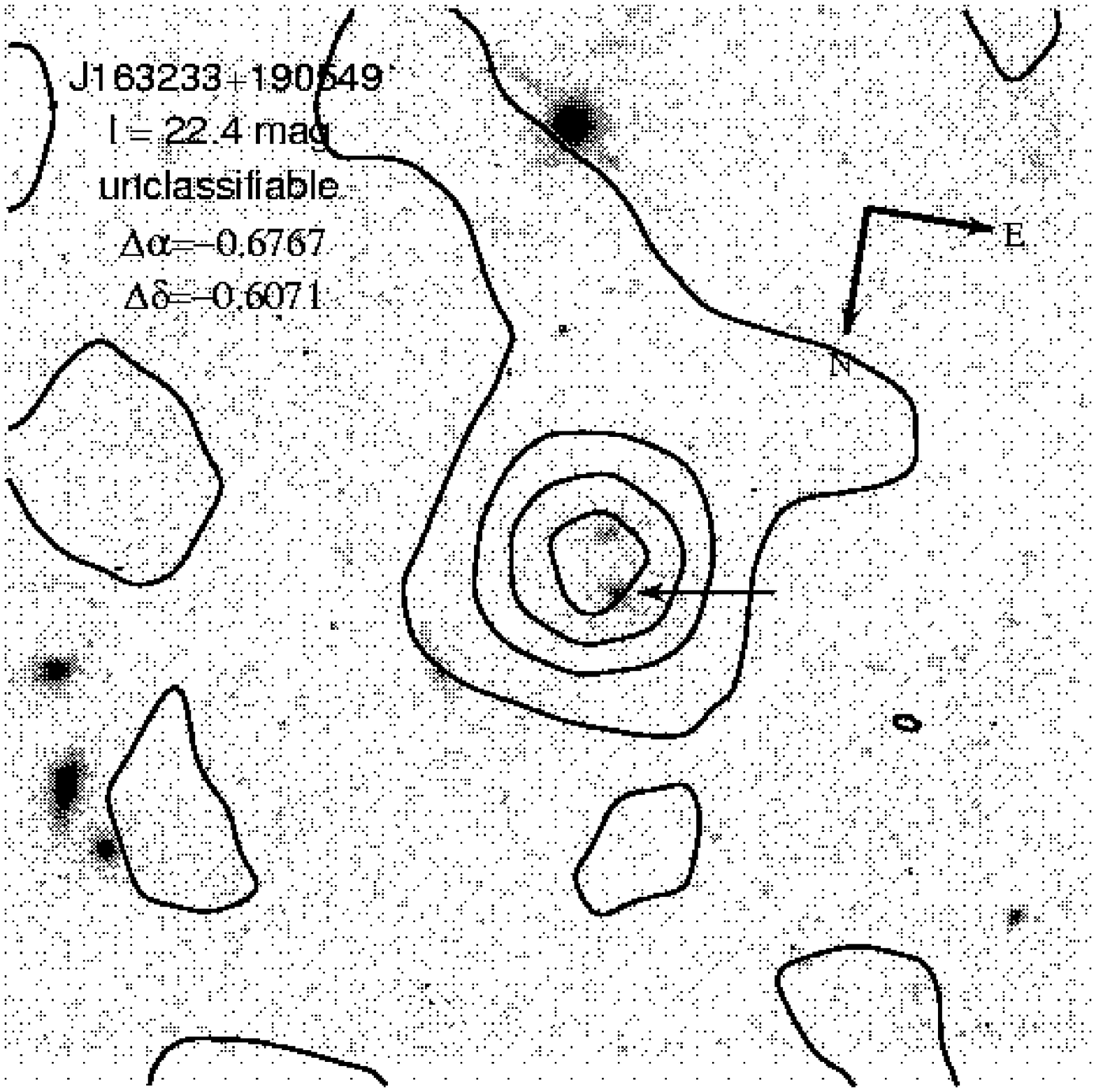 ,width=2.6truein}
\psfig{file=  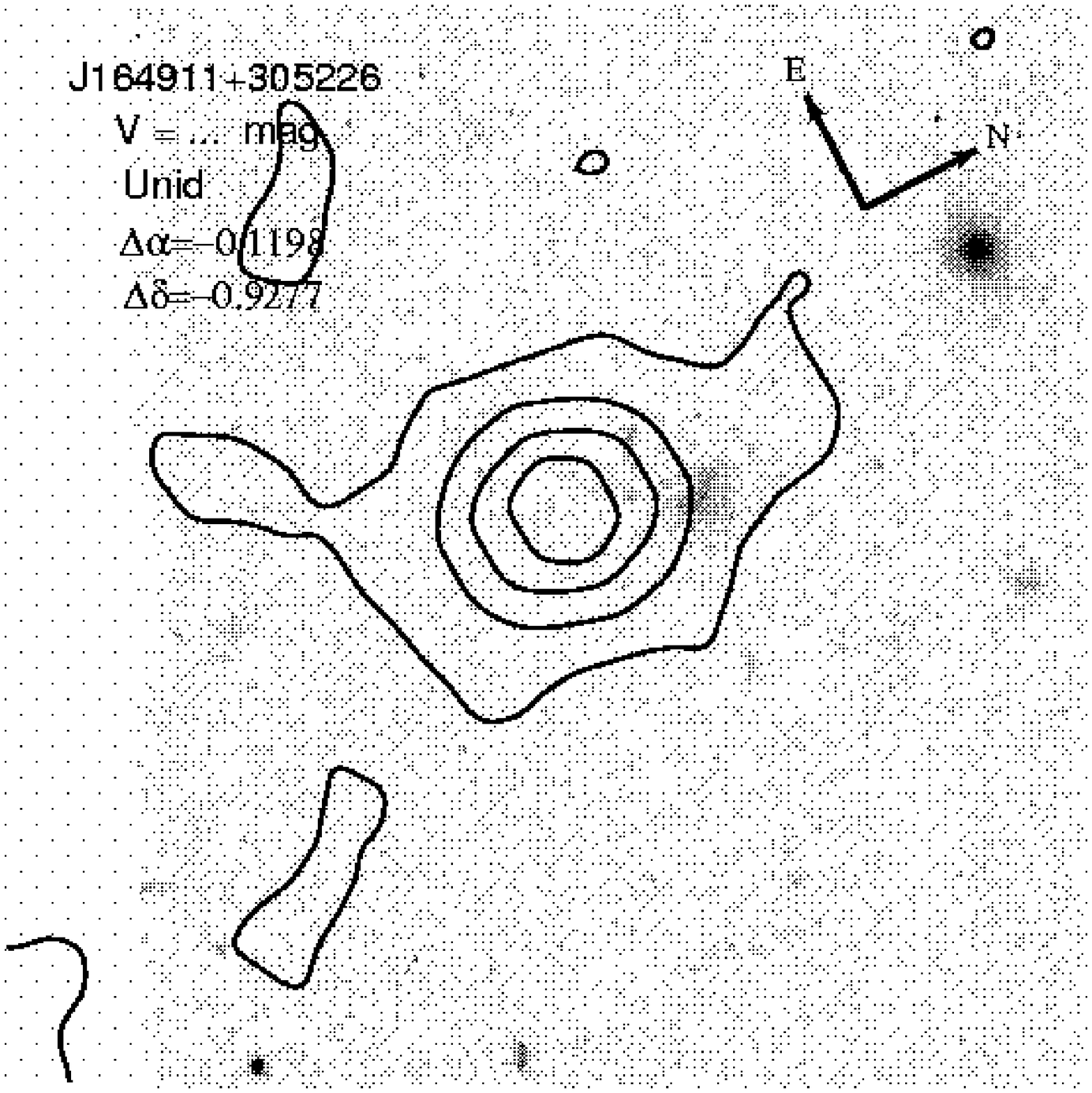 ,width=2.6truein}}}
\end{figure*}
\begin{figure*}
\centerline{\hbox{
\psfig{file=  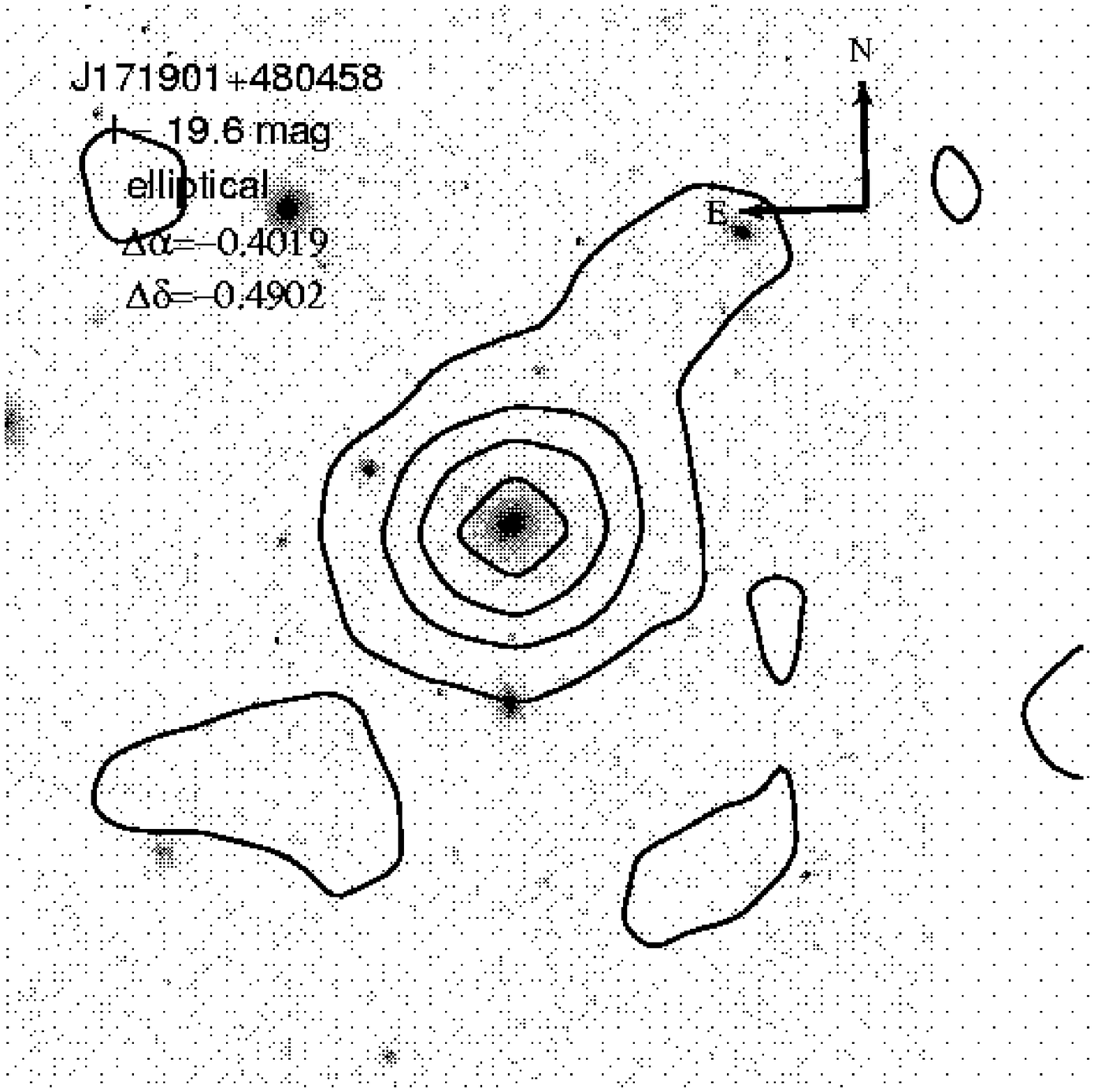 ,width=2.6truein}
\psfig{file=  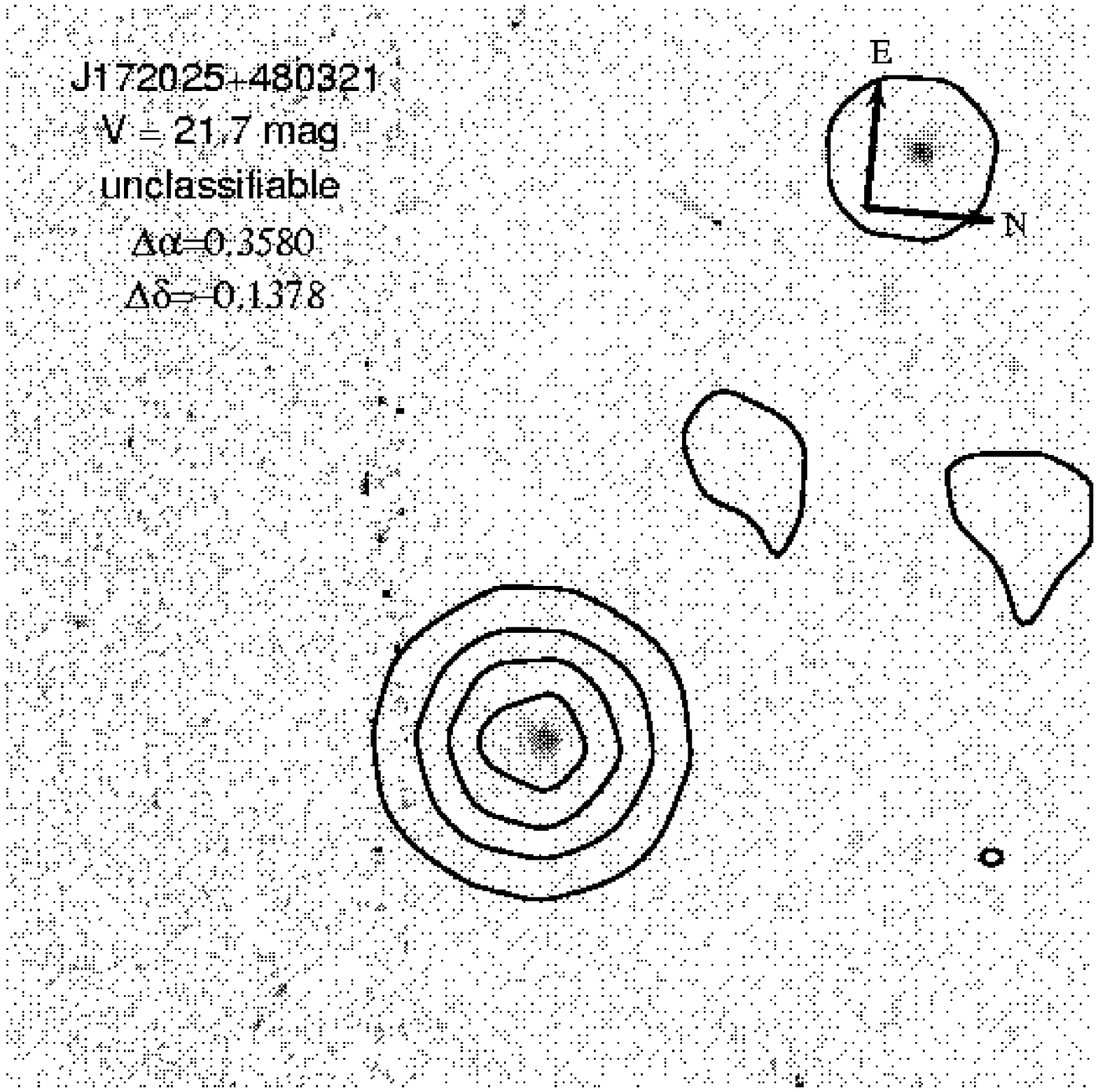 ,width=2.6truein}}}
\centerline{\hbox{
\psfig{file=  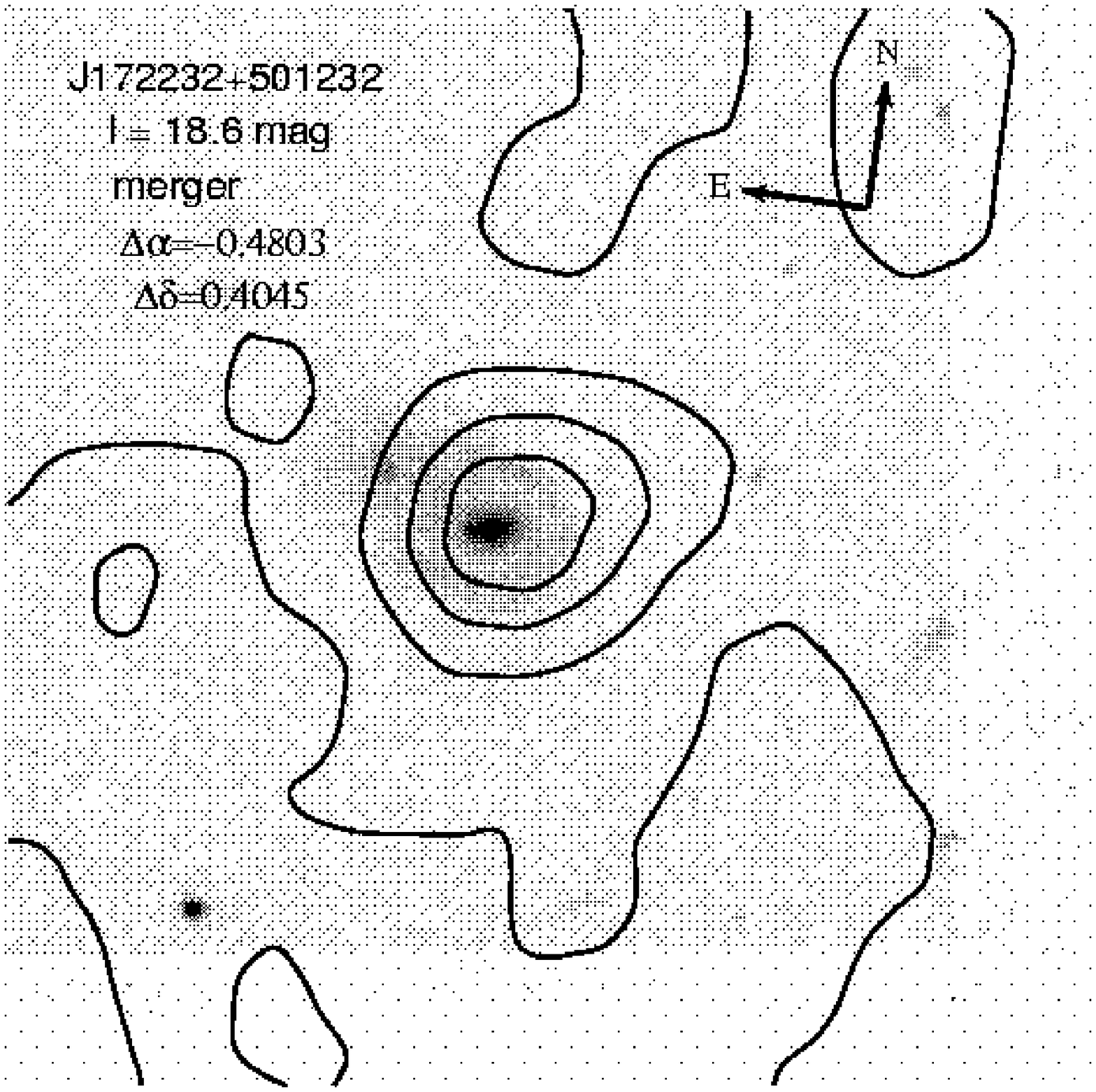 ,width=2.6truein}}}
\end{figure*}

\begin{table}
\caption{Unidentified Radio Sources}\label{unidtab}
\begin{tabular*}{0.48\textwidth}{@{\extracolsep{\fill}}cccr}
\hline
\hline
\multicolumn{1}{c}{Radio Source\tablenotemark{a}}    &    \multicolumn{1}{c}{Passband}   &   \multicolumn{1}{c}{Limit\tablenotemark{b}}   &   \multicolumn{1}{c}{$\log{S_{\rm 1.4}}$}\\
                    \multicolumn{1}{c}{$ $}        &    \multicolumn{1}{c}{$ $}      &   \multicolumn{1}{c}{(mag)}                         &   \multicolumn{1}{c}{(mJy)}\\
\hline
J030237.6+000818    &    F814W     &         24.5      &          0.70  \\
J082828.4+344131    &    F814W     &         24.7      &          1.01  \\ 
J120326.4+443635    &    F606W     &         25.0      &          1.43  \\
J121839.7+295325    &    F606W     &         24.8      &          1.53  \\
J125635.3+215632    &    F814W     &         24.2      &          0.52  \\
J150109.7+225111    &    F814W     &         23.6      &          0.82  \\
J151057.4+312722    &    F606W     &         23.7      &          0.49  \\
J164911.4+305226    &    F606W     &         24.3      &          0.44  \\
\hline
\tablenotetext{a}{The name of the source as per IAU recommendations, as used by FIRST.  FIRST Jhhmmss.s+ddmmss in which the coordinates are equinox J2000.0 and are truncated \citep{bec95}.}
\tablenotetext{b}{The limiting AB magnitude in specified filter given as $50\%$ completeness limit following \citet{sni02}.}
\end{tabular*}
\end{table}

To quantify  the reliability  of the radio-optical  identification, we
use the likelihood ratio  ($LR$) technique described by \citet{der77}.
If  we assume  that the  optical and  radio sources  are intrinsically
located at the  same position, and that potential  clustering does not
lead to a large fraction of false identifications, then the likelihood
that  a radio  source and  its potential  identification are  the same
physical object can be defined as:
\begin{equation}\label{lreqn}
LR(r)=\frac{1}{2\lambda}\exp{\left(\frac{r^2(2\lambda-1)}{2}\right)},
\end{equation}
where
\begin{eqnarray}
r&=&\sqrt{\left(\frac{\alpha^{\rm rad}-\alpha^{\rm opt}}{\sigma_{\alpha}}\right)^2+\left(\frac{\delta^{\rm rad}-\delta^{\rm opt}}{\sigma_{\delta}}\right)^2},\\
\lambda&=&\pi \sigma_{\alpha}\sigma_{\delta}\rho_{BG},\label{bgeqn}\\
\sigma_{\alpha}&=&\sqrt{(\sigma_{\alpha}^{\rm rad})^2+(\sigma_{\alpha}^{\rm opt})^2},\\
\sigma_{\delta}&=&\sqrt{(\sigma_{\delta}^{\rm rad})^2+(\sigma_{\delta}^{\rm opt})^2}.
\end{eqnarray}
where   $\rho_{\rm   BG}$  is   the   surface   density  of   objects,
($\sigma_{\alpha}^{\rm  rad}$,$\sigma_{\delta}^{\rm   rad}$)  are  the
radio  positional uncertainties  \citep{bec95}, $(\sigma_{\alpha}^{\rm
opt}$,$\sigma_{\delta}^{\rm  opt})$ are the  uncertainties in  the WCS
shifts derived from the  fits headers (discussed in \S~\ref{astro} and
\fig{scatter}),   and   $(\alpha^{\rm   rad},\delta^{\rm  rad})$   and
$(\alpha^{\rm opt},\delta^{\rm opt})$ are the celestial coordinates of
the radio and optical sources, respectively.

In  these  relatively  shallow  and  narrow HST  fields,  an  accurate
background density of objects cannot always be determined.  Therefore,
we adopt  the method of \citet{sni02}  to estimate the  galaxy flux at
the 50\% completeness  limit.  This flux is found  by assuming a given
light-profile,  and  using  the  noise  properties  of  the  image  to
determine the  brightness limit to  which 50\% the objects  would have
been recovered.  For a given  total magnitude, pure disk galaxies have
a lower  surface brightness than  pure ellipticals, and  are therefore
somewhat  more difficult  to detect.   However, the  50\% completeness
limit for  an exponential and de~Vaucouleurs typically  differ by only
$\lesssim\!0.1$~mag   for    a   fixed   size    of   $r\!=\!0\farcs3$
\citep{coh03}.  Given this minor  difference in brightness for the two
light-profiles, we  adopt the  limits derived from  the de~Vaucouleurs
profile, which are given  in \tab{tblflux}.  Finally, we integrate the
observed $I$-band  number counts \citep{met96,ode96,cas00,gar00,yas01}
to  these   derived  limits   to  determine  the   background  density
($\rho_{BG}$) in \eqn{bgeqn}.

\begin{table*}
\caption{Flux Results} \label{tblflux}
\begin{tabular*}{0.98\textwidth}{@{\extracolsep{\fill}}lccccccccc}
\hline
\hline
\multicolumn{1}{c}{Radio Source\tablenotemark{a}}  &  \multicolumn{1}{c}{Band\tablenotemark{b}} & \multicolumn{1}{c}{Magnitude} & \multicolumn{1}{c}{SB\tablenotemark{c}} & \multicolumn{1}{c}{$\log{S_{1.4}}$\tablenotemark{d}} & \multicolumn{1}{c}{Depth\tablenotemark{e}}  &  \multicolumn{1}{c}{LR\tablenotemark{f}}  & \multicolumn{1}{c}{$i'$\tablenotemark{g}} & \multicolumn{1}{c}{($g'-r'$)\tablenotemark{g}} & \multicolumn{1}{c}{($r'-i'$)\tablenotemark{g}} \\ 
\multicolumn{1}{c}{$ $}                            &  \multicolumn{1}{c}{$ $}                   & \multicolumn{1}{c}{(mag)}     & \multicolumn{1}{c}{(mag/$\square''$)}   & \multicolumn{1}{c}{ (mJy)}                           & \multicolumn{1}{c}{(mag)}                   &  \multicolumn{1}{c}{$ $}                  & \multicolumn{1}{c}{(mag)}                 & \multicolumn{1}{c}{(mag)}                      &   \multicolumn{1}{c}{(mag)}\\
\hline
J002219.1$-$013030 & $V$ & 24.59 & 23.39 & 0.47 & 25.2 & 11.17 & \nodata &  \nodata &  \nodata\\
J004322.3$-$001343 & $V$ & 18.93 & 16.12 & 0.83 & 23.9 & 10.17 & 18.91$\pm$0.02 &  0.28$\pm$0.02 &  0.12$\pm$0.03\\
J012616.6$-$012126 & $V$ & 19.84 & 22.45 & 1.15 & 23.9 & 2.03 & \nodata &  \nodata &  \nodata\\
J030237.6$+$000818 & $I$ & \nodata & \nodata & 0.70 & 24.5 & 0.0 & \nodata &  \nodata &  \nodata\\
J030249.5$+$000615 & $I$ & 19.79 & 20.33 & 0.15 & 24.7 & 6.42 & 20.13$\pm$0.18 &  0.64$\pm$0.27 &  0.65$\pm$0.22\\
J082820.6$+$344321 & $I$ & 25.59 & 27.21 & 0.69 & 24.9 & 11.65 & \nodata &  \nodata &  \nodata\\
J082828.4$+$344131 & $I$ & \nodata & \nodata & 1.01 & 24.7 & 0.0 & \nodata &  \nodata &  \nodata\\
J084715.5$+$443752 & $I$ & 24.66 & 23.27 & 0.62 & 23.5 & 44.39 & \nodata &  \nodata &  \nodata\\
J084849.5$+$445550 & $I$ & 19.29 & 21.06 & 0.67 & 24.6 & 10.53 & 19.13$\pm$0.06 &  2.40$\pm$0.61 &  1.02$\pm$0.12\\
J091205.2$+$350506 & $V$ & 23.79 & 22.50 & 0.93 & 23.7 & 17.04 & \nodata &  \nodata &  \nodata\\
J091251.0$+$525928 & $I$ & 21.14 & 20.80 & 0.47 & 22.9 & 23.59 & 20.42$\pm$0.12 &  4.79$\pm$0.89 &  1.51$\pm$0.34\\
J094926.6$+$295941 & $V$ & 20.06 & 21.85 & 1.25 & 25.4 & 8.43 & 19.14$\pm$0.07 &  2.08$\pm$0.28 &  0.64$\pm$0.12\\
J094930.7$+$295938 & $V$ & 22.70 & 22.54 & 0.55 & 25.6 & 3.87 & \nodata &  \nodata &  \nodata\\
J100354.5$+$285911 & $V$ & 23.10 & 24.77 & 0.60 & 24.6 & 6.25 & 22.44$\pm$0.44 &  0.96$\pm$1.45 &  1.70$\pm$1.07\\
J102437.2$+$470312 & $I$ & 21.75 & 20.36 & 0.39 & 24.3 & 15.62 & 22.31$\pm$0.31 &  1.36$\pm$0.78 &  0.91$\pm$0.49\\
J102744.6$+$282921 & $I$ & 22.44 & 24.31 & 0.15 & 24.6 & 13.33 & \nodata &  \nodata &  \nodata\\
J103452.3$+$394704 & $I$ & 21.73 & 21.94 & 1.40 & 24.5 & 27.44 & 21.82$\pm$0.48 &  1.92$\pm$1.26 &  0.39$\pm$0.63\\
J104630.8$-$001213 & $I$ & 18.8 & 21.73 & 0.65 & 24.0 & 28.24 & 19.19$\pm$0.08 &  1.91$\pm$0.29 &  0.71$\pm$0.13\\
J104757.0$+$123835 & $V$ & 20.61 & 20.77 & 0.94 & 24.2 & 22.27 & 19.33$\pm$0.06 &  2.04$\pm$0.31 &  0.77$\pm$0.09\\
J111908.6$+$211917 & $V$ & 15.79 & 16.85 & 0.71 & 24.2 & 6.92 & 14.15$\pm$0.00 & --0.08$\pm$0.00 &  0.35$\pm$0.00\\
J112520.7$+$420425 & $I$ & 19.36 & 20.67 & 0.28 & 24.2 & 21.47 & 19.54$\pm$0.06 &  1.76$\pm$0.45 &  1.21$\pm$0.14\\
J114526.3$+$193301 & $V$ & 24.40 & 22.67 & 0.84 & 24.8 & 12.83 & 22.79$\pm$0.36 &  0.30$\pm$0.66 &  0.88$\pm$0.65\\
J114910.5$-$002313 & $V$ & 22.82 & 23.44 & 0.75 & 23.3 & 20.46 & 22.62$\pm$0.95 &  3.97$\pm$1.13 &  0.41$\pm$1.30\\
J114928.3$+$143844 & $V$ & 22.83 & 20.71 & 1.15 & 24.5 & 25.49 & \nodata &  \nodata &  \nodata\\
J115642.8$+$022451 & $V$ & 18.99 & 20.42 & 1.00 & \nodata & \nodata & 18.21$\pm$0.02 &  1.09$\pm$0.03 &  0.32$\pm$0.03\\
$ $                & $I$ & 17.81 & 18.02 & 1.00 & 24.3 & 10.35 & 18.21$\pm$0.02 &  1.09$\pm$0.03 &  0.32$\pm$0.03\\
J120326.4$+$443635 & $V$ & \nodata & \nodata & 1.43 & 25.0 & 0.0 & \nodata &  \nodata &  \nodata\\
J121026.6$+$392909 & $V$ & 19.53 & 16.29 & 1.27 & 25.0 & 12.55 & 18.98$\pm$0.02 &  0.34$\pm$0.03 &  0.45$\pm$0.03\\
J121658.4$+$375439 & $I$ & 20.56 & 19.36 & 0.41 & 23.3 & 20.20 & \nodata &  \nodata &  \nodata\\
J121705.5$-$031137 & $I$ & 23.89 & 22.49 & 1.51 & 24.3 & 11.34 & \nodata &  \nodata &  \nodata\\
J121707.7$-$031127 & $I$ & 15.67 & 17.34 & 0.97 & 23.2 & 2.44 & \nodata &  \nodata &  \nodata\\
J121839.7$+$295325 & $V$ & \nodata & \nodata & 1.53 & 24.8 & 0.0 & \nodata &  \nodata &  \nodata\\
J122331.0$+$155245 & $I$ & 21.02 & 19.03 & 0.64 & 24.9 & 18.67 & 23.91$\pm$2.84 &  0.92$\pm$0.77 & --1.66$\pm$2.87\\
J122624.4$+$173228 & $V$ & 18.82 & 20.38 & 1.15 & 24.5 & 13.56 & \nodata &  \nodata &  \nodata\\
J125029.2$+$302527 & $I$ & 17.83 & 19.93 & 0.44 & 23.9 & 31.38 & 17.85$\pm$0.02 &  1.34$\pm$0.04 &  0.53$\pm$0.03\\
J125635.3$+$215632 & $I$ & \nodata & \nodata & 0.52 & 24.2 & 0.0 & \nodata &  \nodata &  \nodata\\
J125650.0$+$220630 & $V$ & 22.44 & 20.71 & 0.83 & 24.7 & 5.71 & \nodata &  \nodata &  \nodata\\
J131223.6$+$424517 & $I$ & 23.19 & 22.90 & 0.47 & 25.9 & 8.08 & \nodata &  \nodata &  \nodata\\
J131617.8$+$420239 & $V$ & 16.66 & 19.42 & 0.00 & 24.3 & 14.92 & 16.12$\pm$0.01 &  0.78$\pm$0.01 &  0.45$\pm$0.01\\
J134219.9$-$005816 & $I$ & 24.08 & 23.53 & 1.23 & 24.0 & 17.42 & \nodata &  \nodata &  \nodata\\
J140019.9$+$044421 & $I$ & 19.51 & 20.32 & 0.05 & 24.6 & 9.73 & 19.51$\pm$0.10 &  1.47$\pm$0.43 &  1.08$\pm$0.22\\
J143530.0$+$484534 & $V$ & 22.07 & 23.13 & 0.94 & 24.3 & 17.58 & \nodata &  \nodata &  \nodata\\
J150109.7$+$225111 & $I$ & \nodata & \nodata & 0.82 & 23.6 & 0.0 & \nodata &  \nodata &  \nodata\\
J151057.4$+$312722 & $V$ & \nodata & \nodata & 0.49 & 23.7 & 0.0 & \nodata &  \nodata &  \nodata\\
J153955.0$+$342013 & $V$ & 19.49 & 21.28 & 1.11 & 23.7 & 26.62 & 18.21$\pm$0.04 &  1.86$\pm$0.12 &  0.58$\pm$0.06\\
J155938.7$+$473309 & $V$ & 25.37 & 24.08 & 0.49 & 25.0 & 6.78 & \nodata &  \nodata &  \nodata\\
J163141.4$+$373603 & $V$ & 19.87 & 20.87 & 0.41 & 24.7 & 6.43 & 18.57$\pm$0.05 &  1.62$\pm$0.17 &  0.81$\pm$0.08\\
J163233.8$+$190550 & $I$ & 22.36 & 21.26 & 0.26 & 23.4 & 17.40 & \nodata &  \nodata &  \nodata\\
J164911.4$+$305226 & $V$ & \nodata & \nodata & 0.44 & 24.3 & 0.0 & \nodata &  \nodata &  \nodata\\
J171901.1$+$480458 & $V$ & 21.01 & 21.27 & 0.34 & \nodata & \nodata & \nodata &  \nodata &  \nodata\\
$ $                & $I$ & 19.57 & 19.57 & 0.34 & 23.9 & 11.65 & \nodata &  \nodata &  \nodata\\
J172025.4$+$480321 & $V$ & 21.66 & 22.43 & 1.48 & 23.1 & 75.08 & \nodata &  \nodata &  \nodata\\
J172232.9$+$501232 & $V$ & 20.11 & 20.65 & 0.45 & \nodata & \nodata & \nodata &  \nodata &  \nodata\\
$ $                & $I$ & 18.56 & 19.44 & 0.45 & 24.4 & 9.80 & \nodata &  \nodata &  \nodata\\
\hline
\tablenotetext{a}{The name of the source as per IAU recommendations, as used by FIRST.  FIRST Jhhmmss.s+ddmmss in which the coordinates are equinox J2000.0 and are truncated \citep{bec95}.}
\tablenotetext{b}{$V$ and $I$ refer to the HST-WFPC2 bandpasses F606W or F814W, respectively.}
\tablenotetext{c}{Surface brightness in mag$/$arcsec$^2$ within the effective radius ($R_e$).}
\tablenotetext{d}{Integrated flux at 1.4~GHz in mJy provided by the FIRST catalog.}
\tablenotetext{e}{The 50\% completeness limit as determined by the \citet{sni02} method (see \S~\ref{imprep}).}
\tablenotetext{f}{The likelihood ratio as discussed \S~\ref{imprep}.}
\tablenotetext{g}{Obtained from SDSS-dr5 \citep{ade07}.}
\end{tabular*}
\end{table*}

\begin{figure}
\epsscale{1.0}
\plotone{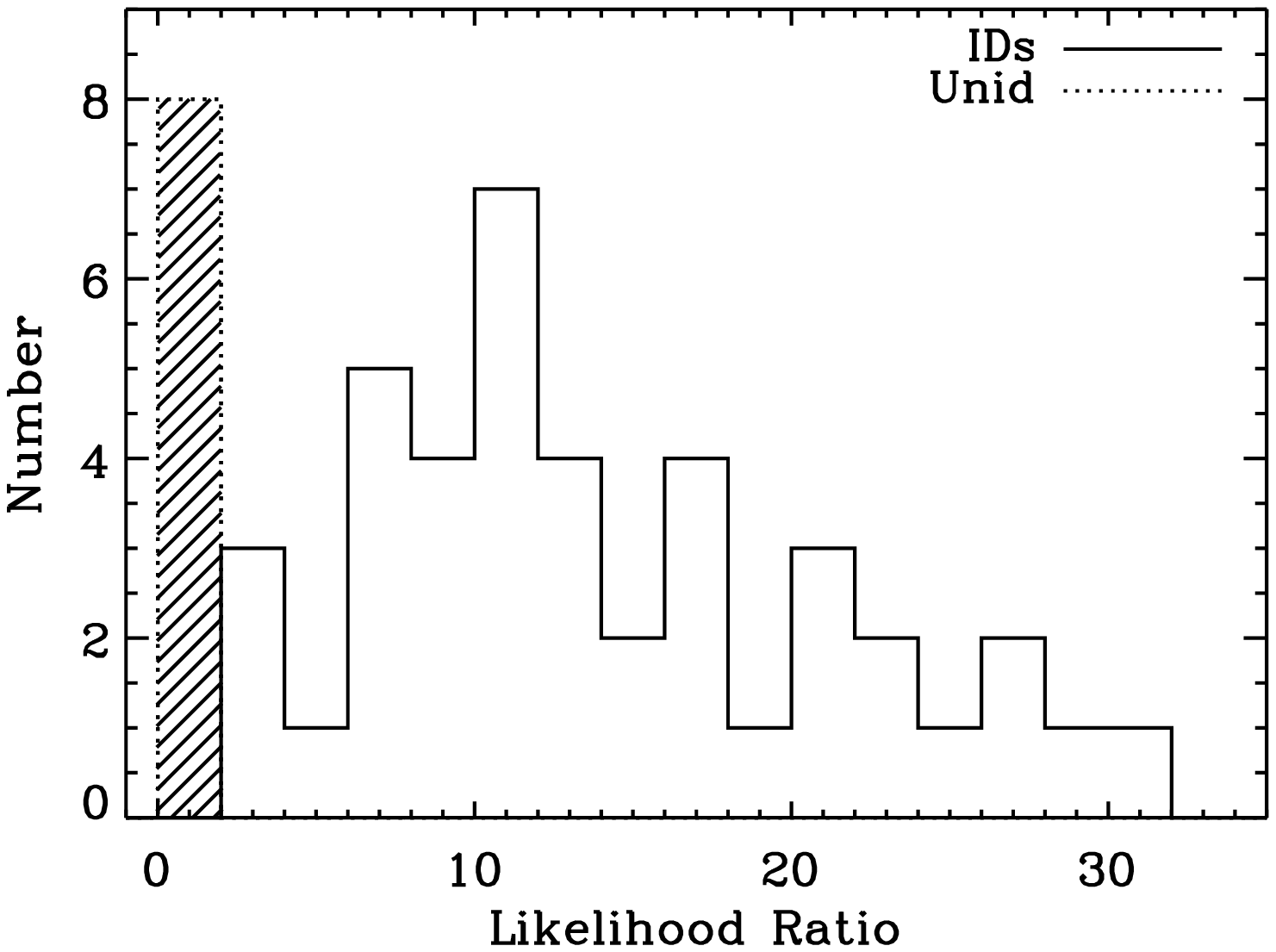}
\caption{Distribution  of  likelihood   ratios,  $LR$  as  defined  by
\eqn{lreqn}.   We compute  the likelihood  that a  radio source  has a
given optical counterpart following \citet{der77}, which is summarized
in \S~\ref{imprep}.   We show all radio sources  which were positively
identified with  an optical counterpart as a  solid distribution.  The
unidentified galaxies  are represented as a  hatched distribution.  As
argued   in  the   text,  every   positively  identified   object  has
$LR\!\geq\!2$,  while  the  unidentified galaxies  have  $LR\!\ll\!2$.
Note     that    two    objects     with    $LR\!>\!40$     are    not
represented.}\label{lrfig}
\end{figure}

The resulting likelihood ratios are listed in \tab{tblflux}, and their
distribution  is plotted in  \fig{lrfig}.  Following  \citet{der77}, a
$LR\!\geq\!2$   provides   a   compromise  between   reliability   and
completeness  for the  radio-optical  identifications.  Therefore,  we
adopt  the  \citet{der77}  likelihood  threshold, where  objects  with
$LR\!\geq\!2$ are  a positive identification.  For  radio sources with
no obvious counterpart (the  Unids), the likelihood ratio was computed
for  the nearest  optical  source, and  all  of these  systems have  a
$LR\!\ll\!2$.

\section{Results} \label{results}

\subsection{Optical Morphologies} \label{morph}

Many of the  counterparts to these faint radio  galaxies are also very
faint     in    the    optical.      Only    2/43     galaxies    have
$I\!\lesssim\!16.0$~mag, the approximate  brightness limit to classify
galaxies from  ground-based imaging in average  seeing conditions.  Of
the 43 radio sources for  which an optical counterpart was found, only
25   can   be  visually   classified   to   a   brightness  limit   of
$AB\!\lesssim\!22$~mag.   The visual  morphologies were  determined by
three  independent  observers  (J.R.,  S.H.C., and  R.A.W.),  and  the
average classification  is listed  for each object  in \tab{tblmorph}.
In order to also provide some {\it quantitative} measure of the galaxy
morphologies, we fit the surface-brightness profiles with the S\'ersic
or $r^{1/n}$ of model:
\begin{equation}\label{sersic}
I(r)=I_e\exp{\left(-b_n\left[\left(\frac{r}{r_e}\right)^{1/n}-1\right]\right)},
\end{equation}
where $I_e$ is  the intensity at the effective  radius $r_e$ and $b_n$
is a constant defined in terms of $n$ which describes the shape of the
profile   \citep[eg.][]{gra05}.    The   S\'ersic  index   ($n$)   can
morphologically classify  galaxies by type,  where elliptical galaxies
have $n\!\simeq\!4$ and  later-type disk galaxies have $n\!\simeq\!1$.
Therefore,  we  have  a  quantitative, morphological  measurements  to
complement  our  visual classifications  for  the 25/43  classifiable,
optical counterparts.  We give these results in \tab{tblmorph}.

For the 25/43  galaxies which are sufficiently bright  and extended to
be  visually classified, there  are 15~ellipticals,  1~S0, 2~late-type
spirals,  3 optical point  sources (these  are generally  quasars, see
\S~\ref{comment}), and 4~mergers.   Therefore, at the millijansky flux
levels, 19/25 galaxies are early-type  or quasars, while only 2/25 are
later-type galaxies.  However, the surface brightness analysis permits
a ``continuum''  of galaxy types based  on the best-fit  value for the
S\'ersic  index.  Therefore,  to produce  a  comparable classification
scheme (such  as early- versus late-type), we  collapse this continuum
using a  threshold value  for the S\'ersic  index of  $n\!\geq\!2$ are
early-types and $n\!<\!2$ are late-types \citep[eg.][]{dri06}.  Of the
34/43~galaxies for which we could reliably measure the S\'ersic index,
20 are early-types  while 14 are later-types, and  seven of these only
have  $n\!<\!2$ at  the $1\sigma$-level  (such  as J002219.1--013030).
Based on  these S\'ersic indices, we find  $\sim\!60$\% of millijansky
radio galaxies  are morphologically early-type  systems.  Furthermore,
these classifications based on the S\'ersic index generally agree with
our  visually  assigned morphologies.   Based  on these  morphological
analyses, the  dominance of red  galaxies in millijansky  radio galaxy
samples \citep{kro85,ham95a} is  now confirmed by robust morphological
observations with high-resolution imaging from HST.

\begin{table*}
\caption{Morphological and Structural Results} \label{tblmorph}
\begin{tabular*}{0.98\textwidth}{@{\extracolsep{\fill}}lcccrrr}
\hline
\hline
\multicolumn{1}{c}{Radio Source\tablenotemark{a}} & \multicolumn{1}{c}{$R_e$\tablenotemark{b}} & \multicolumn{1}{c}{$n$\tablenotemark{c}} & \multicolumn{1}{c}{$\chi^2_{\nu}$\tablenotemark{d}} &  \multicolumn{1}{c}{ANN\tablenotemark{e}} & \multicolumn{1}{c}{Visual\tablenotemark{f}} & \multicolumn{1}{c}{WFPC2\tablenotemark{g}} \\ 
\multicolumn{1}{c}{$ $}                           & \multicolumn{1}{c}{(arcsec)}               & \multicolumn{1}{c}{$ $}                  & \multicolumn{1}{c}{$ $}                             &  \multicolumn{1}{c}{Type}                 & \multicolumn{1}{c}{Type}                    & \multicolumn{1}{c}{Target} \\
\hline
J002219.1$-$013030 & 0.23    & 1.56$\pm$0.82 & 0.30    &    2.5   & x & ANY  \\
J004322.3$-$001343 & 0.11    & \nodata       & \nodata &  --4.1   & ps & ANY  \\
J012616.6$-$012126 & 1.33    & 4.95$\pm$0.74 & 0.11    &  5.0     & SBb-SBc & ANY  \\
J030237.6$+$000818 & \nodata & \nodata       & \nodata &  \nodata & UNID & FIELD-030239+00065  \\
J030249.5$+$000615 & 0.51    & 4.39$\pm$0.41 & 0.01    &  --2.3   & E & FIELD-030251+00071  \\
J082820.6$+$344321 & 0.84    & \nodata       & \nodata &    5.5   & x & 6C0825+34  \\
J082828.4$+$344131 & \nodata & \nodata       & \nodata &  \nodata & UNID & 6C0825+34  \\
J084715.5$+$443752 & 0.21    & 1.00$\pm$0.00 & 1.84    &    8.5   & xE & GAL-084720+443739  \\
J084849.5$+$445550 & 0.90    & 3.98$\pm$0.03 & 0.19    &    0.6   & E (cluster) & LYNX-E  \\
J091205.2$+$350506 & 0.22    & 1.24$\pm$0.75 & 0.05    &    7.6   & xE & ANY  \\
J091251.0$+$525928 & 0.34    & 1.80$\pm$0.70 & 0.18    &    5.2   & xE & SBS0909+523  \\
J094926.6$+$295941 & 0.91    & 3.46$\pm$0.46 & 0.01    &  --3.1   & E & ANY  \\
J094930.7$+$295938 & 0.37    & 2.00$\pm$0.54 & 0.02    &  1.1     & S0 & ANY  \\
J100354.5$+$285911 & 0.86    & \nodata       & \nodata &  7.2     & merger & ANY  \\
J102437.2$+$470312 & 0.21    & 3.77$\pm$0.02 & 1.19    &  --4.1   & xE & PAR  \\
J102744.6$+$282921 & 0.94    & \nodata       & \nodata &  6.8     & x & HIGH  \\
J103452.3$+$394704 & 0.44    & 2.66$\pm$0.67 & 0.20    &  2.1     & xE & PAR  \\
J104630.8$-$001213 & 1.57    & \nodata       & \nodata &  3.0 & merger & 10HR-B  \\
J104757.0$+$123835 & 0.43    & 2.58$\pm$0.41 & 0.01    &  --2.5 & E & ANY  \\
J111908.6$+$211917 & 0.65    & 2.63$\pm$0.23 & 0.11    &  --0.9 & ps & PG1116+215  \\
J112520.7$+$420425 & 0.73    & 1.72$\pm$0.44 & 0.01    &  3.0 & E/S0 & HI-LAT  \\
J114526.3$+$193301 & 0.18    & 0.73$\pm$0.24 & 1.08    &  3.0 & xC & ANY  \\
J114910.5$-$002313 & 0.53    & 3.66$\pm$0.04 & 0.50    &  8.3 & xC & ANY  \\
J114928.3$+$143844 & 0.15    & 2.17$\pm$0.30 & 0.25    &  0.9 & E & ANY  \\
J115642.8$+$022451 & 0.44    & 1.76$\pm$0.60 & 0.04    &  1.9 & E & PAR  \\
J120326.4$+$443635 & \nodata & \nodata       & \nodata &  \nodata & UNID & ANY  \\
J121026.6$+$392909 & 0.09    & 4.28$\pm$0.02 & 0.73    &  --2.8 & ps & ANY  \\
J121658.4$+$375439 & 0.23    & 5.45$\pm$0.33 & 0.06    &  --2.6 & xE & MS1214.3+3811  \\
J121705.5$-$031137 & 0.21    & 0.50$\pm$0.17 & 0.00    &  6.4 & x & 2MASSW-J1217-03  \\
J121707.7$-$031127 & 0.86    & 2.62$\pm$0.39 & 0.03    &  --3.5 & E & 2MASSW-J1217-03  \\
J121839.7$+$295325 & \nodata & \nodata       & \nodata &  \nodata & UNID & ANY  \\
J122331.0$+$155245 & 0.16    & \nodata       & \nodata &  --1.7 & xC & PAR  \\
J122624.4$+$173228 & 0.82    & 3.08$\pm$0.39 & 0.04    &  --2.2 & E (companion) & ANY  \\
J125029.2$+$302527 & 1.05    & 1.96$\pm$0.48 & 0.09    &  0.4 & E (companions) & HI-LAT  \\
J125635.3$+$215632 & \nodata & \nodata       & \nodata &  \nodata & UNID & ANYWHERE  \\
J125650.0$+$220630 & 0.18    & 3.81$\pm$0.07 & 1.88    &  --1.2 & xC & ANY  \\
J131223.6$+$424517 & 0.35    & 1.54$\pm$0.70 & 0.10    &  3.1 & E & SSA13E  \\
J131617.8$+$420239 & 1.42    & 1.48$\pm$0.34 & 0.54    &  0.2 & SBc & ANY  \\
J134219.9$-$005816 & 0.31    & \nodata       & \nodata &  8.8 & x & HI-LAT  \\
J140019.9$+$044421 & 0.58    & 5.43$\pm$0.33 & 0.06    &  --0.5 & E (companions) & BIG2  \\
J143530.0$+$484534 & 0.65    & \nodata       & \nodata &  6.4 & merger & NGC5689  \\
J150109.7$+$225111 & \nodata & \nodata       & \nodata &  \nodata & UNID & TVLM513-46546  \\
J151057.4$+$312722 & \nodata & \nodata       & \nodata &  \nodata & UNID & WAS96  \\
J153955.0$+$342013 & 0.91    & 2.68$\pm$0.54 & 0.01    &  3.6 & E & ANY  \\
J155938.7$+$473309 & 0.22    & 0.47$\pm$0.19 & 0.86    &  8.6 & xC & ANY  \\
J163141.4$+$373603 & 0.63    & 2.81$\pm$0.41 & 0.04    &  --3.3 & E & PAR  \\
J163233.8$+$190550 & 0.24    & 1.88$\pm$1.72 & 0.03    &  3.4 & xC & 2MASSW1632+1904  \\
J164911.4$+$305226 & \nodata & \nodata       & \nodata &  \nodata & UNID & HIGH  \\
J171901.1$+$480458 & 0.40    & 3.10$\pm$0.51 & 0.01    &  --0.8 & E & HIGH  \\
J172025.4$+$480321 & 0.57    & 0.68$\pm$0.28 & 0.26    &  6.1 & xE & ANY  \\
J172232.9$+$501232 & 0.60    & \nodata       & \nodata &  --2.3 & merger & PAR  \\
\hline
\tablenotetext{a}{The name of the source as per IAU recommendations, as used by FIRST.  FIRST Jhhmmss.s+ddmmss in which the coordinates are equinox J2000.0 and are truncated \citep{bec95}.}
\tablenotetext{b}{Half light radius in arcsec as measured by LMORPHO \citep{ode96}.}
\tablenotetext{c}{S\'ersic index ($n$) defined by the \eqn{sersic}.}
\tablenotetext{d}{Reduced $\chi^2_{\nu}$ between the observed profile and \eqn{sersic}.}
\tablenotetext{e}{Artificial Neural Network type classification as measured by LMORPHO\citep{ode96}}
\tablenotetext{f}{Morphology of optical counterpart. `x' used to classify objects too small to assign a reliable morphology. `xC' indicates the object is compact. `xE' indicates the object is extended.  `*' indicates a unique note for that source.}
\tablenotetext{g}{This is the header keyword for the field target from the WFPC2 images, which helped us assess the ``randomness" of each WFPC2 field.}
\end{tabular*}
\end{table*}

\vspace*{24pt}

\subsection{Optical Magnitude Distribution} \label{mdist}

\begin{figure}
\epsscale{1.0}
\plotone{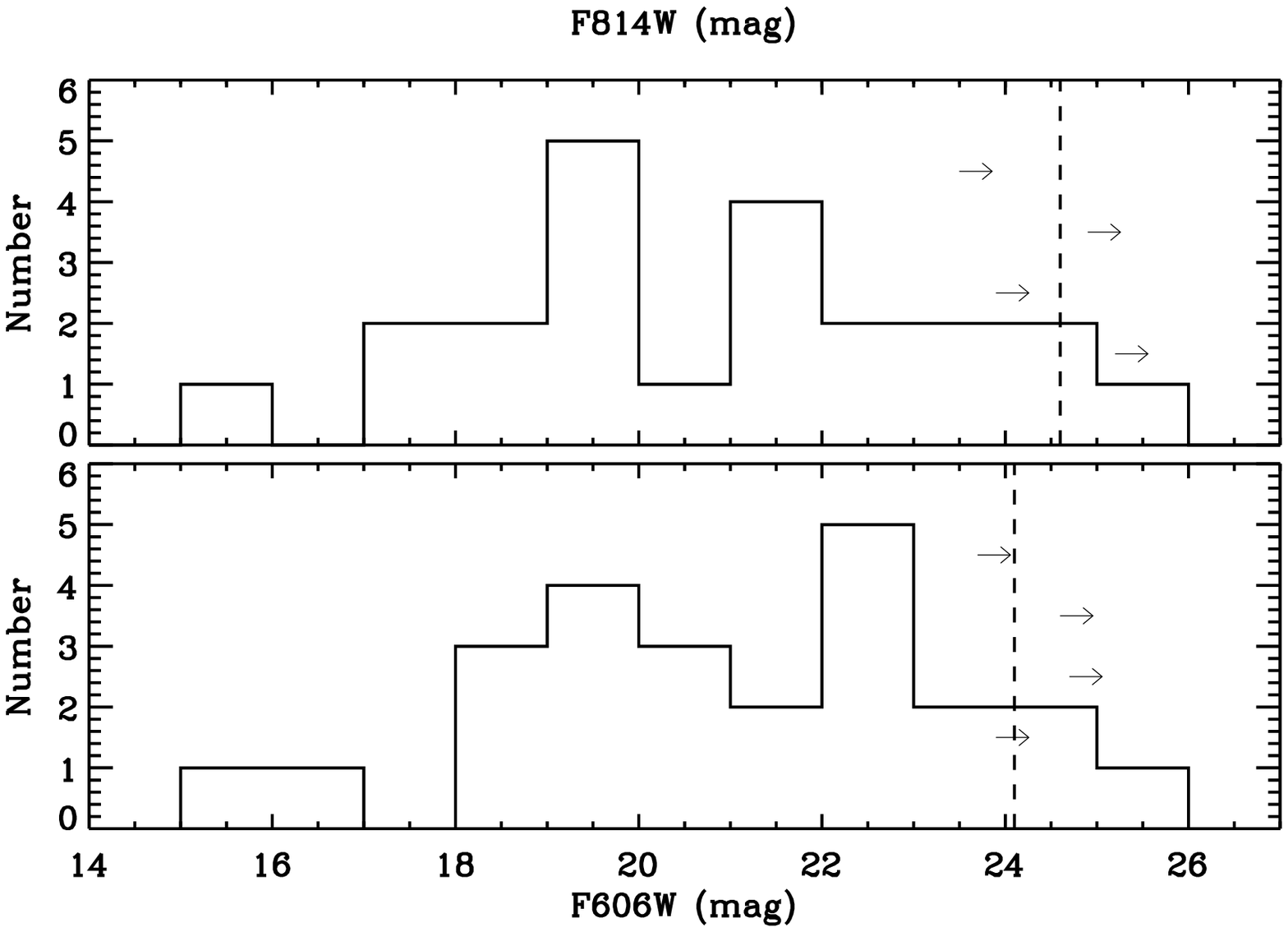}
\caption{Optical WFPC2 magnitude  distributions, separated by bandpass
(top: $V$-band, bottom: $I$-band).  The arrows represent the depth for
fields where no optical  counterpart could be identified. The vertical
dashed line shows the  flux-averaged, 50\% completeness limit for each
bandpass.   These magnitude  distributions peak  $\gtrsim\!2$~mag {\it
brighter} than  the average completeness  limit for the  average WFPC2
field.  While the statistics may be relatively low, this suggests that
the  turn-over in  the  optical magnitude  distributions from  earlier
studies was not due to their shallow imaging. }\label{origmagdist}
\end{figure}

We  measure  optical  fluxes  from  the HST-WFPC2  images,  using  the
software   package   LMORPHO   \citep{ode96},   which   we   list   in
\tab{tblflux}.  In \fig{origmagdist}, we show the distributions of the
$V$- and  $I$-band apparent magnitudes.   While the statistics  may be
relatively    low,    these     distributions    clearly    peak    at
$V\!\simeq\!22$~mag  and $I\!\simeq\!26$~mag,  which is  in  each plot
several magnitudes brighter than the typical field depth.  These peaks
are similar to  the what was observed in  the LBDS \citep{win84b}, and
its  Hercules  subfield  \citep{wad00},  which are  only  complete  to
$V\!\simeq\!22.7$~mag  and  $I\!\simeq\!21$~mag, respectively.   Since
our compendium of HST-WFPC2  fields are generally $\sim\!2$~mag deeper
than the  optical imaging  in the LBDS,  these peaks in  the magnitude
distributions are  more secure than  reported in those  earlier works,
which  were based  on  ground-based photographic  or  CCD imaging  and
poorer   seeing.   Furthermore,   in  the   microjansky   flux  range,
\citet{ham95a}   find  similar   peaks  in   the   apparent  magnitude
distributions in the Canada-France Redshift Survey (CFRS).

\begin{figure}
\epsscale{1.0}
\plotone{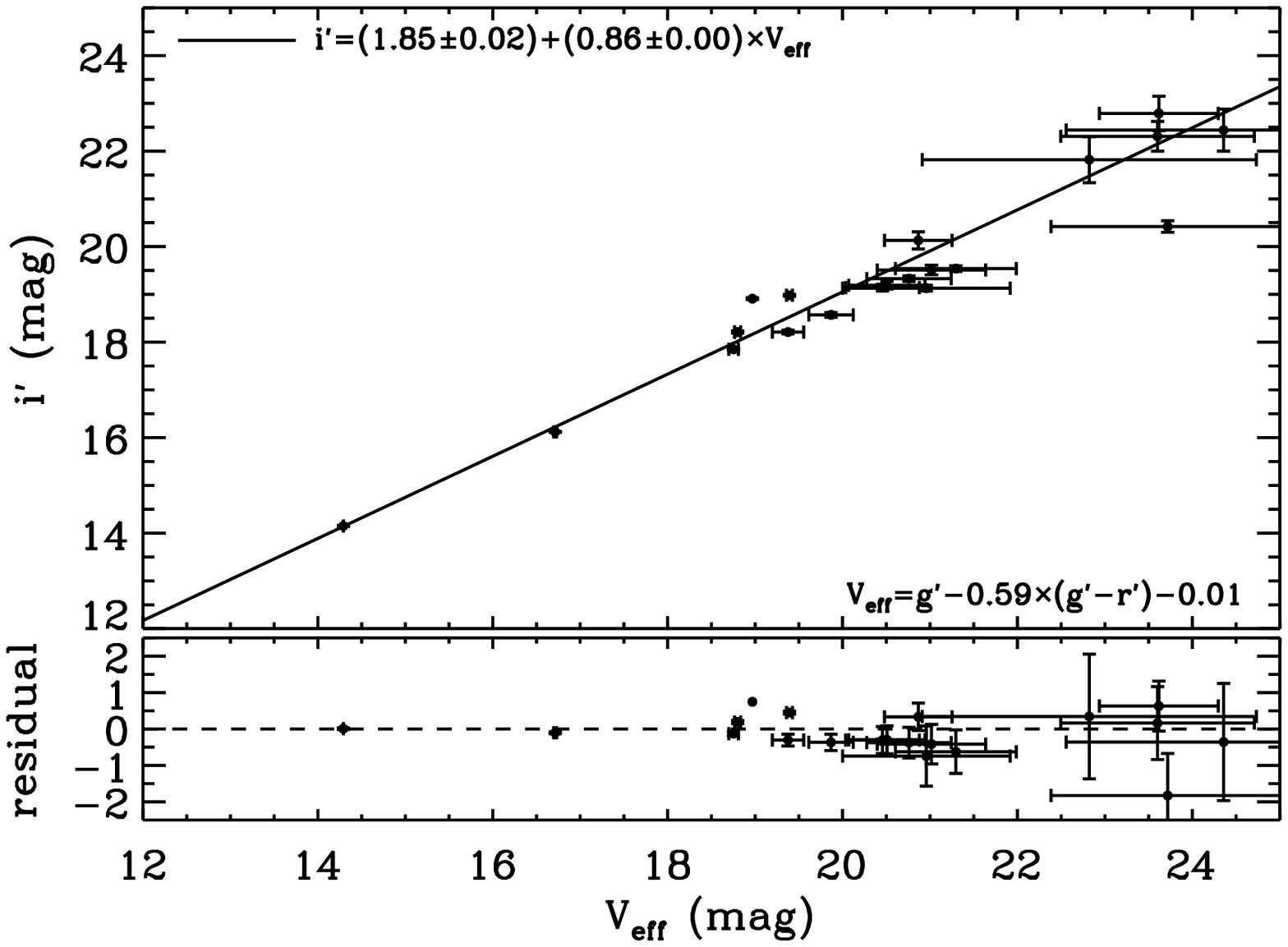}
\caption{Colors of  FIRST radio galaxies in our  sample from SDSS-DR5.
The {\it effective} $V$-band is constructed following \citet{jes05} as
a combination of the SDSS  $g'$- and $r'$-bandpasses, which we give in
the lower-right corner of the  upper panel.  The solid line represents
the  best-fit  linear  model  to  the  data, which  is  given  in  the
upper-left  corner of  the upper  panel.   The lower  panel shows  the
residuals, as the observations minus the model.  The average color for
these  radio  galaxies from  the  SDSS  database is  $\left<(i'-V_{\rm
eff})\right>\!\simeq\!1.8$~mag.}\label{sdss1}
\end{figure}

\begin{figure}
\epsscale{1.0}
\plotone{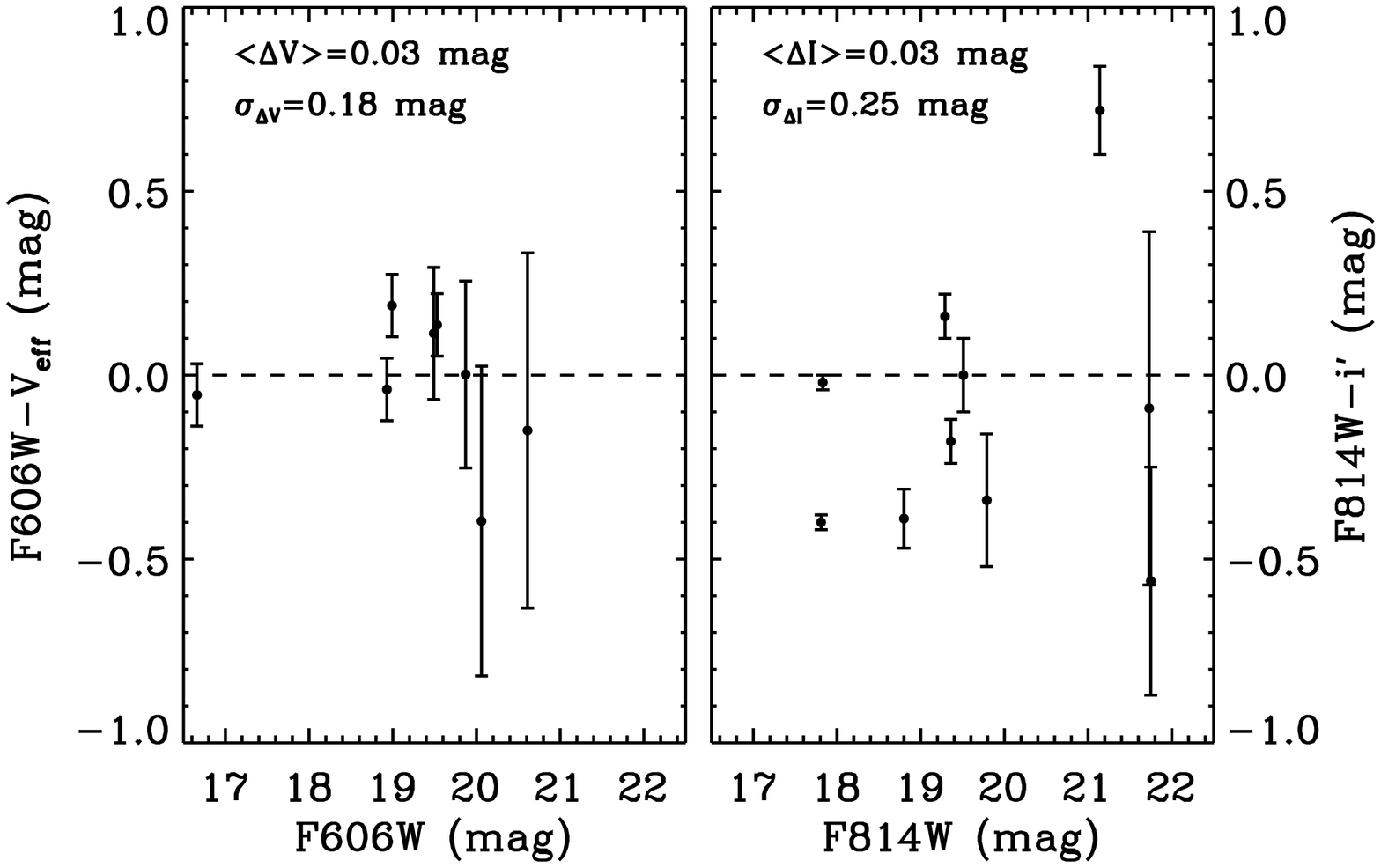}
\caption{Comparison  of   the  fluxes  from   SDSS-DR5  and  HST-WFPC2
observations.  These  panels show  the ratio between  the HST  and the
SDSS fluxes as a function of the HST flux.  The average difference and
standard deviation  are listed in  the lower-right of each  panel.  In
the left panel,  we show the $({\rm F606W}-V_{\rm  eff})$ residual for
the  objects which were  only observed  in the  $V$-band and  had SDSS
data.  Since  the average  difference is only  $-0.08\pm0.79$~mag, the
conversion  from  $g'$  and  $r'$  to $V_{\rm  eff}$,  and  hence  the
correlation shown in \fig{sdss1} is relatively robust. The right panel
shows that  the once  the $V$-band measurements  are converted  to the
$i'$-band, there  is relatively little additional  error introduced by
assuming that $i'\!\approx\!I$ for these radio galaxies.}\label{sdss2}
\end{figure}

Since for  a given object only  HST/WFPC2 $V$- or  $I$-band imaging is
available, we require an algorithm to convert the observed flux in one
bandpass to what is expected in the other.  Therefore, we retrieve the
$g'$-, $r'$-,  and $i'$-fluxes for  22/43 galaxies available  from the
Sloan Digital  Sky Survey, Data  Release~5 \citep[SDSS-DR5;][]{ade07}.
The $g'$- and $r'$-band fluxes  are converted to an effective $V$-band
for  a typical  galaxy  spectral energy  distribution (SED)  following
\citet{jes05}:
\begin{equation}\label{vtrans}
V_{\rm eff}=g'-0.59\times (g'-r')-0.01~{\rm (mag)}.
\end{equation}
In \fig{sdss1}, we  show the SDSS $i'$-band flux as  a function of the
effective  $V$-band  flux, as  synthesized  from  the  SDSS $g'$-  and
$r'$-band  observations  and the  best-fit,  linear  model  to fit  of
$i'\!=\!(1.85\pm0.02)+(0.86\pm0.00)\times  V_{\rm  eff}$.   To  verify
these conversions from $g'$ and $r'$ to $V_{\rm eff}$ and from $i'$ to
$I$,   we  show  $\Delta   V\!\equiv\!(V-V_{\rm  eff})$   and  $\Delta
I\!\equiv\!(I-i')$ as  a function  the respective HST  observations in
\fig{sdss2}.  For  galaxies which only  have $V$-band WFPC2  data, the
difference   between  the   HST-WFPC2   and  the   effective  $V$   is
$\left<\Delta  V\right>\!=\!0.03\pm0.18$~mag, while  the corresponding
difference       in      the      $I$-band       is      $\left<\Delta
I\right>\!=\!0.03\pm0.25$~mag.  This indicates that the correlation in
\fig{sdss1} can  be used to systematically convert  the WFPC2 $V$-band
fluxes  to  corresponding  $I$-band  fluxes  with  an  uncertainty  of
$\lesssim\!0.3$~mag.  This  uncertainty is not large  enough to affect
any of our conclusions in a significant way.

\begin{figure}
\epsscale{1.0}
\plotone{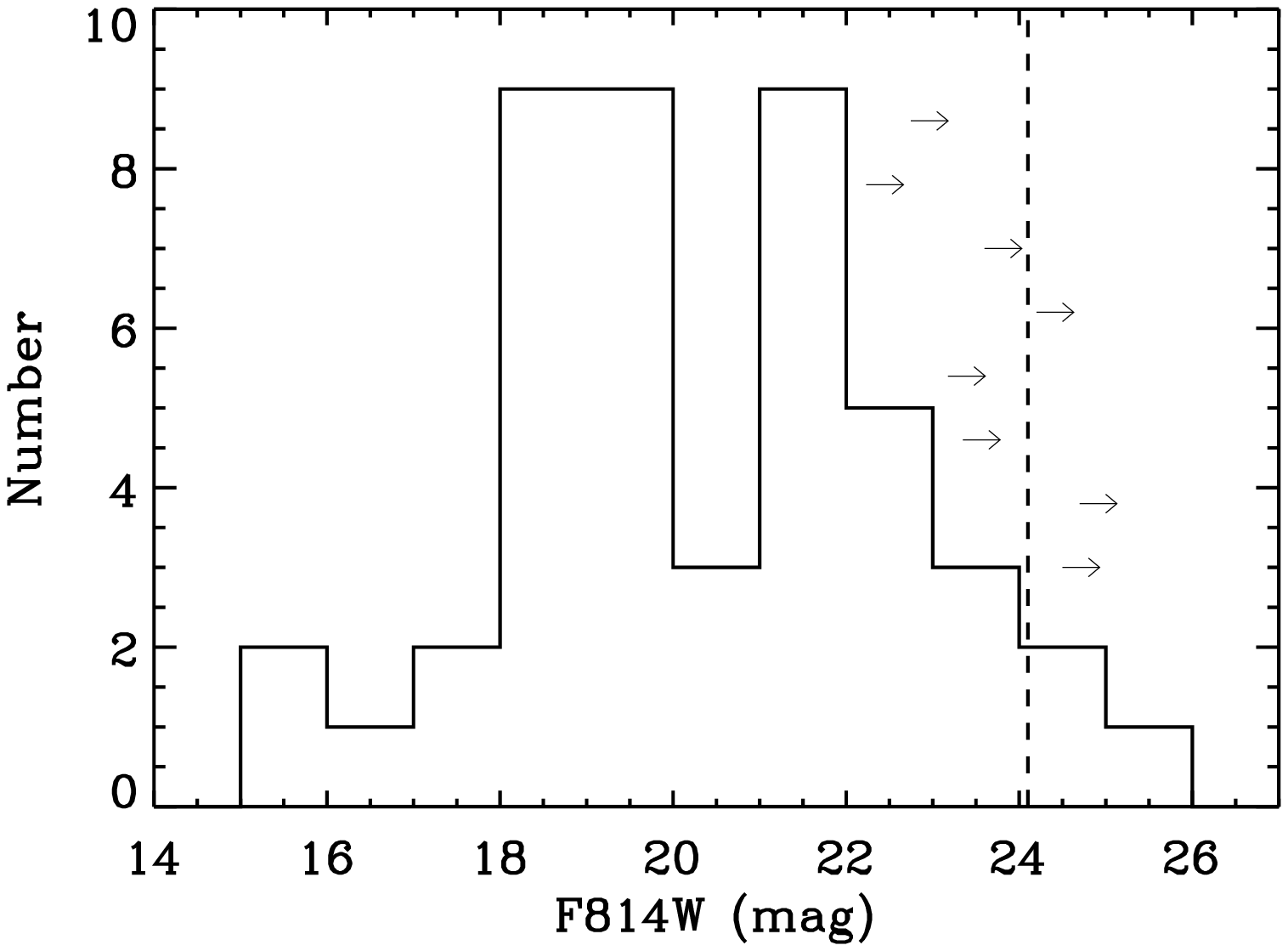}
\caption{$I$-band  magnitude  distribution  for all  identified  radio
galaxies.  The fluxes of  sources where only $V$-band observations are
available  are converted  assuming  the SDSS-DR5  colors described  in
\S~\ref{mdist}, \fig{sdss1}, and \fig{sdss2}. The vertical dashed line
shows the  flux-averaged, 50\% completeness  depth of the  the fields,
while the  arrows represent  the depths of  the eight fields  where no
optical counterpart  was identified to their  respective field limits.
This peak at $I\!\simeq\!\alli$~mag  is nearly 3~mag brighter than the
depth  of the  typical WFPC2  field.  The  source counts  of  a random
population of  field galaxies  should increase as  a power-law  to the
completeness  limit of the  field, contrary  to what  is seen  for the
radio selected galaxies.  }\label{cuml}
\end{figure}

With  these   converted  fluxes,   we  show  the   $I$-band  magnitude
distribution for the 43 identified optical galaxies in \fig{cuml}.  To
determine  the location  of  the peak,  we  fit a  simple Gaussian  to
magnitude  distribution.    The  peak  at   $I\!\simeq\!\alli$~mag  is
significantly brighter ($\sim\!3$~mag)  than the completeness limit of
our typical WFPC2  field, which suggests that this peak  is not due to
incompleteness.  This peak may  reflect the redshift distribution, and
possibly strong cosmological evolution of the radio source population.
\citet{wad00} point out that if  the radio galaxies had the same space
density at  all redshifts, then we  expect the number  of galaxies per
magnitude interval  would increase toward the  observational limit, as
is roughly  the case  for the optical  field galaxy counts.   With the
exceptional resolution  and increased depth of  the HST-WFPC2 dataset,
we    extend   the    typical   limit    of    $AB\!\sim\!22$~mag   to
$AB\!\sim\!24$~mag for  the millijansky population  over a significant
field-of-view.   This magnitude distribution  is quite  different than
what  is  expected  from   a  general  population  of  field  galaxies
\citep{tys88}.   The  radio-selected  galaxy  counts  likely  peak  at
$I\!\simeq\!\alli$~mag since the  majority of millijansky sources have
$L^*$-type luminosities at $z\!\sim\!0.8$ \citep{con89}.

\subsection{Radio-to-Optical Spectral Index} \label{index}
\begin{figure}
\epsscale{1.0}
\plotone{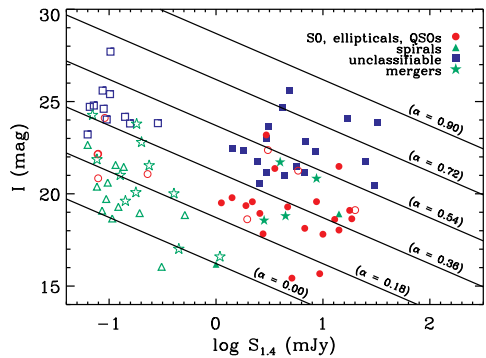}
\caption{Integrated radio flux versus total $I$-band magnitude for the
complete FIRST+WFPC2 sample.  The  $V$-band flux has been converted to
an  effective $I$-band  as described  in \S~\ref{mdist}.   Objects are
separated into  early, late,  merger, and unclassifiable  types, while
the  filled and  open symbols  are from  this work  and \citet{afo06},
respectively.   The lines  are contours  of  constant radio-to-optical
flux  ratios as defined  by \eqn{alphadef}.   In the  millijansky flux
range, the  majority of the galaxies have  early-type morphologies and
shallow     spectral    slopes     of    $\alpha_{ro}\!\lesssim\!0.4$.
Additionally, there  is only one late-type spiral  and 4~mergers which
span      a       large      range      of       spectral      indices
($0\!\lesssim\!\alpha_{ro}\!\lesssim\!0.5$).    However,  lower  radio
fluxes ($S_{1.4}\!\ll\!1$~mJy), there  are very few early-type systems
where    late-type   systems    dominate   the    counts    with   the
$\alpha_{ro}\!\lesssim\!0.4$.    This  dramatic   change   in  optical
morphologies at $S_{1.4}\!\simeq\!1$~mJy may be related to a change in
the source of the radio emission, such as primarily starburst galaxies
($S_{1.4}\!\lesssim\!1$~mJy)        versus        primarily        AGN
($S_{1.4}\!\gtrsim\!1$~mJy).      Additionally,     the    line     of
$\alpha_{ro}\!\simeq\!0.18$ marks a fiducial point, below which nearly
all galaxies are late-type.}\label{alpha}
\end{figure}

\citet{kro85}   show   that   the  radio-to-optical   spectral   index
($\alpha_{ro}$)  can  be used  to  distinguish  between  red and  blue
galaxies.  Therefore, we perform a similar analysis with morphological
types determined from the high-resolution imaging.  In \fig{alpha}, we
show  the $I$-band  flux  as a  function  of the  radio  flux for  the
43~optically  identified galaxies.  Objects  are separated  into three
classes by  their optical morphology (discussed  in \S~\ref{morph}) of
early,  late,   and  unclassifiable  as  filled   red  circles,  green
triangles, and blue squares,  respectively.  We show lines of constant
radio-to-optical spectral index defined by:
\begin{eqnarray}\label{alphadef}
\alpha_{ro}&=&\frac{\log{(S_{1.4}/S_{I})}}{\log{(20.4~{\rm cm}/8012~{\rm \AA})}}\\
&=&0.185  \times\log{(S_{1.4}/S_{I})},
\end{eqnarray}
where      $\log{(S_I/1{\rm      Jy})}\!=\!-0.4\times(I-8.72)$     mag
\citep{oke83}.   The solid  lines denote  logarithmic radio-to-optical
flux ratios  from 1  to 10$^5$.  In  general, the  elliptical galaxies
(filled red circles) have $I\!\lesssim\!20$~mag and shallower spectral
indices    of     $\alpha_{ro}\!\lesssim\!0.4$.     Conversely,    the
unclassifiable     galaxies     (filled     blue     squares)     have
$\alpha_{ro}\!\gtrsim\!0.4$,   which  is   primarily   due  to   their
brightnesses   ($I\!\gtrsim\!21$~mag).    Furthermore,   since   these
galaxies  span the  entire  range  of radio  fluxes,  they are  likely
representative of the general radio sample. Their lower optical fluxes
could be  caused by  an increased dust  obscuration associated  with a
surrounding  starburst and/or  enhanced/beamed  radio emission.   Deep
infrared  observations  would  be  capable  distinguishing  these  two
scenarios,  and likely  detect the  8/43 optically  unidentified radio
galaxies.

In the GOODS-S field  \citep{gia04}, \citet{afo06} publish the optical
and  radio brightnesses  for their  sample of  43 optically-identified
microjansky  radio  galaxies  with morphological  classifications.   In
\fig{alpha}, we show their microjansky radio galaxies as open symbols,
where late-type and star-forming  galaxies generally constitute a much
larger  fraction of  radio  galaxies than  at  the millijansky  level.
Moreover,     these      late-type     galaxies     generally     have
$\alpha_{ro}\!\lesssim\!0.2$, which  is a region  of ($\alpha_{ro}-I$)
parameter  space  where  few  early-type  galaxies are  found  in  our
millijansky sample.   For galaxies  sufficiently resolved and  given a
reliable  optical  classification, spirals  and  mergers dominate  the
radio  source counts  at  the microjansky  level  (25/43, 53\%)  while
early-type  galaxies  are  the  dominant type  at  millijansky  fluxes
(19/25, 76\%).

\subsection{Optically Unidentified Radio Sources (Unids)} \label{unids} 
\begin{figure}
\epsscale{1.0}
\plotone{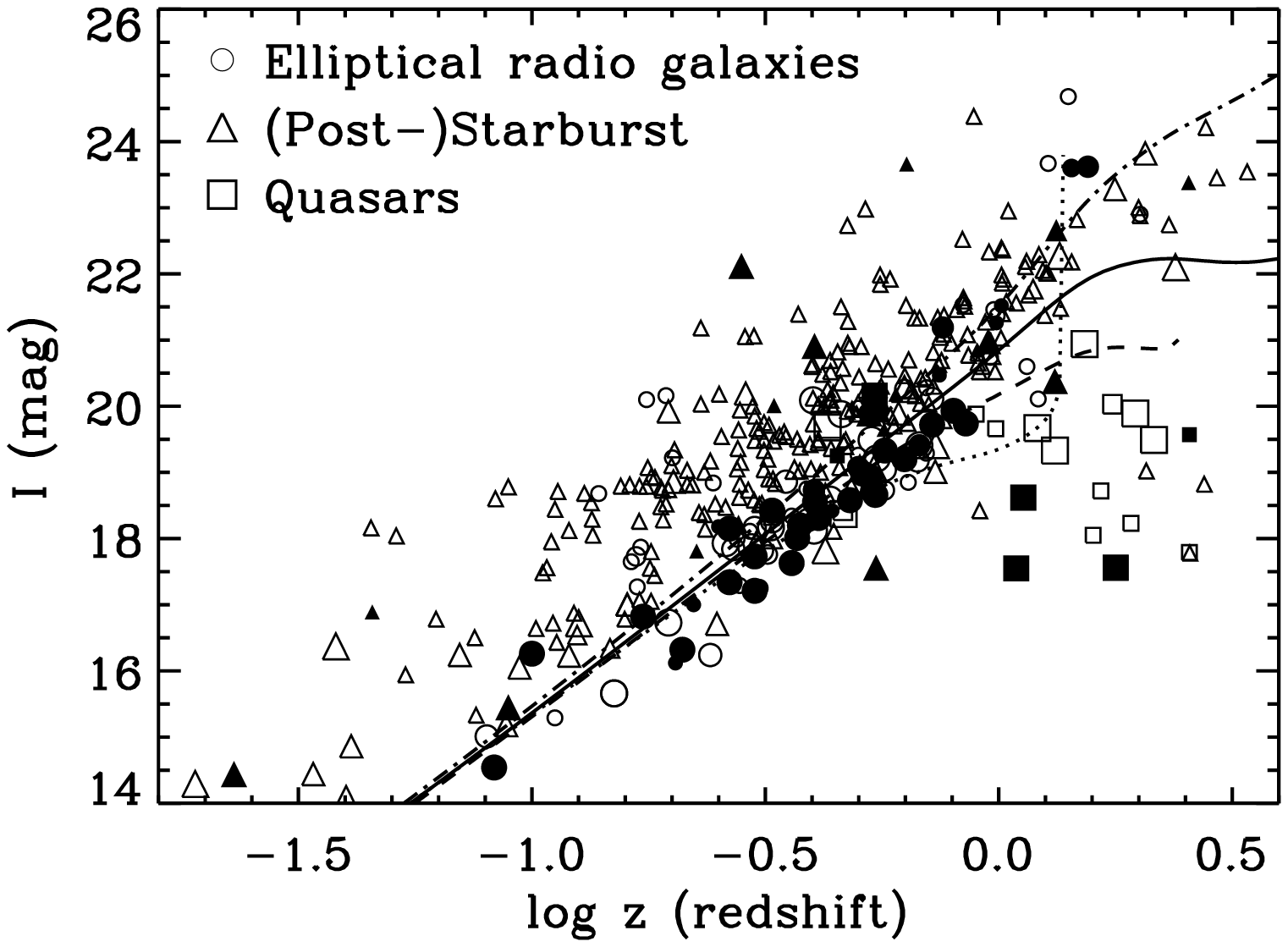}
\caption{The  Hubble  diagram  for  faint  radio  galaxies.   Data  is
compiled               from              numerous              sources
\citep{win84b,kro85,win85,fom97,ric98,wad01,roc02,cha03}.           The
circles, triangles, and squares represent ellipticals, starbursts, and
quasars, respectively.  The open symbols represent point-like or small
radio  sources,  while  filled  symbols  show  (classical)  double  or
extended   radio  sources;   the  size   of  the   plotted   point  is
logarithmically    proportional    to     the    radio    flux    from
$10-10^5$~microjansky  \citep{win03}.   The  lines  are  from  stellar
population synthesis models of  \citep{bru03}.  The dot-dashed line is
a constant  star-formation model and  solid, dashed, and  dotted lines
are  an exponential  star-forming history  with  ages of  13, 11,  and
9~Gyr,  respectively.  The  $e$-folding timescale  in  the exponential
models   is  1.3~Gyr.    All   models  assume   a   stellar  mass   of
10$^{11}$~M$_{\odot}$,  a   Salpeter  IMF  \citep{sal55},   and  solar
metallicity.   The  obvious  trend   in  this  diagram  suggests  that
early-type radio  galaxies are roughly standard candles,  and that the
unidentified millijansky objects  with $I\!\gtrsim\!22$~mag are likely
at $z\!\gtrsim\!1$.  NOTE: The  $I$-band magnitudes in this figure are
in  the Vega  system,  when necessary  we  assume $(I_{\rm  AB}-I_{\rm
Vega})\!\simeq\!0.48$~mag.}\label{hubble}
\end{figure}

For  the sample  of 51  FIRST  sources, eight  were found  to have  no
optical  counterpart  identifiable  to $AB\!\sim\!23.6-25.0$~mag  (see
\tab{tblflux} for  50\% depths of  each field from the  Snigula method
described in \S~\ref{imprep}).   Like the unclassifiable galaxies, the
unidentified objects  have radio fluxes which span  a comparable range
to  the  entire radio  sample  (see  \tab{unidtab})  and thus  may  be
representative  of  the   general  population.   However,  unlike  the
unclassifiables, their lack of any optical flux may indicate that they
are are  considerably higher redshift.   In \fig{hubble}, we  show the
Hubble diagram ($I$-band flux as a function of redshift) for all faint
radio   galaxies,  where   the   size  of   the   plotted  symbol   is
logarithmically  proportional to the  radio flux  \citep[see][for more
details]{win03}.  From \fig{hubble}, millijansky radio sources with an
optical  counterpart  with a  brightness  of $I\!\gtrsim\!22$~mag  are
likely  at  $z\!\gtrsim\!1$  \citep[eg.][]{dragons},  indicating  that
these  galaxies may  be  standard candles.   However, the  significant
spread  in redshift  for galaxies  with  $I\!\gtrsim\!22$~mag suggests
than an additional effect may  be present, which could be an increased
optical extinction, different stellar populations at high redshift, or
significant luminosity evolution.   Without a robust optical detection
of  these galaxies, we  can only  speculate about  their counterparts.
Based  on the  Hubble diagram,  these optically  unidentified galaxies
could     be     similar    to     the     distant    red     galaxies
\citep[DRGs;][]{fs04,vd06,g06,dragons}  at  $z\!\simeq\!2$.  The  DRGs
are generally old, dusty, and  massive galaxies with very red infrared
colors      $(I-J)\!\gtrsim\!2$~mag,     $(J-H)\!\sim\!2$~mag,     and
$(H-K)\!\sim\!1$~mag.  Therefore, near-infrared observations are ideal
for identifying the optical/infrared counterpart.

\vspace*{24pt}
\section{Comments on Individual Radio Sources}\label{comment}
We give brief  comments on each of the 51 radio  sources in the sample
and their  most likely optical identification  (where applicable).  We
note the  overlap with previous  works where available found  from the
NASA/IPAC Extragalactic  Database\footnote{NED is operated  by the Jet
Propulsion  Laboratory,  California  Institute  of  Technology,  under
contract with the National Aeronautics and Space Administration.}.

\begin{enumerate}
\item{{\bf    J002219.1$-$013030}:   Unclassifiable.     The   optical
counterpart is  entirely visible  despite being near  the edge  of the
WFPC2 field.   The object is also  within $18\arcsec$ of  a NVSS radio
source \citep{con98}.}
\item{{\bf   J004322.3$-$001343}:  Point-source.    This   object  was
identified as a quasar in the SDSS \citep{ric04}.}
\item{{\bf   J012616.6$-$012126}:  Late-type,   barred   spiral  (type
SBb-SBc).   This  object   is  a  bright  ($V\!\simeq\!19.8$~mag)  and
well-resolved mid-late type spiral galaxy.}
\item{{\bf J030237.6$+$000818}: Unidentified with $I\!\geq\!24.5$~mag.
There are four CFRS objects within $30\arcsec$ \citep{ham95b}.}
\item{{\bf  J030249.5$+$000615}: Elliptical.   There is  no noticeable
optical elongation to indicate any alignment of the radio source within
the galaxy.  \citet{ham95b} classified this source as an E0.}
\item{{\bf J082820.6$+$344321}:  Unclassifiable. The optical candidate
is very faint ($I\!\simeq\!25.6$~mag).}
\item{{\bf        J082828.4$+$344131}:        Unidentified        with
$I\!\geq\!24.7$~mag.}
\item{{\bf J084715.5$+$443752}:  Unclassifiable.  The radio  source is
clearly extended and was found by \citet{win85}.}
\item{{\bf J084849.5$+$445550}:  Elliptical. There is  no evidence for
any optical elongation with the  resolved radio source.  The object is
located in  a group or poor  cluster, and the shape  of radio contours
were  used to  determine  the identification.   This  object was  also
identified by \citet{oor87}.}
\item{{\bf J091205.2$+$350506}: Unclassifiable.  The optical source is
extended and very faint  ($V\!\simeq\!23.8$~mag), and is near a bright
star.}
\item{{\bf    J091251.0$+$525928}:   Unclassifiable.     The   optical
counterpart is  small, and possibly extended.  The  radio counters are
clearly elongated.}
\item{{\bf  J094926.6$+$295941}: Elliptical.   The  optical and  radio
emission is not aligned.}
\item{{\bf  J094930.7$+$295938}:  S0.   The  radio source  is  clearly
double-lobed and the optical  identification is slightly off the radio
axis, the  radio contours  appear to be  somewhat ``bent''  toward the
identified object.}
\item{{\bf  J100354.5$+$285911}: Merger.  The radio  source  is likely
double-lobed  with the  merging system  located at  the center  of the
radio axis.}
\item{{\bf  J102437.2$+$470312}:  Unclassifiable.  The optical object 
is marginally extended.}
\item{{\bf J102744.6$+$282921}:  Unclassifiable.  This object  is very
faint ($I\!\simeq\!22.4$~mag) and on  the outskirts of a bright spiral
galaxy, which  is $\sim\!7\arcsec$ to the west.   Therefore, the spiral
galaxy  center is likely  {\it not}  the optical  identification.  The
radio    source   may    be   caused    by   a    supernova   remnant
\citep[eg.][]{win84b,kro85}  or  a  background  galaxy, which  may  be
somewhat extincted by the foreground spiral galaxy.}
\item{{\bf J103452.3$+$394704}: Unclassifiable.  The optical source is
extended, likely a galaxy in a group.}
\item{{\bf  J104630.8$-$001213}:  Merger.    The  merger  pair  is  an
elliptical and  early-type disk  galaxy, however neither  show obvious
signs  of interaction.   The optical  candidate was  in the  sample of
\citet{gla95}.}
\item{{\bf J104757.0$+$123835}: Elliptical. The radio source is fairly
round and was identified by \citet{con98}.}
\item{{\bf   J111908.6$+$211917}:  Point   source.  This   object  was
identified as a QSO by \citet{bar01}.}
\item{{\bf J112520.7$+$420425}: Elliptical$/$S0.   The radio source is
round and was identified by \citet{con98}.}
\item{{\bf J114526.3$+$193301}:  Unclassifiable.}
\item{{\bf J114910.5$-$002313}:  Unclassifiable.}
\item{{\bf   J114928.3$+$143844}:   Elliptical.    This   galaxy   was
identified by \citet{con98}.}
\item{{\bf J115642.8$+$022451}:  Elliptical.  The optical  emission is
roughly  perpendicular  to  the   radio  contours.   This  object  was
identified by \citet{con98}.}
\item{{\bf J120326.4$+$443635}: Unidentified with $V\!\geq\!25.0$~mag.}
\item{{\bf J121026.6$+$392909}:  Point source.  It is  located in group
and was classified as a QSO \citep{hew93}.}
\item{{\bf J121658.4$+$375439}: Unclassifiable.  The optical source is
marginally  extended.  The radio  source may  be double-lobed,  but of
very unequal lobe flux.}
\item{{\bf  J121705.5$-$031137}: Unclassifiable.  The radio  source is
resolved.}
\item{{\bf J121707.7$-$031127}:  Elliptical.  The optical  emission is
round,  while  the radio  counters  are  elongated.   This object  was
classified as an AGN in the SDSS \citep{ric04}.}
\item{{\bf        J121839.7$+$295325}:        Unidentified        with
$V\!\gtrsim\!24.8$~mag.  The  morphology of the  nearby optical source
suggests that  it could be gravitationally lensed  by the unidentified
radio galaxy \citep{ryan08b}.}
\item{{\bf J122331.0$+$155245}: Unclassifiable.  This radio source was
 identified by \citet{con98}.}
\item{{\bf  J122624.4$+$173228}: Elliptical.  This  galaxy may  have a
companion to the east-north-east, and is in a small group.  This radio
galaxy was identified by \citet{con98}.}
\item{{\bf J125029.2$+$302527}:  Elliptical. This galaxy is like in a 
cluster or group.  The radio source has round inner contours, but appears 
to be resolved on larger scales.}
\item{{\bf J125635.3$+$215632}: Unidentified with $I\!\geq\!24.2$~mag.
There are three relatively bright objects within $\sim4\arcsec$ of the
radio source. The radio source was identified by \citet{con98}.}
\item{{\bf J125650.0$+$220630}:  Unclassifiable. The radio  source was
identified by \citet{con98}.}
\item{{\bf J131223.6$+$424517}:  Elliptical.}
\item{{\bf  J131617.8$+$420239}: Late-type  barred spiral  (SBc). This
object  is $\gtrsim\!10\arcsec$ in  diameter and  is also  an infrared
source,   identified    in   the    Two   Micron   All    Sky   Survey
\citep[2MASS;][]{twomass}.}
\item{{\bf J134219.9$-$005816}:  Unclassifiable.} 
\item{{\bf J140019.9$+$044421}: Elliptical.   The object is located in
the  center  of  a  small  group.  The  radio  contours  are  slightly
elongated, while the optical emission is fairly round.}
\item{{\bf  J143530.0$+$484534}: Merger. The  optical candidate  is on
innermost radio contour.}
\item{{\bf J150109.7$+$225111}: Unidentified with $I\!\geq\!23.6$~mag.
There is a disk galaxy $\simeq\!3\arcsec$ from the radio position that
could be  the optical  counterpart.  Since the  astrometric correction
for  this  field was  small,  the nearby  disk  galaxy  is likely  not
related,  owing  to  the  large  separation  and  the  relatively  low
background object density.}
\item{{\bf        J151057.4$+$312722}:        Unidentified        with
$V\!\geq\!23.7$~mag.}
\item{{\bf J153955.0$+$342013}:  Elliptical.  The optical  emission is
fairly elongated, while the radio contours are round.  This system may
have several companion systems to the south and west.}
\item{{\bf  J155938.7$+$473309}: Unclassifiable.  This  object may  be
located in a  dense group of faint, small  objects.  This radio source
was identified by \citet{con98}.}
\item{{\bf  J163141.4$+$373603}: Elliptical.   The  optical and  radio
emission appear to be aligned. The radio emission is part of a larger,
low SB complex.}
\item{{\bf    J163233.8$+$190550}:    Unclassifiable.   The    optical
counterpart  is the  northern of  the two  small objects.   This radio
source was  identified by \citet{con98}.}
\item{{\bf J164911.4$+$305226}: Unidentified with $V\!\geq\!24.3$~mag.
There is  a late type  galaxy approximately $4\arcsec$ from  the radio
position,  but given that  the astrometric  corrections are  less than
$2\arcsec$, and  the $LR\!\sim\!0.0$, this is  considered an unidentified
source.}
\item{{\bf J171901.1$+$480458}:  Elliptical.  This radio  source found
in \citet{con98}.}
\item{{\bf  J172025.4$+$480331}: Unclassifiable.  This  object may  be
marginally extended.}
\item{{\bf J172232.9$+$501232}: Merger.  The primary galaxy is clearly
a  late-type,  possibly  barred  spiral  which shows  clear  signs  of
interaction.   The secondary  galaxy is  considerably smaller  and may
also be a late-type disk system.}
\end{enumerate}

\section{Summary and Discussion}\label{sum}

We have determined the optical morphology and structure from HST-WFPC2
imaging of a randomly-selected  sample of 51~millijansky radio sources
from the FIRST survey.   Earlier studies of millijansky radio galaxies
have often  distinguished early-  versus late-types based  on observed
optical colors  or spectra.  However,  we have extended such  works by
providing optical  classifications based  on HST imaging,  and confirm
that elliptical  galaxies constitute  the majority of  the millijansky
source population.

We find that the optical  flux distribution of these galaxies peaks at
$I\!\simeq\!\alli$~mag, which is consistent with evolutionary model of
\citet{con89}: faint radio  galaxies generally have $L^*$-type optical
luminosities  with  a  median  redshift  of  $z\!\simeq\!0.8$  and  an
effective  redshift decline  for $z\!\gtrsim\!1.5$.   At  low redshift
($z\!\lesssim\!1$),     most    radio     sources     brighter    than
$S_{1.4}\!\gtrsim\!1$~mJy are generally  found in massive ellipticals,
and  there   is  a  clear   deficit  of  these  massive   galaxies  at
$z\!\gtrsim\!3$   \citep[eg.][]{dri98}.    \citet{del06}  argue   that
$\sim\!50$\%  of the  stars which  will  likely end  up in  elliptical
galaxies in the  local Universe are formed at  $z\!\simeq\!3$, and are
not finally  assembled into  the galaxy until  $z\!\simeq\!0.8$.  This
down-sizing  picture  requires an  increased  major  merger rate  from
$z\!\sim\!0$ to  $z\!\sim\!3$, which has  been observed in  the Hubble
Ultra  Deep  Field  \citep[eg.][]{ryan08}.   Finally,  this  effective
redshift  cut-off may  arise, since  $\sim\!1$~Gyr may  be  on average
required to trigger an active  nucleus after the major merger based on
hydrodynamical models \citep{spi05}.  This  must be further studied in
the  critical redshift  range of  $1\!\lesssim\!z\!\lesssim\!3$, which
may  prove  useful  in  tracing  the high-redshift  evolution  of  the
supermassive black hole--bulge mass relation \citep[eg.][]{fer05}.  At
these  redshifts, the  optical  counterparts will  be extremely  faint
(${\rm  AB}\!\gtrsim\!25$~mag), and may  have a  high dust  content or
their luminosity  function may  evolves strongly or  episodically with
redshift.   Consequently, dedicated observations  with the  Wide Field
Camera~3 (WFC3) or the {\it James Webb Space Telescope} will be needed
for  the  continued  study  of  the population  of  millijansky  radio
sources.

\acknowledgements  The  authors  would  like to  thank  the  anonymous
referee and Rolf Jansen for  their constructive input and Tom Keck for
his efforts. This work was supported  in part by the NASA Space Grant,
NSF  grant AST-9802963,  and NASA  grants AR-8357.01A  and AR-8768.01A
from STScI under NASA contract NAS5-26555.

\end{document}